# $[C_2mim][CH_3SO_3]$ – A Suitable New Heat Transfer Fluid? Part 2. Thermophysical Properties of Its Mixtures with Water

Francisco E.B. Bioucas[a,§], Carla S.G.P. Queirós[a], Daniel Lozano-Martín[b], M.S. Ferreira[a,#], Xavier Paredes[a], Ângela F. Santos[a], Fernando J.V. Santos[a], Manuel L.M. Lopes[a], Isabel M. S. Lampreia[a], Maria José V. Lourenço[a], Carlos A. Nieto de Castro[a,]* and Klemens Massonne[,¶]

[a]Centro de Química Estrutural, Institute of Molecular Sciences, Faculdade de Ciências, Universidade de Lisboa, Campo Grande, 1749-016 Lisboa, Portugal

[b]Grupo de Termodinámica y Calibración (TERMOCAL), Research Institute on Bioeconomy, Escuela de Ingenierías Industriales, Universidad de Valladolid, Paseo del Cauce, 59, E-47011 Valladolid, Spain

[c]BASF SE, GC/I - M300, 67056 Ludwigshafen, Germany

*Corresponding author: cacastro@ciencias.ulisboa.pt

[§] Present address: Department of Chemical and Biological Engineering, Lehrstuhl für Advanced Optical Technologies, Thermophysical Properties, FAU, Erlangen.
E-mail: francisco.bioucas@fau.de
[#] E-mail: marianasferreira9@gmail.com
[¶] Formerly at BASF, Ludwigshafen, Germany

## Abstract

Ionic liquids have proved to be excellent heat transfer fluids and alternatives to common HTF's used in industry for heat exchangers and other heat transfer equipment. However, its industrial utilization depends on the cost/kg of its production, to be competitive for industrial applications with biphenyl and diphenyl oxide, alkylated aromatics and dimethyl polysiloxane oils, which degrade above 200ºC and possess some environmental problems. The efficiency of a heat transfer fluid depends on the fundamental thermophysical properties influencing convective heat transfer (density, heat capacity, thermal conductivity, and viscosity), as these properties are necessary to calculate the heat transfer coefficients for different heat exchanger geometries.

In part 1, the thermophysical properties of pure 1-ethyl-3-methylimidazolium methanesulfonate [$C_2$mim][$CH_3SO_3$] (CAS no. 145022-45-3), (ECOENG™ 110), produced by BASF, under the trade name of Basionics® ST35, with an assay ≥ 97% with ≤ 0.5% water and ≤ 2% chloride ($Cl^-$), were presented, for temperatures slightly below room temperature and up to 355 K.

In this paper we report the thermophysical properties of mixtures of [$C_2$mim][$CH_3SO_3$] with water, in the whole concentration range, at $P$ = 0.1 MPa. The properties measured were density and speed of sound (293.15 < $T$/K < 343.15), viscosity, electrical, thermal conductivity and refractive index (293.15 < $T$/K < 353.15), and infinite dilution diffusion coefficient of the ionic liquid in water (298.15 K).

Properties for the mixture like the isobaric expansion coefficient, the isentropic compressibility, apparent molar volumes, apparent molar isentropic compressions, and the thermodynamic excess properties, like excess molar volume, excess molar isobaric expansion and isentropic compression, excess viscosity, molar refraction, and infinite dilution diffusion coefficients of the cation, the anion and other common anions in ionic liquids were obtained, within this temperature range. The validity of the Walden relation for this ionic liquid was also determined. This amount of experimental information, in addition to recent molecular simulation studies available in the literature permit a clear picture of the structure of these binary mixtures, the influence of composition and temperature, paving the way to a technological discussion of the possible application of these mixtures as new heat transfer fluids or battery electrolytes.

## 1. INTRODUCTION

The thermophysical properties of ionic liquids and their water mixtures are very important from a scientific and industrial points of view. In fact, scientifically, they can be used for a better understanding of the interactions between the cation and anion of the ionic liquid with a hydrogen bonding molecule like water, their influence in the structure of the mixture, the predominance of free ions, ion clusters or aggregates, hydration and percolation. Application wise, it is fundamental to understand the effect of the properties in the heat transfer coefficients necessary for the design and operation of chemical plants which use them[1], analyzing how we can influence the choice of a given heat transfer fluid, by optimizing its composition and operational ranges.

In Part I paper[2] we reported the thermophysical properties of one of the possible new heat transfer fluids, 1-ethyl-3-methylimidazolium methanesulfonate [$C_2$mim][$CH_3SO_3$] (CAS Nr. 145022-45-3), (ECOENG™ 110), produced by BASF, under the trade name of Basionics® ST35, and a brief study of its toxicological properties was also presented.

ST35 has significant properties that make it useful for many applications, not only as a heat transfer fluid. The study of this fluid, raised, as explained therein, the question: What is a good alternative heat transfer fluid, capable of replacing environmentally unfriendly compounds today used in the chemical and allied industries?

These new fluids must be better than the ones currently in use. In what sense they should be better may be highly dependent on the application. In some cases, all that matters is an improved heat transfer efficiency; in other cases, superior compatibility with current equipment (in the form of lower corrosion and greater compatibility with metals and sealants/gaskets), leading to lower maintenance downtime and reduced costs, might be preferred; yet in other cases, priority could be given to biodegradability and non-toxicity, or good thermal stability for example. For many applications, the right compromise between all these characteristics and of course price, is the answer. In addition, the use of an ionic liquid, their mixtures with water and possible use of the dispersion of nanomaterial particles to enhance thermal properties, like thermal conductivity, requires that the ionic liquid must be biodegradable and nontoxic[3].

Water is, by far, the best heat transfer fluid, in many engineering applications. However, its corrosive behavior at moderate and high temperatures, led engineers to look for other heat transfer fluids, either by replacing water completely, or by decreasing their corrosive capabilities, adding other chemical compounds as additives/lubricants.

It is the purpose of this paper to report the thermophysical properties of the mixtures of [$C_2$mim][$CH_3SO_3$] with water, in the whole concentration range, at $P$ = 0.1 MPa. The properties measured were density and speed of sound (293.15 < $T$/K < 343.15), viscosity, electrical, thermal conductivity and refractive index (293.15 < $T$/K < 353.15), and infinite dilution diffusion coefficient of the ionic liquid in water (298.15 K). Additionally, the structure of the mixtures will be discussed, using our experimental results and the data available based on correlation spectroscopy and computer simulations for this and similar mixtures of hydrophilic ionic liquids with water, like, $[C_2mim][C_2H_5SO_4]$[4], and $[C_4mim][BF_4]$[5,6] and $[C_2mim][CH_3COO]$[7,8] .

This study will be complemented with heat transfer studies in pilot heat exchangers, to decide about its suitability as a new and more efficient heat transfer fluid, the ionic liquid possibly also promoting better lubrication and protection/pickling effect on the surface of the metallic plates (usually stainless steel) of the heat exchanger. Preliminary results show that selected $[C_2mim][CH_3SO_3]$ + water mixtures have thermophysical properties that contributes to energy and capital savings in cylindrical tube heat exchangers, results that will be presented in Paper 3 of this series, as their use in low temperature applications is highly promising.

## 2. EXPERIMENTAL SECTION

2.1. **Materials**. $[C_2mim][CH_3SO_3]$ used in this study, with a CAS no. 145022-45-3, was obtained from BASF, under the trade name of Basionics® ST35, with an assay ≥ 97 wt%, with ≤0.5 wt% water and chloride ($Cl^-$) ≤ 2 wt%, as described in Bioucas et al. (2018)[2], and supported by $^1$HNMR and IR spectra. Water used for solutions was ion exchanged ultrapure water (resistivity, 18 MΩ cm) from a Milli-Q water purification system. The electrical conductivity of pure water used to prepare the samples was monitored and found to be smaller than $4x10^{-4}$ $S{\cdot}m^{-1}$, possibly to slight contamination during handling. However, the contribution of other ions to these measurements was assumed negligible, not affecting the measured properties. For temperatures above 323.15 K, the mobility of the ions increases, the measurements discussed in section 3.3 showing that it can be as great as $2.3x10^{-3}$ $S{\cdot}m^{-1}$, nevertheless much smaller than the uncertainty of the data.

2.2. **Sample Preparation**. As reported in Part 1, ST35 MSDS[9], shows the conditions to be obeyed on handling this ionic liquid, accordingly to hygiene and safety practice, namely that contact with the skin, eyes, and clothing must be avoided. Handling in the laboratory was done under protected atmosphere. For the preparation of the mixtures, samples of ST35 taken form the barrel were dried in vacuum for more than 24h. Their water content was measured with Coulometric Karl Fischer titration (Metrohm 831), and the mass was measured using a Kern AEJ balance with an accuracy of $1\times10^{-5}$ g. All the samples were prepared under a dried flux of nitrogen. The samples used for the different property measurements, namely as a starting point for mixtures preparation, are displayed in Table 1. For the measurement of viscosity, electrical conductivity and refractive index, a stream of dry nitrogen was used to minimize the absorption of water, as these properties are strongly sensitive to the amount of water present in the samples[2,8]. All mixtures were prepared by weight, taking into account the water content of the "pure" ST35, and the relative combined standard uncertainty in the mixtures composition, expressed in mole fraction, taking into account the water content of [$C_2$mim][$CH_3SO_3$] previously reported[8,10] , was found to be $u_{r,c}(x_{IL})$ = 0.002. Buoyancy corrections were made. Please note that in terms of mass fractions, a 50/50 mass composition corresponds to a $x_w$ ~ 0.9.

### 2.3. **Methods of Property Measurement**

*2.3.1. Density and speed of sound*. Density, $\rho$, and speed of sound, $c$, measurements at temperatures for 293.15 < $T$/K < 343.15 were simultaneously carried out with an *Anton Paar* density and sound velocity meter DSA 5000 M, with an automatic temperature control within ±0.001 K and a standard uncertainty $u(T)$ = 0.01 K. Ultrasound transducer experimental frequency was close to 3 MHz. The densimeter cell has been calibrated in the temperature interval used with high quality Millipore® water (resistivity 18 MΩ·cm) and tetrachloroethylene (H&D Fitzgerald, density certified, (https://density.co.uk/products/liquid-standards/), and a new calibration methodology previously proposed by Lampreia and Nieto de Castro (2011) was used[11]. Due to the high viscosity of the mixtures, the density values were corrected for this extra damping, as recommended by the densimeter manufacturers. Calibration of speed of sound cell was made at 293.15 K with Milli-Q water, according to the manufacturer recommendation. Accuracy has been assessed by comparison with literature values of speed of sound in toluene and cyclohexane at 298.15 K, as previously described[8,10,12]. Degassed ultrapure water at $T$ = 293.15 K were regularly measured to check the equipment. As described by Nobre et al. (2020)[13], before measuring the speed of sound of any mixture, the speed of sound of water at each temperature, $c_{0,exp}$, has been measured, at the operating frequency of the instrument, 3 MHz. As these values do not always reproduced the reference value $c_{0,ref}$ at each temperature, differences ($c_{0,exp}$ - $c_{0,ref}$) were estimated and $c_{0,ref}$ was added to that difference, thus eliminating any systematic deviation[13,14]. The true $c$ value (reported in Table 2) is then $c\ (T,\ x_w) = (c_{exp} - c_{0,exp}) + c_{0,ref}\ (T,\ x_w)$. The possible existence of ultrasonic absorption in the measurement of the speed of sound of ionic liquids with Anton Paar DSA 5000 sound velocity meter was analysed, due to the strong absorption in the low frequency range (less than 10 MHz) discussed by Dzida et al (2017)[15]. The experience in this laboratory with amphiphilic compounds in water has shown in fact that reproducibility between measurements could not be achieved at this frequency for such water mixtures. However, any sign of similar phenomena with ionic liquids was never noticed, for temperatures greater than 293 K [8,10,16]. Following the results obtained for [$C_2$mim][$CH_3SO_3$] by Musial et al (2019)[17], we have compared all the deviations of the available experimental data form the correlations proposed by Bioucas et al (2018)[2] in Fig. S7. This figure shows that there is a very good agreement with the data of Marium et al (2017) [18] (± 1.5 $m\cdot s^{-1}$; ± 0.06%), and those of Musial et al (2019)[17] deviate around -1.3 $m\cdot s^{-1}$ (-0.07%), within the mutual uncertainty of data. Data of Sing et al (2014)[19] have greater deviations, between 41 and 25 $m\cdot s^{-1}$ (≈ 2 %). This agreement with the data of Musial et al (2019)[17] let us to consider $c$ as a pure thermodynamic property, as discussed by Rowlinson and Swinton[20]. If this valid for the pure ionic liquid, the stronger absorber, it can be generalized for the mixture, were less molecules of [$C_2$mim][$CH_3SO_3$] are present. The reference density and speed of sound values for water were taken from the data in references[21,22], respectively. Standard uncertainties of density and speed of sound measurements are then found to be $u$ ($\rho$) = 0.30 kg·m$^{-3}$ and $u$ ($c$) = 0.5 m·s$^{-1}$, respectively. The expanded global uncertainty was found to be $U$ ($\rho$) = 0.6 kg·m$^{-3}$ and $U$ ($c$) = 1 m·s$^{-1}$.

*2.3.2 Viscosity.* Kinematic viscosity, $v$, was measured between for 293.15 K < $T$/K < 353.15 K, at $P$ = 0.1 MPa, under a stream of dry nitrogen, using Ubbelohde micro viscometers of SCHOTT Instruments GmbH (538 10 I, 538 13 IC, 538 20 II, 538 23 IIC), and were calibrated by a step-up procedure based on water satisfying ISO3696[23] requisites for grade 1, using kinematic viscosity values from ISO/TR3666[24]. The measuring procedure was previously described in reference[2]. For the measurement of the flow time, a ViscoClock unit from Schott-Gerätte with an uncertainty of ± 0.01s was used. Additional contributions for the uncertainty in the viscosity measurements were considered due to the step-up procedure for the higher viscosities and higher temperatures. Standard uncertainty of the temperature, pressure and viscosity measurement were, respectively, $u(T)$ = 0.01 K, $u(P)$ = 1kPa and $u_r$ ($\eta$) = 0.6 %. The expanded global relative uncertainty of viscosity was found to be $U_r$ ($\eta$) = 1.2 % ($k$ = 2).

**Table 1 - Water content in [$C_2$mim][$CH_3SO_3$] samples used for mixture preparation[a]. Confidence intervals are reported standard uncertainties**

| **Sample** | $w_w$ **/%** | $w_w$ **/%**[b] |
|---|---|---|
| Density | 0.0715 ± 0.0068 | NA |
| Speed of sound | 0.0715 ± 0.0068 | NA |
| Kinematic viscosity | 0.597 ± 0.026 | NA |
| Electrical Conductivity | 0.615 ± 0.026 | NA |
| Thermal Conductivity | 0.422 ± 0.028 | NA |
| Refractive Index[c] | 0.997 ± 0.098[d] | 1.945 ± 0.328[d] |
| Infinite dilution diffusion coefficient | 0.253 ± 0.042 | NA |

[a] Mole fraction of water in the samples varies from 0.0081 (0.0715 %) to 0.1851 (1.945 %)

[b] After measurements

[c] Also used for pure [$C_2$mim][$CH_3SO_3$] refractive index determination. An averaged value of 1.499 % was used for water free values determination

[d] Average for three different samples

*2.3.3 Electrical conductivity.* The electrical conductivity measurements were made with a 5-pole (SI-Analytics, LF 913 T) conductivity cell calibrated using two certified standards from Radiometer Analytical (KCl 1D and KCl 0.1D) at 298.15 K and 323.15 K. The conductance and capacitance were measured as a function of frequency between 250 Hz to 4 kHz using an impedance analyser (HP 4129A LF), the apparatus previously described[2,8,25,26], for 293.15 < $T$/K < 353.15 and 11 frequencies between 250 Hz and 4 kHz, at each temperature, measured with a Pt100 with an accuracy of 0.01 K, as described previoiusly[2]. For each frequency 5 measurements were made after allowing some time for frequency stabilization of the impedance analyzer internal generator. For frequencies above 1000 Hz it was verified that there was no change in the response function with frequency, as found for the pure ionic liquid[2], so it can be assumed that only the ion conduction part is involved to respond to the applied electrical field. Values for the frequency of 3000 Hz were chosen and used for the current work. The electrical conductivity expanded global relative uncertainty, $U_r$ ($\kappa$), was found to be better than 2 % ($k = 2$).

*2.3.4 Thermal conductivity*. Thermal conductivity, $\lambda$, measurements were performed at temperatures 293.15 < $T$/K < 353.15, for all composition range, using commercial equipment from Hukseflux Thermal Sensors TPSYS02 and a Non-Steady-State Probe (NSSP) TP08 with and estimated uncertainty by the manufacturers of ± (3% + 0.02) $W \cdot m^{-1} \cdot K^{-1}$ and a temperature accuracy of $u(T) = 0.02$ K, as described by Bioucas et al.[2]. The correct operation of this instrument was assessed by measurements with water, an IUPAC standard reference liquid, as fully described in our recent publication, Lozano et al.[27]. Data obtained for the thermal conductivity of water (using a MilliQ® sample) between 283.6 K and 344.8 K did not deviate by more than 0.8% from the IUPAC SRD correlation[28], which proved the correct operation of the probe. However, and as discussed before[27], for temperatures above 323.15 K the probe had to be removed and cleaned and the sample had to be equilibrated for longer times to reduce convection currents. In addition, to detect the possible onset of convection, detailed measurement with water around room temperature were performed, with three different power inputs in the transient hot-wire probe, giving temperature rises between 0.25 K and 0.70 K. No curvature in the straight-line fits $\Delta T_{\mathrm{rise}}$ vs. ln $t$, were found, and no power dependence existed. These results also support that there is no convection in these measurements and that the working model for the transient hot-wire is valid, capable of producing results at room temperature within 0.55% standard deviation, confirming our previous statements[2] of $U_r(\lambda) = 2$ %, at a95 % confidence level ($k = 2$). As the viscosity of [$C_2$mim][$CH_3SO_3$] at room temperature is twenty five times greater than that of water at 343.15 K, it is expected that no convection exists on the measurements reported here for the whole concentration range, as already discussed also for [$C_6$mim][($CF_3SO_2)_2$N][29].

*2.3.5 Refractive index*. Refractive index, $n^D$, for sodium D line (wavelength 589.26 nm) of [$C_2$mim][$CH_3SO_3$] and its mixtures with water were measured at atmospheric pressure under a small nitrogen flush ($P$ varying between 99.93 and 100.97 kPa, relative humidity, $H$ around 65%), at temperatures for 293.15 <$T$/K< 353.15 K, with a Anton Parr (Abbemat 500) refractometer. The refractometer was checked according to the procedure described by Queirós et al. (2020)[8], and the experimental values were corrected by the calibration. These corrections, at the highest temperature do not amount by more than – 0.0004 RI units, a value greater than data uncertainty. Temperature was measured with an uncertainty of 0.03 K and the expanded uncertainty in the refractive index is $U$ ($n$) = 0.0002, at 95% confidence level ($k = 2$), except for two points at the higher temperatures, for which $U$ ($n$) = 0.0005, due to experimental fluctuations.

*2.3.6 Infinite dilution coefficient in water*. The present measurements were obtained with the Taylor dispersion technique[30], applied to measure the diffusion coefficients in many systems[31,32,33,34]. Recent application to ionic liquids[8,35] showed that the present equipment can only be applied to infinite dilution measurements in water and other solvents, due to the high viscosities of the ionic liquids and the design constraints of the equipment. A number of 25 replicate measurements for the diffusion coefficient were determined for this system, at 298.15 K, at infinite dilution, with different flow rates. Temperature was measured with an uncertainty of 0.02 K. The relative expanded uncertainty, at a 95% confidence level, for 20 experimental measurements ($k$ = 2.086), was found to be $U_r$ ($D$) = 4.8 %.

## 2. RESULTS AND DISCUSSION

**3.1. Density, speed of sound and derived properties.** The experimental density, $\rho$, and speed of sound, $c$, at atmospheric pressure, for the aqueous [$C_2$mim][$CH_3SO_3$] system, at the seven temperatures, are summarized in Table 2 and shown in Fig. 1 and Fig. 2. Density varies monotonically with the mole fraction of [$C_2$mim][$CH_3SO_3$], having first a plateau, where the density of the mixture is almost insensitive to water content, up to $x_w \sim 0.3$, then decreasing very fast up to pure water. The isobaric thermal expansion coefficient, at a given composition, is always positive, as density decreases with the increase of the temperature in all the range of mole fraction studied.

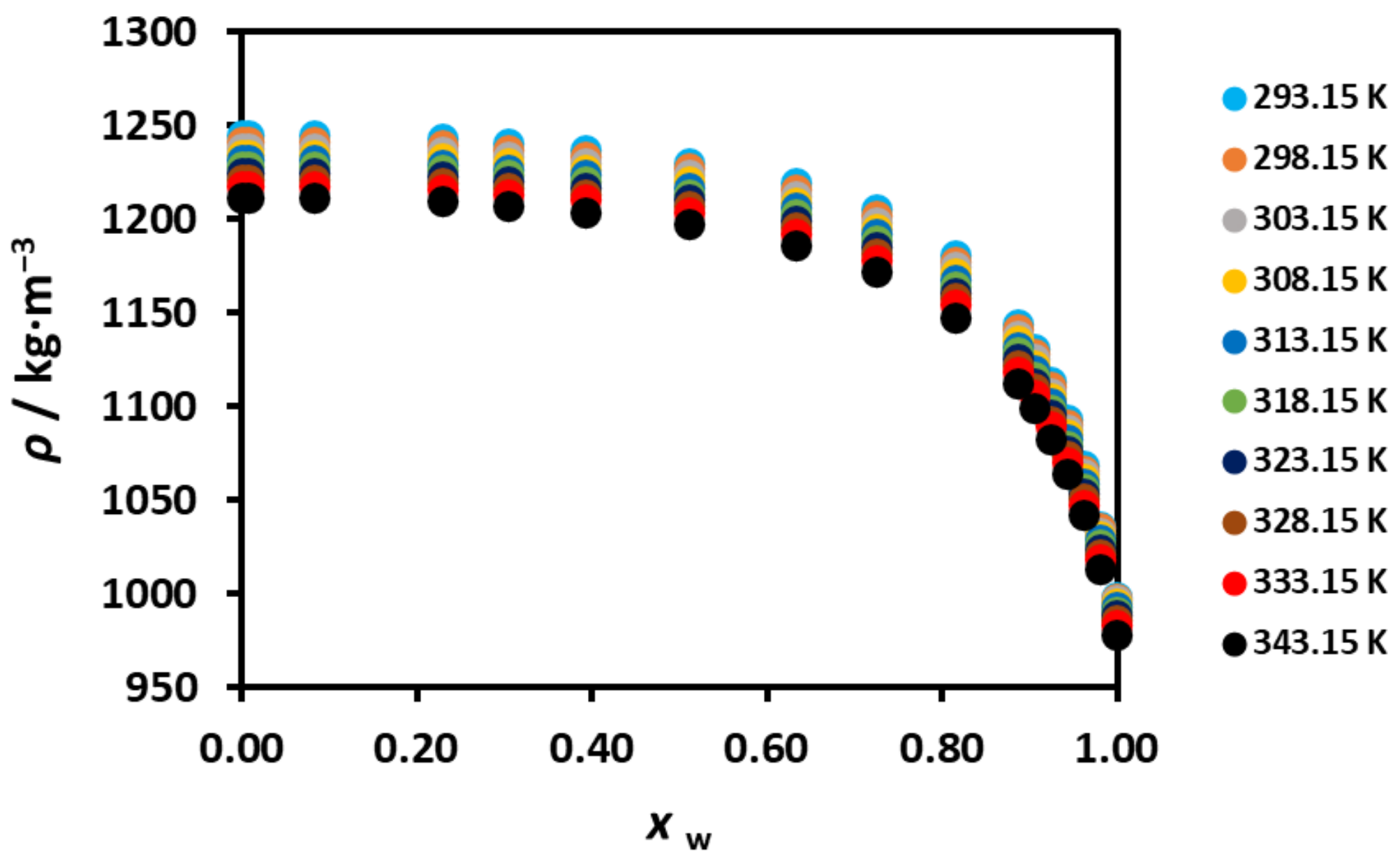

Figure 1 - Density, $\rho$, for the aqueous [$C_2$mim][$CH_3SO_3$] system, at studied temperatures, as a function of $x_w$, the mole fraction of water.

Extrapolated water-free values were obtained, starting from the initial water content of our "pure" liquid samples ($x_w$ = 0.0082), fitting the density as a quadratic function of the mole fraction of the ionic liquid, for the five compositions closer to $x_{IL}$ = 1, and calculating $\lim_{x_w \to 0} \left(\frac{\partial \rho}{\partial x_w}\right)_T$ from the analytical equation found for each temperature. The procedure is fully described in the viscosity section. The extrapolated water-free values of the density of the ionic liquid are also presented in the last line of Table 2, and it can be seen that the corrections are very small, a result of the less sensitivity of density of [$C_2$mim][$CH_3SO_3$] to water impurities[36]. These values were fitted to quadratic polynomials as a function of temperature, as done previously[2] for the initial dried sample ($x_w$ = 0.0082), with an uncertainty of 0.033 kg·m$^{-3}$, at a 95% confidence level ($k$ = 2), no point deviating from the fit by more than 0.016 kg·m$^{-3}$. These new values agree with the values previously reported, for a sample with $x_w$ = 0.0042[2] to within 0.76 kg·m$^{-3}$ at 293.15 K, decreasing to 0.23 kg·m$^{-3}$ at 343.15 K, slightly higher than the mutual uncertainty of the data (0.28 kg·m$^{-3}$), namely at the lower temperatures. No explanation was found for this discrepancy

The speed of sound has a different behavior than density, having a smooth increase as the mole fraction of water increases, passing through a maximum around $x_w$ ~ 0.82, and decreasing sharply, with a sigmoid shape. As expected, the speed of sound decreases as the temperature increases, since at higher temperatures the density of the medium is lower, and the molecules of water and ionic liquid are further apart. The maximum possibly corresponds to a particular state of maximum interaction between the ionic liquid cation and anions and the water molecules. It is interesting, as already noted for the aqueous systems of [$C_4$mpyr][$N(CN)_2$] and [$C_2$mim][$CH_3COO$], that there is an inversion of the temperature coefficient of the speed of sound, at very dilute IL mixtures ($x_w$ ~ 0.975, in this case)[8,10], a very important sign of the change of the structure of the mixtures by the addition of water molecules, which we will discuss below.

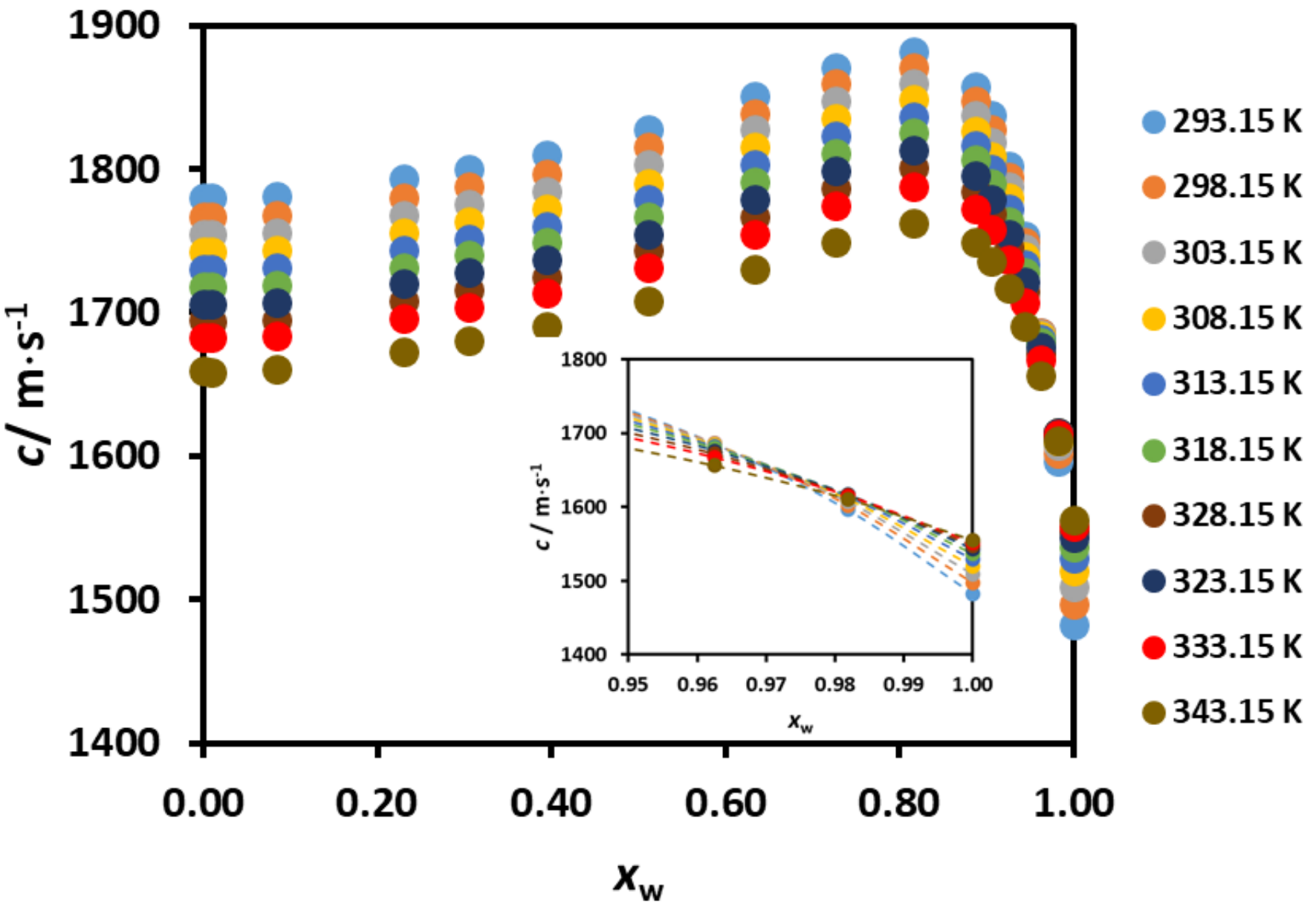

Figure 2 – Speed of sound, $c$, for the aqueous [$C_2$mim][$CH_3SO_3$] system, at the studied temperatures, as a function of $x_w$. The inset shows the temperature inversion at $x_w$ ~ 0.975.

Extrapolated water-free values for the speed of sound were obtained, starting from the initial water content of our "pure" liquid samples ($x_{\mathrm{w}}$ = 0.082), fitting the speed of sound as a quadratic function of the mole fraction of the ionic liquid for the first 5 mole fractions, and calculating $\lim_{x_w \to 0} \left(\frac{\partial c}{\partial x_w}\right)_T$.

Values for the extrapolated water-free values of speed of sound of the ionic liquid presented in the last line of Table 2, and it can be seen that the corrections amount to -0.3 $\mathrm{m \cdot s^{-1}}$, or 0.017%, also a result of the low sensitivity of the speed of sound of $[C_2mim][CH_3SO_3]$ to water impurities[37]. These values were fitted to quadratic polynomials as done previously[2] for a dried sample ($x_{\mathrm{w}}$ = 0.042), with an uncertainty of 0.6 $\mathrm{m \cdot s^{-1}}$, at 95% confidence level ($k$ = 2), no point deviating from the fit by more than 0.13%. These new values agree with the values previously reported[2] to within -0.020%, well within the mutual uncertainty of the data. Comparison of the previous values with all the available data for the speed of sound of the pure ionic liquid has been presented elsewhere[2] and will not be repeated here.

**Table 2 – Density, $\rho$, and speed of sound, $c$, for [$C_2$mim][$CH_3SO_3$] + water, in the temperature range $T$ = (293.15 to 343.15) K[a] and $P$ = 0.1 MPa[a], as a function of the molar fraction of water, $x_w$. Water values[29].**

| $x_w$[b] | $T$ = 293.15 K | | $T$ = 298.15 K | | $T$ = 303.15 K | | $T$ = 308.15 K | | $T$ = 313.15 K | | $T$ = 318.15 K | | $T$ = 323.15 K | | $T$ = 328.15 K | | $T$ = 333.15 K | | $T$ = 343.15 K | |
|---|---|---|---|---|---|---|---|---|---|---|---|---|---|---|---|---|---|---|---|---|
| | $\rho$[c] /kg·m$^{-3}$ | $c$[d] /m·s$^{-1}$ | $\rho$ /kg·m$^{-3}$ | $c$ /m·s$^{-1}$ | $\rho$ /kg·m$^{-3}$ | $c$ /m·s$^{-1}$ | $\rho$ /kg·m$^{-3}$ | $c$ /m·s$^{-1}$ | $\rho$ /kg·m$^{-3}$ | $c$ /m·s$^{-1}$ | $\rho$ /kg·m$^{-3}$ | $c$ /m·s$^{-1}$ | $\rho$ /kg·m$^{-3}$ | $c$ /m·s$^{-1}$ | $\rho$ /kg·m$^{-3}$ | $c$ /m·s$^{-1}$ | $\rho$ /kg·m$^{-3}$ | $c$ /m·s$^{-1}$ | $\rho$ /kg·m$^{-3}$ | $c$ /m·s$^{-1}$ |
| 1.0000 | 998.2 | 1482.7 | 997.0 | 1497.0 | 995.6 | 1509.4 | 994.0 | 1520.1 | 992.2 | 1529.2 | 990.2 | 1536.7 | 988.0 | 1542.9 | 985.7 | 1547.7 | 983.2 | 1551.3 | 977.8 | 1555.1 |
| 0.9819 | 1035.9 | 1596.2 | 1034.3 | 1602.4 | 1032.4 | 1607.4 | 1030.3 | 1611.3 | 1028.2 | 1614.1 | 1025.9 | 1615.9 | 1023.4 | 1616.5 | 1020.8 | 1616.2 | 1018.1 | 1615.2 | 1012.4 | 1610.8 |
| 0.9625 | 1067.8 | 1686.1 | 1065.6 | 1686.1 | 1063.3 | 1685.5 | 1060.9 | 1684.1 | 1058.5 | 1682.2 | 1055.8 | 1679.5 | 1053.1 | 1676.1 | 1050.3 | 1672.1 | 1047.4 | 1667.4 | 1041.3 | 1656.3 |
| 0.9440 | 1092.2 | 1753.1 | 1089.7 | 1749.0 | 1087.1 | 1744.4 | 1084.5 | 1739.2 | 1081.7 | 1733.8 | 1078.9 | 1727.8 | 1076.0 | 1721.3 | 1073.0 | 1714.4 | 1070.0 | 1707.0 | 1063.6 | 1690.9 |
| 0.9251 | 1112.5 | 1801.7 | 1109.8 | 1794.6 | 1107.0 | 1787.3 | 1104.1 | 1779.7 | 1101.2 | 1771.7 | 1098.2 | 1763.5 | 1095.1 | 1754.8 | 1092.0 | 1745.8 | 1088.8 | 1736.4 | 1082.3 | 1716.8 |
| 0.9055 | 1130.6 | 1837.1 | 1127.7 | 1828.1 | 1124.7 | 1818.7 | 1121.7 | 1809.3 | 1118.6 | 1799.6 | 1115.5 | 1789.7 | 1112.3 | 1779.5 | 1109.1 | 1769.0 | 1105.8 | 1758.3 | 1099.1 | 1736.2 |
| 0.8870 | 1143.9 | 1857.6 | 1140.9 | 1847.4 | 1137.8 | 1837.3 | 1134.7 | 1827.0 | 1131.5 | 1816.6 | 1128.3 | 1806.2 | 1125.1 | 1795.3 | 1121.8 | 1783.9 | 1118.4 | 1772.4 | 1111.6 | 1748.7 |
| 0.8167 | 1180.5 | 1882.3 | 1177.3 | 1870.9 | 1173.9 | 1859.5 | 1170.6 | 1848.4 | 1167.3 | 1836.9 | 1164.0 | 1825.5 | 1160.6 | 1813.2 | 1157.2 | 1800.7 | 1153.8 | 1788.1 | 1146.9 | 1762.7 |
| 0.7259 | 1205.0 | 1871.1 | 1201.7 | 1859.1 | 1198.3 | 1847.2 | 1195.0 | 1835.3 | 1191.7 | 1823.4 | 1188.3 | 1811.2 | 1184.9 | 1799.0 | 1181.6 | 1786.7 | 1178.2 | 1774.2 | 1171.4 | 1749.3 |
| 0.6329 | 1218.9 | 1850.8 | 1215.6 | 1839.0 | 1212.2 | 1827.1 | 1208.9 | 1815.2 | 1205.6 | 1803.3 | 1202.2 | 1791.3 | 1198.9 | 1779.2 | 1195.6 | 1767.1 | 1192.2 | 1754.9 | 1185.5 | 1730.3 |
| 0.5110 | 1230.0 | 1827.5 | 1226.7 | 1815.1 | 1223.3 | 1802.7 | 1220.0 | 1790.4 | 1216.7 | 1778.5 | 1213.3 | 1766.7 | 1210.0 | 1754.8 | 1206.7 | 1743.0 | 1203.4 | 1731.1 | 1196.8 | 1707.6 |
| 0.3936 | 1236.6 | 1809.5 | 1233.3 | 1797.0 | 1229.9 | 1784.7 | 1226.5 | 1772.5 | 1223.2 | 1760.4 | 1219.9 | 1748.6 | 1216.6 | 1736.7 | 1213.3 | 1725.0 | 1209.9 | 1713.3 | 1203.4 | 1690.0 |
| 0.3049 | 1240.0 | 1800.2 | 1236.6 | 1787.9 | 1233.2 | 1775.8 | 1229.9 | 1763.8 | 1226.6 | 1751.7 | 1223.2 | 1739.8 | 1219.9 | 1727.9 | 1216.6 | 1716.1 | 1213.3 | 1704.2 | 1206.7 | 1680.7 |
| 0.2292 | 1242.5 | 1793.0 | 1239.2 | 1780.2 | 1235.8 | 1767.9 | 1232.4 | 1755.8 | 1229.1 | 1743.9 | 1225.7 | 1731.9 | 1222.4 | 1720.0 | 1219.1 | 1707.9 | 1215.8 | 1696.1 | 1209.2 | 1672.8 |
| 0.0843 | 1244.5 | 1780.8 | 1241.2 | 1767.6 | 1237.8 | 1755.3 | 1234.4 | 1743.1 | 1231.1 | 1731.0 | 1227.8 | 1719.1 | 1224.4 | 1707.2 | 1221.1 | 1695.4 | 1217.8 | 1683.7 | 1211.2 | 1660.7 |
| 0.0082 | 1244.5 | 1779.6 | 1241.2 | 1766.9 | 1237.8 | 1754.4 | 1234.4 | 1742.1 | 1231.1 | 1730.0 | 1227.8 | 1717.9 | 1224.4 | 1705.9 | 1221.1 | 1694.1 | 1217.8 | 1682.3 | 1211.3 | 1659.0 |
| 0.0000[e] | 1244.4 | 1779.3 | 1241.1 | 1766.6 | 1237.8 | 1754.1 | 1234.4 | 1741.8 | 1231.0 | 1729.6 | 1227.7 | 1717.6 | 1224.4 | 1705.6 | 1221.1 | 1693.8 | 1217.8 | 1682.0 | 1211.2 | 1658.6 |

[a] Standard uncertainties, $u(T)$ = 0.01 K, $u(P)$ = 0.002 MPa

[b] Relative combined standard uncertainty, $u_{r,c}(x_w)$ = 0.002

[c] Expanded global uncertainty, $U(\rho)$ = 0.60 kg·m$^{-3}$, at a 95 % confidence level ($k$=2)

[d] Expanded global uncertainty, $U(c)$ = 1 m·s$^{-1}$, %, at a 95 % confidence level ($k$=2)

[e] Extrapolated water-free value

Data on density and speed of sound permit the calculation of other thermodynamic coefficients, like the isobaric thermal expansion coefficient, $\alpha_{\mathrm{P}}$ ($T$, $x$) and the isentropic compressibility $\kappa_{\mathrm{S}}$ ($T$, $x$), given by the relations:

$$\alpha_{\mathrm{P}}(T,x) = -\frac{1}{\rho(T,x)}\left(\frac{\partial\rho(T,x)}{\partial T}\right)_{P,x}$$
$$\kappa_{\mathrm{S,x}}(T,x) = \frac{1}{\rho(T,x)c(T,x)^2} \qquad (1)$$

Values of these coefficients were calculated for the mixture from density and speed of sound data obtained in this work, and are displayed, respectively, in Figs. 3 and 4, and Table S1. With respect to the thermal expansion, it is constant with temperature up to $x_{\mathrm{w}} \sim 0.5$ (view inset in Fig 3.), but discrimination arises for the low ionic liquid content mixtures, as we approach pure water ($x_{\mathrm{w}} > 0.82$), where a change in structure occurs. In the case of the isentropic compressibility, Fig. 4 shows that it decreases slightly as the mole fraction of water increases, nearly temperature independent, but around $x_{\mathrm{w}} \sim 0.82$ starts to increase sharply (almost doubling), until it reaches the value for pure water. As found for the speed of sound, the temperature derivative changes sign, and a crossing of the extreme temperatures was found. These results might reflect the change in the structure of the mixture, as suggested by the apparent molar volume of $[C_2mim][CH_3SO_3]$ in water, as will be discussed below.

Values of the density of the mixtures, as a function of the mole fraction of water and temperature were fitted to polynomials of the type:

$$\rho(x_w,T) = \sum_{i=0}^{i=8} a_i(T)x_w^i \qquad (2)$$

Values of the coefficients in kg·m$^{-3}$, their respective standard deviations and standard deviation of the fits are presented in Table S2. Figure S1 shows the deviations of the experimental data from the correlation are well within the expanded global uncertainty of the measurements at the 95% confidence level, except for 2 points at the highest temperature. Comparison of the previous values with all the available data for the density of the pure ionic liquid has been presented elsewhere[2] and will not be repeated here. Additionally, the values of the density of pure water do not deviate from accepted reference values[38] by more than 0.001% in the whole temperature range, an expected result as the densimeter was calibrated with pure water, as described before[11].

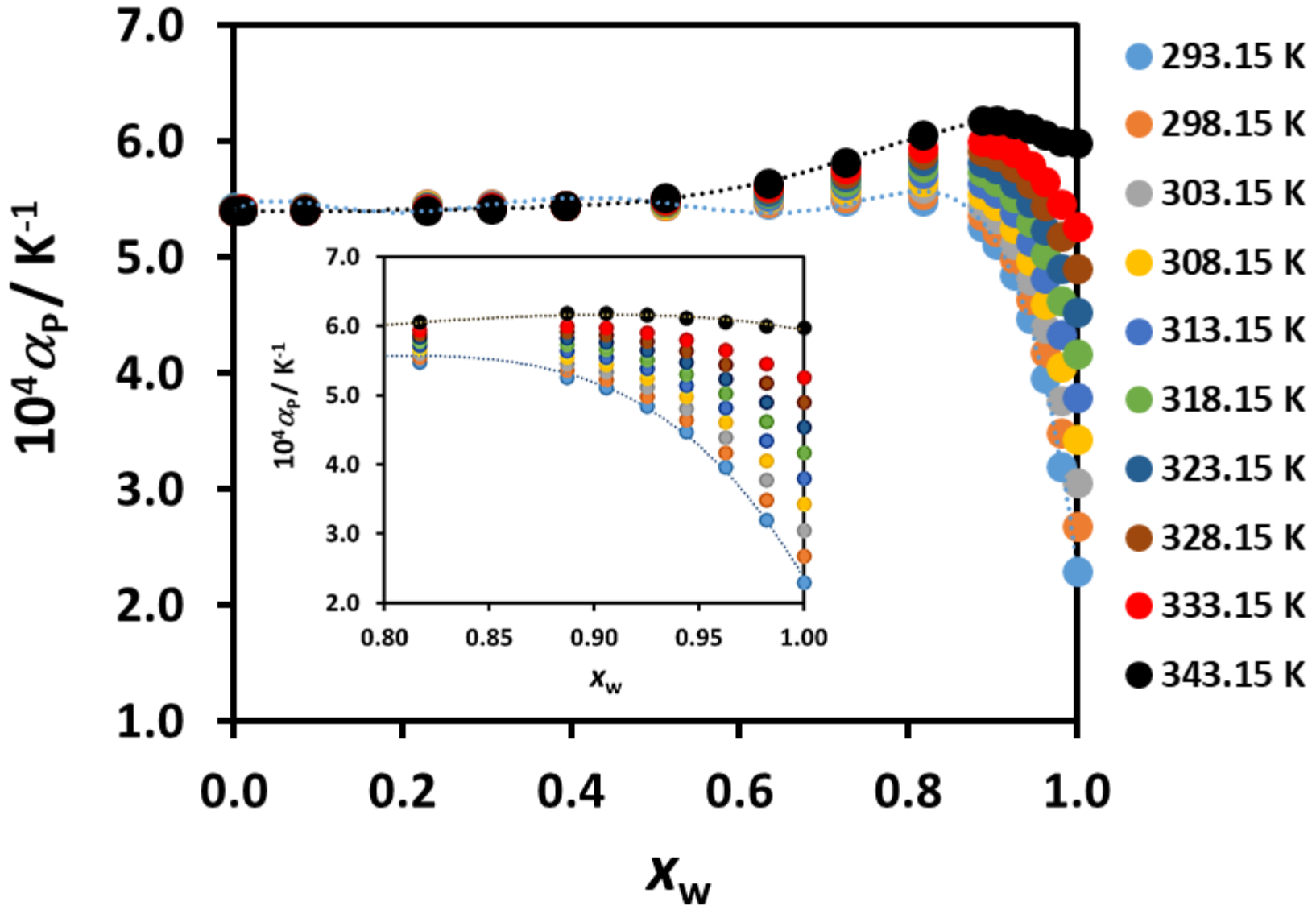

Figure 3 – The isobaric thermal expansion, $\alpha_P$ for the aqueous [$C_2$mim][$CH_3SO_3$] system, at studied temperatures, as a function of $x_w$. The inset shows the sensitive influence of temperature for $0.8 < x_w < 1.0$. The dashed lines are only guiding for the extreme temperatures, 293.15 and 343.15 K.

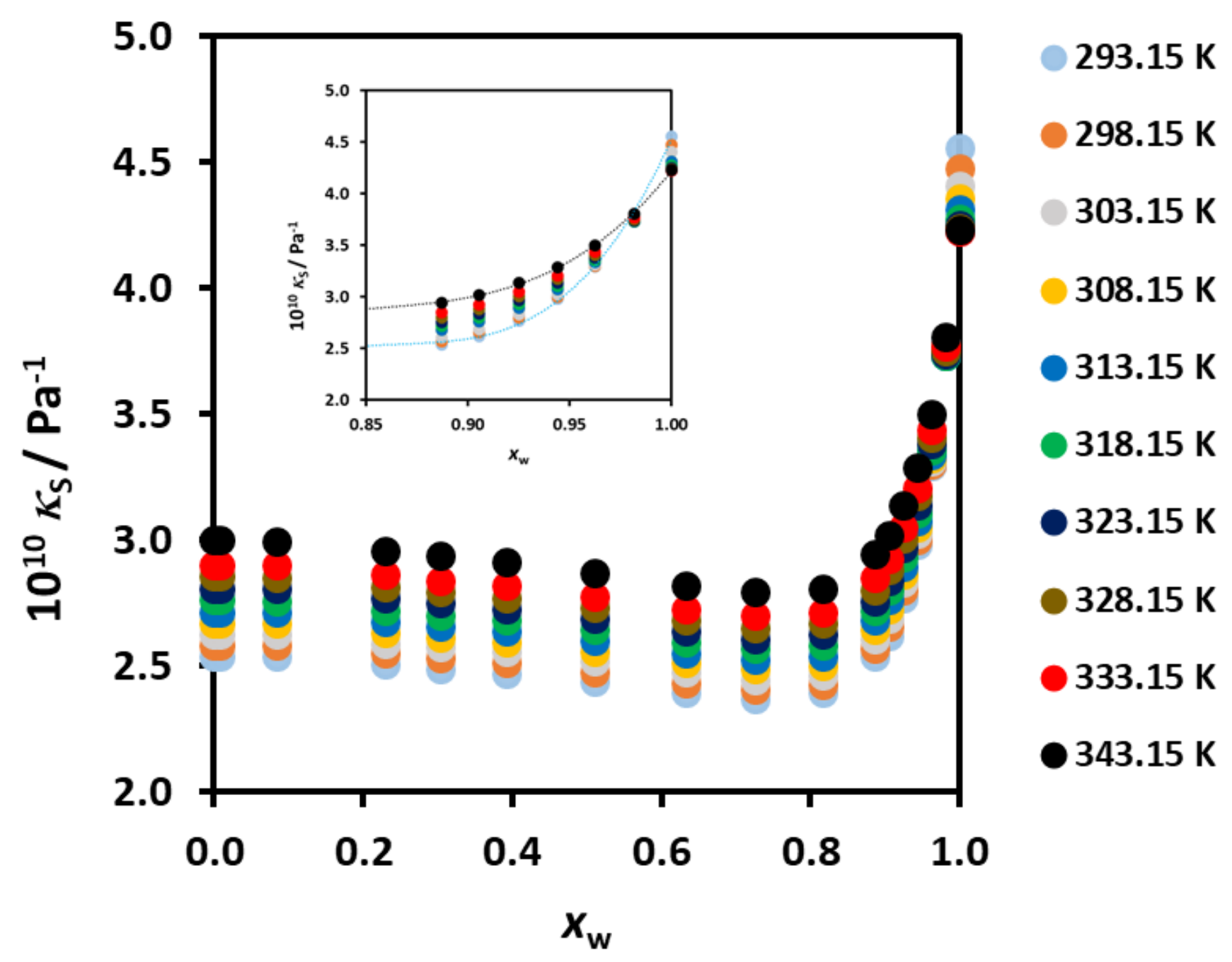

Figure 4 – The isentropic compressibility, $\kappa_S$, for the aqueous [$C_2$mim][$CH_3SO_3$] system, at studied temperatures, as a function of $x_w$. The inset shows the smaller influence of temperature for $0.85 < x_w < 1.0$, including the inversion in temperature variation, at $x_w \sim 0.975$. The dashed lines are only guiding for the extreme temperatures, 293.15 and 343.15 K.

The values of the density were used to calculate the excess molar volume of the mixture, using eq. (3):

$$V_m^{\mathrm{E}} = \frac{\bar{M}}{\rho} - x_w \frac{M_w}{\rho_w} - x_{IL} \frac{M_{IL}}{\rho_{IL}} \qquad (3)$$

where $\overline{M}$ is the mean molecular weight of the mixture, $M_i$ and $\rho_i$ are the molecular weights and densities of the pure components, w for water and IL for $[C_2mim][CH_3SO_3]$.

The values obtained are displayed in Fig. 5 and in Table S3. The excess volume is always negative, decreasing, as expected when the temperature decreases (more negative. However, for the low water content mixtures, up to $x_w \sim 0.3$, the excess volume is not very sensitive to the change in temperature. The minimum is achieved for $x_w \sim 0.6$, and again, it is only weakly temperature dependent. The estimated standard uncertainties of the mole fractions and the excess volumes are, respectively, $u_{r,c}(x_{IL}) = 0.002$ and $u(V_m^E) = 0.02$ cm$^3$·mol$^{-1}$.

The Redlich–Kister (R-K) fitting method in the form of eq. (4), was applied to fit the values obtained experimentally. Results are also displayed in Figure 5.

$$V_m^E = x_w(1 - x_w) \sum_{k=0}^{k=5} A_k (1 - 2x_w)^k \quad (4)$$

The number of $A_k$ parameters has been optimized using the F-test and their values are given in Table S3, together with the respective standard uncertainties and the standard deviation of the fits. The standard deviation of the fits does not exceed 0.01 cm$^3$·mol$^{-1}$, and no point deviates by more than 0.02 cm$^3$·mol$^{-1}$.

The excess partial molar volumes at infinite dilution for each component, $V_w^{E,\infty}$ and $V_{IL}^{E,\infty}$ can be obtained from these R-K fits, and were found slightly increasing with temperature for water, while for the ionic liquid the variation was found to be stronger.  This behavior can be seen in also in Table S4 and Fig. S2.

Comparison of present data with available data was done using the excess volumes curves obtained from the density data of each author. All available data sets for the density of $[C_2mim][CH_3SO_3]$ and their mixtures with water, at $P = 0.1$ MPa are presented in Table S5. In this table methods of measurement, temperature ranges, quoted uncertainty, sample purity and water content are identified. There are 5 batches of data, all obtained with vibrating tube instruments, as our data. Data of Stark et al. (2011)[39], obtained for $315.15 < T/K < 358.15$, have a claimed expanded global uncertainty of 2 kg·m$^{-3}$, data of Anantharaj and Banerjee (2013)[40], obtained for $298.1 < T/K < 373.1$ with a claimed expanded global uncertainty of 1.1 kg·m$^{-3}$, those of Marium et al. (2017)[18], for 293.15 < $T$ /K < 313.15, with a claimed expanded global uncertainty of 1.2 kg·m$^{-3}$, that of Amado-Gonzalez et al. (2017)[41], at 298.15 K, with a claimed uncertainty of 0.02 kg·m$^{-3}$ and those of Chereches et al. (2019)[42], obtained for 288.15 < $T$/K < 313.15, with a claimed expanded global uncertainty of 1.1 kg·m$^{-3}$, but only in graph form, not comparable. No proof exists that the molar fractions of the mixtures were corrected for the initial amount of water of the "pure" ionic liquid, and especially the data of Anantharaj and Banerjee (2013)[40] used samples with $w_w$ ~ 1.22%.

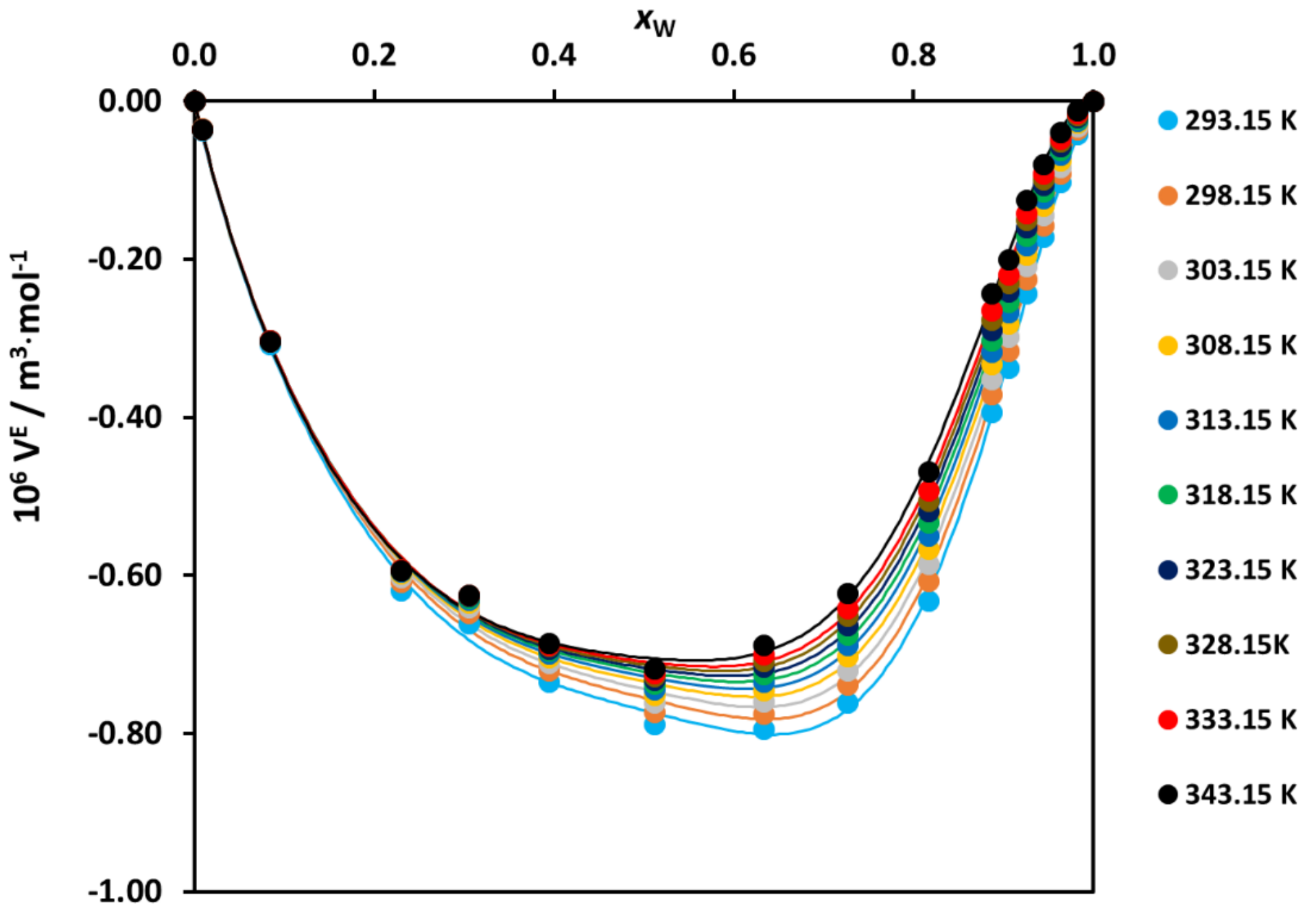

Figure 5. Excess molar volumes, $V_m^E$ , for the aqueous [$C_2$mim][$CH_3SO_3$] system, at the studied temperatures, as a function of $x_w$, the mole fraction of water. Full lines stand for 5$^{th}$ degree Redlich-Kister fitting curves.

Table S5 also displays the excess volume and mole fractions of water corresponding to the minimum ranges of the data. Data of Amado-Gonzalez et al. (2017)[41] is very close the pure water side, and always positive, contrary to the other sets of data. It can be seen that all other data agrees reasonably well with ours with respect to the minimum of the $V_{min}^{E}$ ($x_w$), around 0.62 at the lower temperatures and 0.58 at the higher temperatures, but that the minimum values of the excess volumes vary from -0.47 to -0.80 $m^3 \cdot mol^{-1}$ around room temperature, and from -0.41 to -0.69 at the higher temperatures. Additionally, data of Stark et al. (2011)[39] and Marium et al. (2017)[18] originate points with positive excess volume for $x_w >$ 0.96, while our data, with a standard uncertainty $u(V_m^E)$ = 0.04 $cm^3 \cdot mol^{-1}$, show for $x_w$ = 0.9625, $V_m^E$ = -0.103 ± 0.02 $m^3 \cdot mol^{-1}$, a negative value. This might be due to the values used for the density of the pure ionic liquid, which in the case of Marium et al. (2017)[18] deviated -0.17%, and data of Anantharaj and Banerjee (2013)[40] deviated -0.57%. In the case of Stark et al. (2011)[39] data agreed with data of Bioucas et al. (2018)[2] to within 0.02%, and therefore this might be due to them not having used the pure ionic liquid but a mixture with $x_w$ = 0.001, for the excess volume calculations. Figure 6 shows the deviations of all data, including present work, from the Redlich-Kister eq. (4). It can be seen that the deviations are high, amounting to a big fraction of the excess volume, and much greater than the mutual uncertainty of the data, confirming the discussion performed above, for the density of the liquid mixtures.

One interesting way of trying to understand the interactions in binary mixtures, by observing the effect of the water molecules in the structure of the ionic liquid with more detail, to explain the changes in the properties with the water content, is the calculation of "apparent" properties of the ionic liquid. One of the most interesting properties is the apparent molar volume of $[C_2mim][CH_3SO_3]$, $V_{\varphi,2}$ , defined as:

$$V_{\varphi,IL} = \frac{V_m - x_w V_w}{x_{IL}} \qquad (5)$$

where $V_w$ is the molar volume of pure water, at the same temperature of the mixture. $V_{\varphi,2}$ carries all the "non-idealities" of the mixture. Equation 5 can easily be transformed to:

$$V_{\varphi,IL} = \left[\frac{x_w M_w (\rho_w - \rho)}{x_{IL} \rho \rho_w}\right] + \frac{M_{IL}}{\rho} \qquad (6)$$

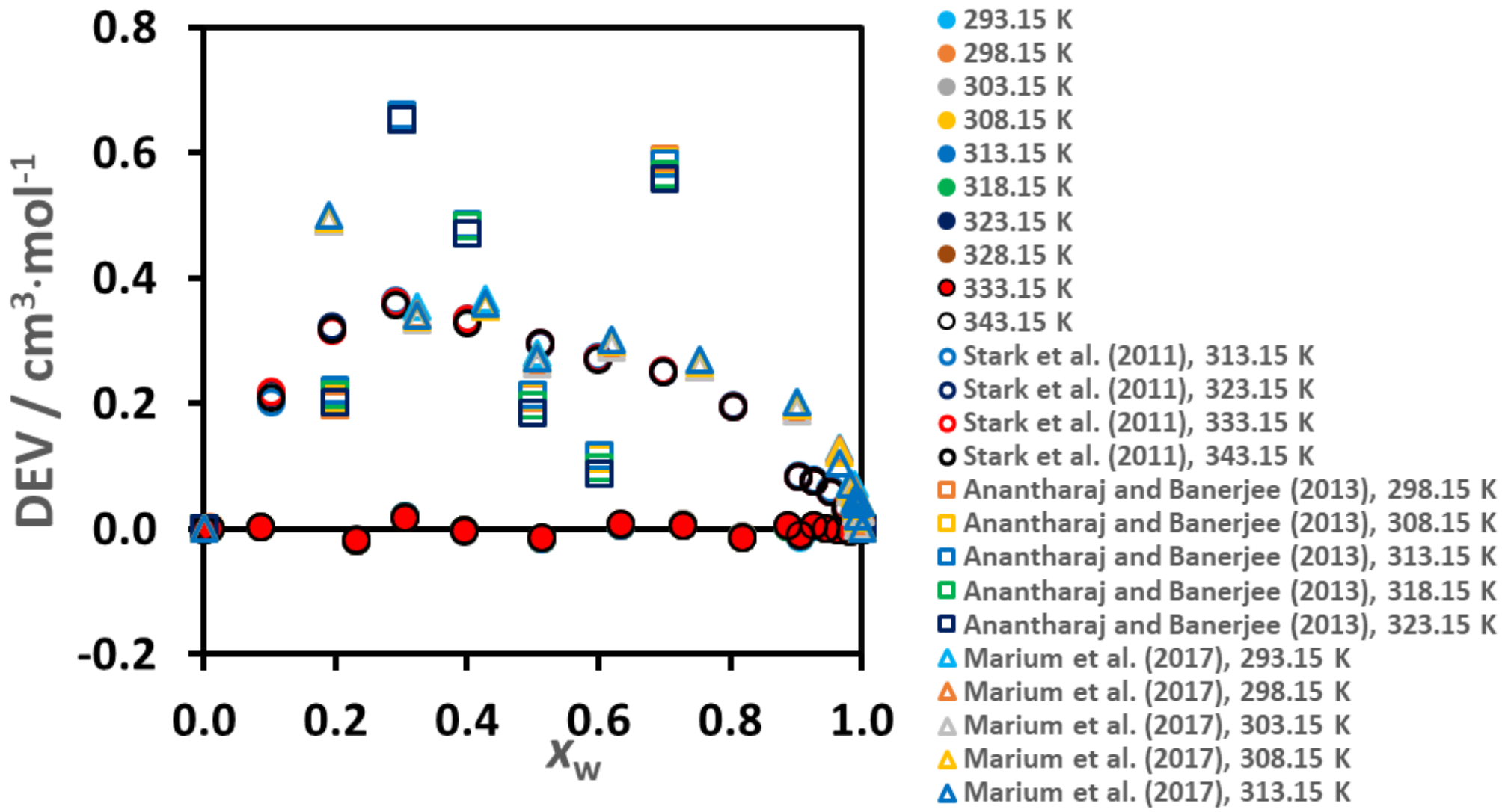

Figure 6 – Deviation, DEV = $V_m{}^E{}_{ThisWork}$ - $V_m{}^E{}_{Lit}$, from R-K fits, at each temperature. Filled circles, our data. Open circles, Stark et al. (2011)[39]; open squares, Anantharaj and Banerjee (2013)[36]; open triangles, Marium et al. (2017)[18].

Figure 7 shows the variation of the the apparent molar volume of [$C_2$mim][$CH_3SO_3$] (values in Table S6), as a function of the mole fraction of water. It is clear that there is a smooth decrease of this property with the increase of the mole fraction of water, but, continuing the addition of water molecules, the apparent molar volume starts to increase, passing through a minimum. This seems to support the change of structure of the mixture to a less packed one, the ionic liquid accepting very well the water molecules in its structure up to $x_w \sim 0.85$. This effect is the same in the whole temperature range studied. Additionally it was possible to evaluate the infinite dilution apparent volume of [$C_2$mim][$CH_3SO_3$], $V_{\varphi,IL}^{\infty}$ , by fitting the values for $x_w > 0.91$, with a linear variation as a function of the mole fraction of water, and extrapolating for infinite dilution, $x_{IL} = 0.000$. Values obtained are also presented in Table S6 and Fig. S2. Expanded global uncertainties in these values at a 95 % confidence level varied between 0.34 and 0.16 cm$^3$·mol$^{-1}$, at 293.15 and 343.15 K, respectively.

Regarding the speed of sound of [$C_2$mim][$CH_3SO_3$] + water mixtures, only one set of data is available, that of Marium et al. (2017)[18], for 293.15 < $T$/K < 313.15, in the whole composition range. Comparison with our data for the two extreme temperatures of these data is shown in Fig. 8, where it can be seen that the two sets of data agree within mutual uncertainty up to $x_w$ = 0.60, but deviate considerably, specially between 0.8 and 0.97, up to -3 %, at both temperatures. No reason was found for these discrepancies.

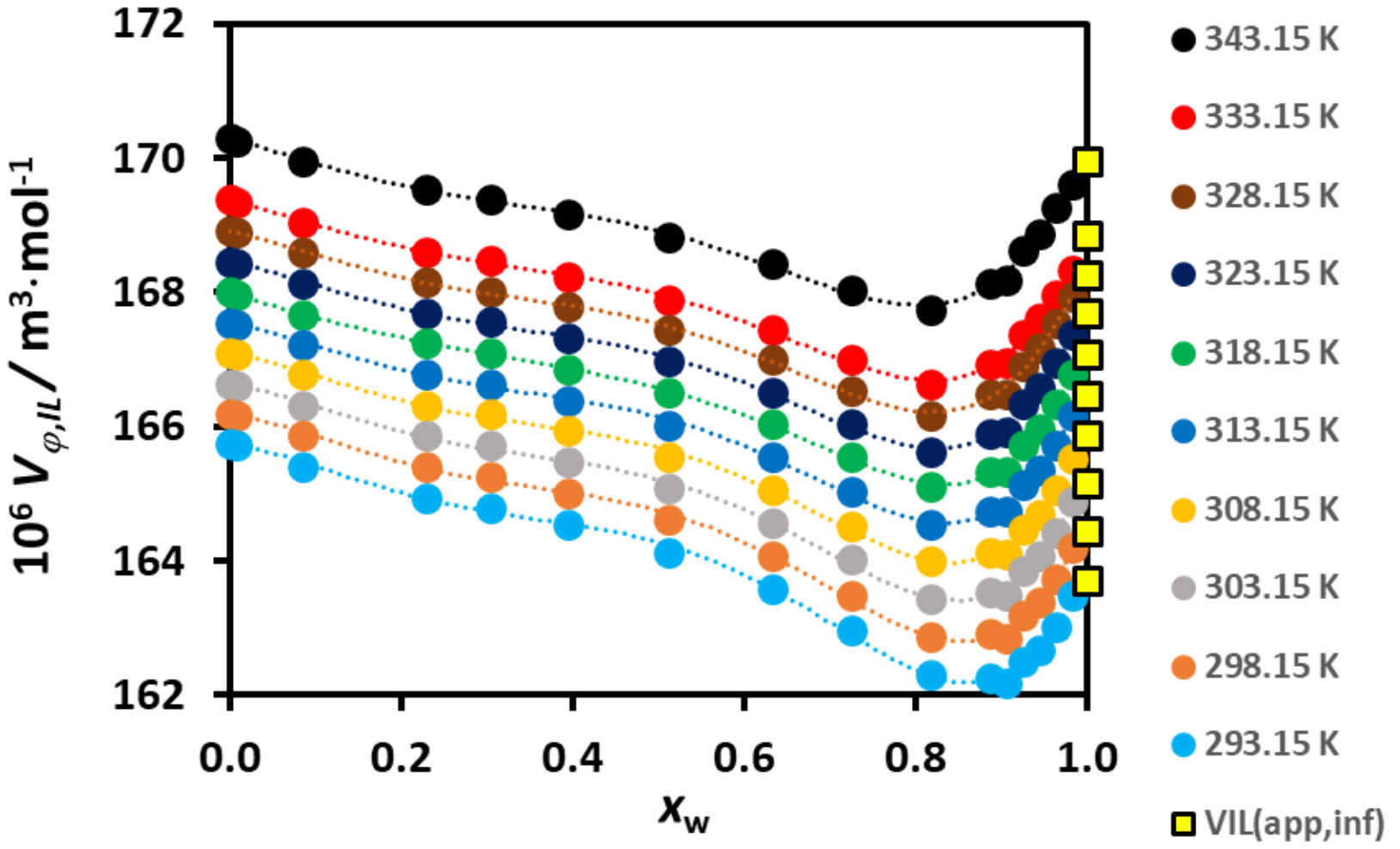

Figure 7. The apparent molar volume of [$C_2$mim][$CH_3SO_3$], $V_{\varphi,IL}$, for the aqueous [$C_2$mim][$CH_3SO_3$] system, at the studied temperatures, as a function of $x_w$, the mole fraction of water. Dotted lines stand for trend line polynomials, to better view the parallel behavior with temperature. The values marked with yellow squares correspond the infinite dilution apparent volume of [$C_2$mim][$CH_3SO_3$], $V_{\varphi,IL}^{\infty}$ , at each temperature.

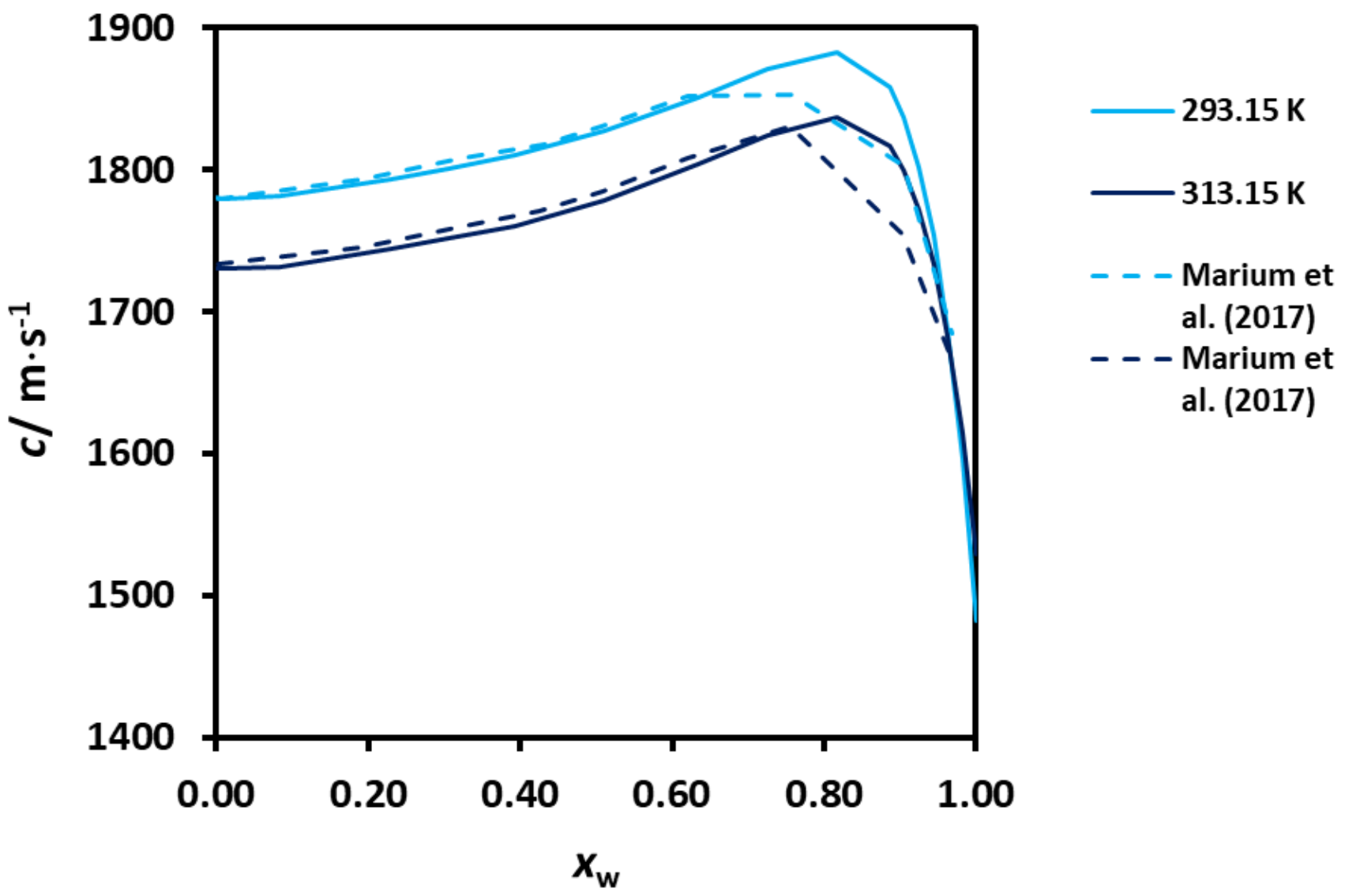

Figure 8 – Comparison between the speed of sound data of Marium et al. (2017)[18] with present work at 293.15 K and 313.15 for the aqueous [$C_2$mim][$CH_3SO_3$] system, as a function of $x_w$. Lines are just trend lines for visibility. Light blue lines, 293.15 K and dark blue lines, 313.15 K.

The experimental results obtained allow also the calculation of the apparent molar isentropic compression, $K_{s,\varphi,IL}$, of [$C_2$mim][$CH_3SO_3$] as a function of the water content, calculated from the following relation:

$$K_{S,\varphi,IL} = \frac{M_2\kappa_S}{\rho} - \left[\frac{M_w\,(1-x_{IL})\left(\kappa_{S,w}\rho - \kappa_S\rho_w\right)}{x_{IL}\rho\rho_w}\right] \quad (7)$$

where $\kappa_{S,w}$ and $\kappa_S$ are, respectively, the isentropic compressibility of water and of the mixture, at the respective temperature. Calculated values are shown also in Table S6. Figure 9 show its variation with the mole fraction of water, for the different temperatures studied. Although this property does not vary very much with the mole fraction of water and temperature, it is clear that there is a change of positive to negative apparent molar isentropic compression for $0.95 < x_w < 1.0$, at the 4 highest temperatures, an additional sign of change in the mixture structure around 0.95, as observed with the other properties.

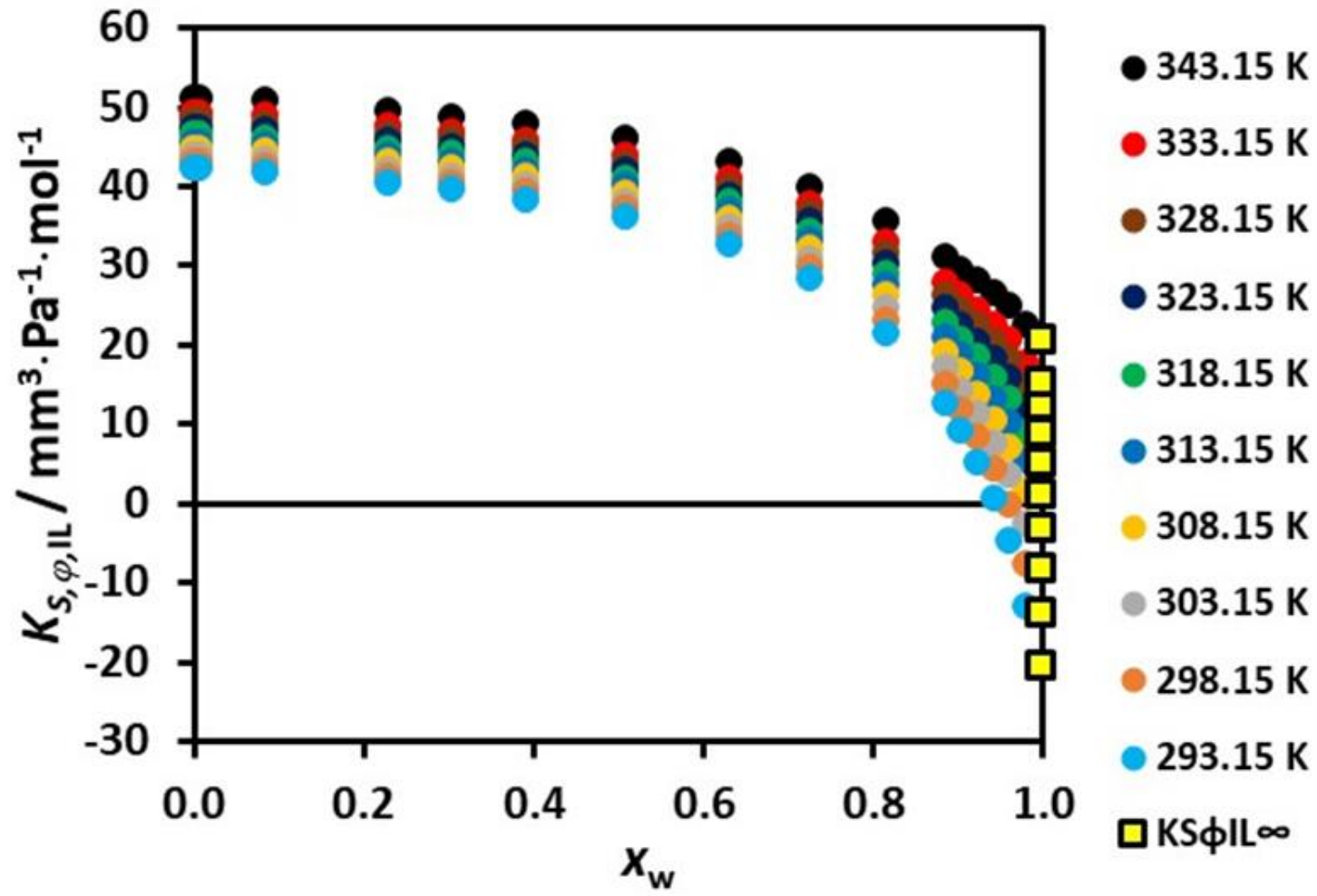

Figure 9 – The apparent molar isentropic compression, $K_{S,\varphi,IL}$, of $[C_2mim][CH_3SO_3]$ in the aqueous $[C_2mim][CH_3SO_3]$ system, at the studied temperatures, as a function of $x_w$. For $0.95 < x_w < 1.0$, there is a change of sign, for the highest temperatures. The values marked with yellow squares correspond the infinite dilution apparent volume of $[C_2mim][CH_3SO_3]$, $K^{\infty}_{S,\varphi,IL}$ , at each temperature.

It was also possible to evaluate the infinite dilution apparent molar isentropic compression for $[C_2mim][CH_3SO_3]$, $K^{\infty}_{S,\varphi,IL}$ , by fitting the values for $x_w > 0.89$, with a quadratic variation as a function of the mole fraction of water, and extrapolating for infinite dilution, $x_{IL} = 0.000$. Values obtained are also presented in Table S6 and Fig. S3. Expanded global uncertainties in these values at a 95 % confidence level varied between 1.0 and 0.3 $mm^3 \cdot MPa^{-1} \cdot mol^{-1}$, at 293.15 and 343.15 K, respectively.

As already applied above for the excess volumes, other excess thermodynamic functions can be calculated from the present data and can give further information about the structure of the mixture. In general, excess molar properties are defined by the equation:

$$Z_m^{\mathrm{E}} = Z_m - Z_m^{\mathrm{id}} \qquad (8)$$

where $Z_m^{\mathrm{E}}$ is the difference between values for the molar property of the real mixture, $Z_m$, and those for an ideal mixture, at the same temperature, pressure and composition of the real mixture, $Z_m^{\mathrm{id}}$ . In this equation, $Z_m$ can be: the molar volume, $V_m$ , the molar isobaric expansion, $E_{P,m} = (\partial V_m/\partial T)_P$; the molar isobaric heat capacity, $C_{P,m}$; the molar isothermal compression, $K_{T,m} = V_m \times \kappa_T$ ; and the molar isentropic compression $K_{S,m} = V_m \times \kappa_S$ [10]. For the Gibbsian properties[43], such as $V_m$, $E_{P,m}$, and $C_{P,m}$, values for the ideal mixture can be calculated by the following additive rule,

$$Z_m^{\mathrm{id}} = x_w Z_{m,w} + x_{IL} Z_{m,IL} \qquad (9)$$

where $Z_{m,i}$ stands for any of the Gibbsian molar properties of the respective component in its pure state, at the same temperature and pressure of the real mixture. In addition, for the case of $K_{S,m}$ , which is not a Gibbsian property, $K_{S,m}^{\mathrm{id}}$ is calculated using equation:

$$K_{S,m}^{\mathrm{id}} = K_{T,m}^{\mathrm{id}} - T\frac{\left(E_{P,m}^{\mathrm{id}}\right)^2}{C_{P,m}^{\mathrm{id}}} \qquad (10)$$

$K_{T,m}^{\mathrm{id}}$, $E_{P,m}^{\mathrm{id}}$ and $C_{P,m}^{\mathrm{id}}$ values were estimated using eq. (9) and data for pure components displayed in Table S7 of the Supplementary Information. Excess molar isobaric expansions and isentropic compressions calculated using eqs. (8,9), are reported in Table S6 and displayed in figures 10 and 11. Present values were used for the density and speed of sound for pure water and [$C_2$mim][$CH_3SO_3$], while $C_{P,m}$ values for water were obtained from NIST SRD69[38] and for [$C_2$mim][$CH_3SO_3$] from Ficke et al (2010)[44].

Redlich-Kister (eq. 4) was also applied to the excess molar isobaric expansions and excess molar isentropic compressions, and coefficients obtained, their standard deviations, and standard deviations of the fits (6$^{\mathrm{th}}$ – 6D and 4$^{\mathrm{th}}$ - 4D order, respectively) are shown in Table S9. Values of the infinite dilution excess isentropic molar compression for the ionic liquid, $K_{\mathrm{S,IL}}^{\mathrm{E},\infty}$, and for water, $K_{\mathrm{S,w}}^{\mathrm{E},\infty}$, were obtained from $K_{S,\varphi,IL}^{\infty}$ and $K_{S,\mathrm{IL}}^{\mathrm{id}}$ given by:

$$K_{S,\text{IL}}^{\text{id}} = K_{S,m,\text{IL}} + TC_{P,m,\text{IL}}\left(\frac{E_{P,m,\text{IL}}}{C_{P,m,\text{IL}}} - \frac{E_{P,m,\text{w}}}{C_{P,m,\text{w}}}\right)^2 \qquad (11)$$

and are displayed in Table S9 also. The value for water is almost zero within the uncertainty of the calculation, but for the ionic liquid is very negative and as the temperature increases, the values increase also (lower negative values).

As it can be seen, the excess molar isobaric expansion is highly structured, with a maximum for $x_{\text{w}} \sim 0.82$ for all temperatures, but with a negative portion in the lower water content zone, agreeing with the result shown in Table S6 for the apparent molar volume of [$C_2$mim][$CH_3SO_3$], in the aqueous system. This negative portion, for 343.15 K lasts up to $x_{\text{w}} \sim 0.50$, a sign that temperature is having an effect on the structure of the mixture. The inset in Fig. 10 shows the detailed behavior, experimental and the R-K fits, predicting negative values for the excess molar isobaric expansion at very low water mole fractions. The excess molar isentropic compression has a regular behavior, always negative, with a minimum for $x_{\text{w}} \sim 0.79$ at 293.15 K, decreasing to $x_{\text{w}} \sim 0.72$ for 343.15 K. The minimum in this zone is highly correlated with the results found for the maximum in the speed of sound (Fig. 2), as expected.

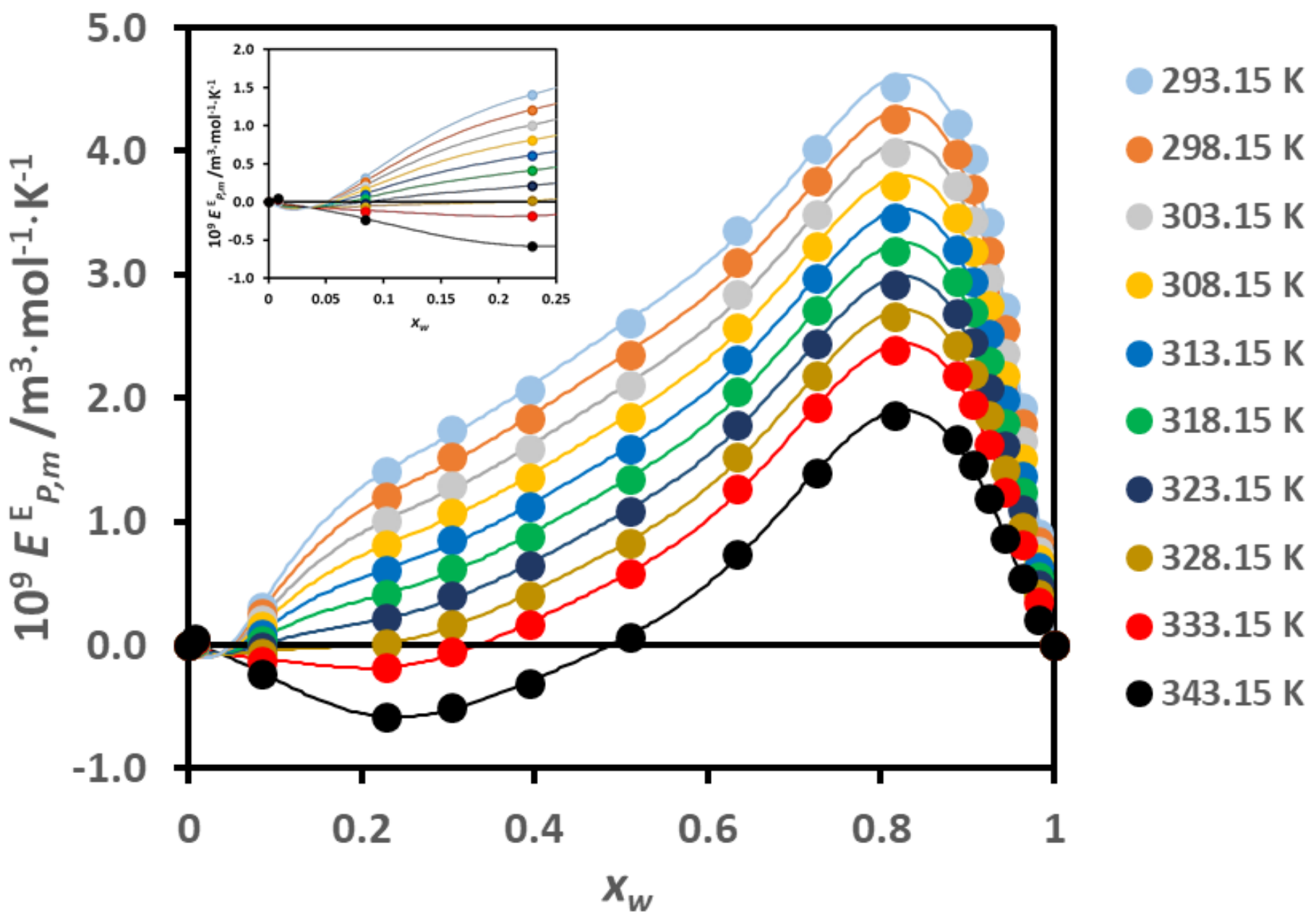

Figure 10 – Excess molar isobaric expansion, $E^{\mathrm{E}}_{P,m}$ , in the aqueous [$C_2$mim][$CH_3SO_3$] system, at the studied temperatures, as a function of $x_{\mathrm{w}}$. Lines represent Redlich-Kister (R-K) 6G fits. The inset shows the behavior for $x_{\mathrm{w}} < 0.25$.

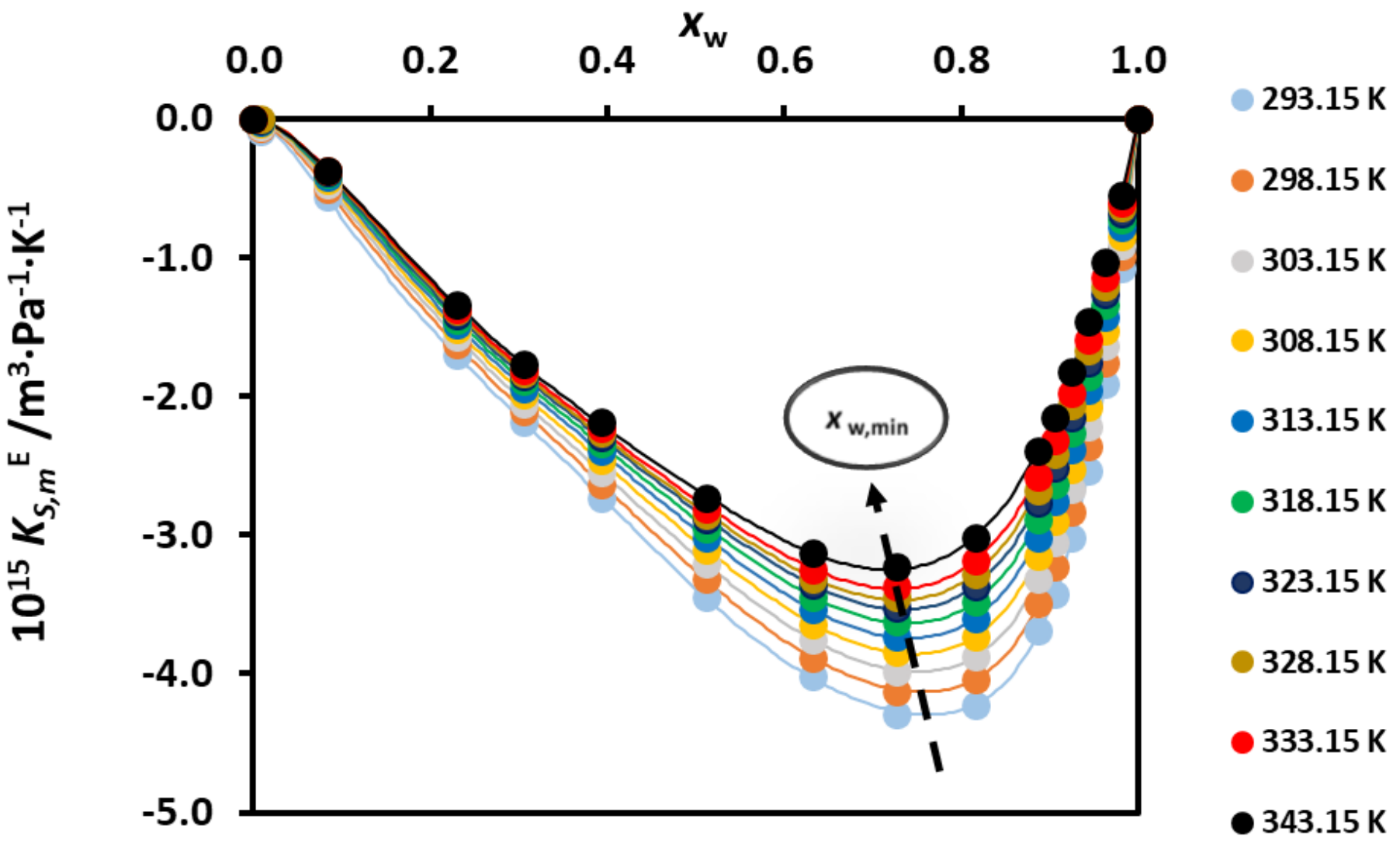

Figure 11 – Excess molar isentropic compression, $K_{S,m}^{E}$ , in the aqueous [$C_2$mim][$CH_3SO_3$] system, at the studied temperatures, as a function of $x_w$. Lines represent Redlich-Kister (R-K) 4G fits. The arrow illustrates the displacement of the minimum of the function to lower mole fractions of water, as the temperature increases.

**3.2. Viscosity.** The measured values for the kinematic viscosity of the [$C_2$mim][$CH_3SO_3$] + water, as a function of the mole fraction of the ionic liquid mixtures and temperature are displayed in Table 3. Values for the molar fraction $x_w$ = 0.0643 correspond also to previous measurements[2], but calculated with new calibration constants, for the higher viscosity capillary viscometers, reflecting changes for the temperatures smaller than 333.15 K. Values of the dynamic viscosity were calculated from the density values reported above. For completeness, values for the pure water property values are also displayed, obtained from reference[38]. The mixtures were found to be Newtonian, like the pure ionic liquid, previously reported[2]. The viscosity values are shown in Fig. 12, as a function of the mole fraction of water, $x_w$, for the different temperatures studied.

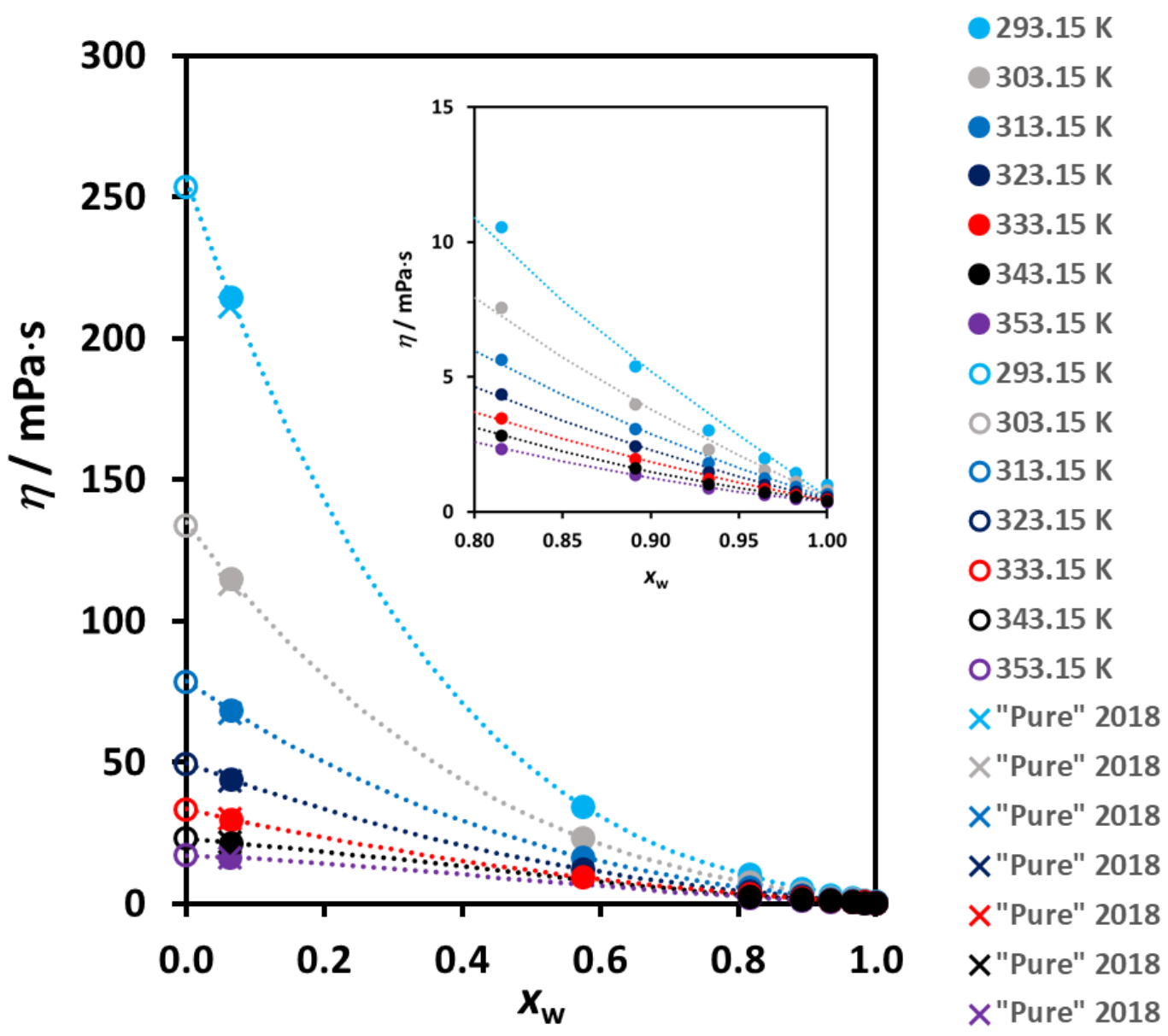

Figure 12 – Viscosity of the aqueous [$C_2$mim][$CH_3SO_3$] system, at the studied temperatures, as a function of the mole fraction of water, $x_w$, full dots. For comparison, the extrapolated water-free values of viscosity, as described in the text, open dots, and the results published by the authors, Bioucas et al. (2018)[2], crosses, at the corresponding temperatures, are shown. Dotted lines are polynomial fits reported in Table S10 and eq. (S2), at the respective temperatures. The inset displays the values of viscosity of the mixture for $x_w > 0.8$.

Extrapolated water-free values for the viscosity of [$C_2$mim][$CH_3SO_3$] can be obtained with the general procedure applied before [2,8,10,16,26,29,45], by using the equation:

$$\eta_{wf} = \eta_{exp}(x_w, T) - \left(\frac{\partial \eta}{\partial x_w}\right)_T x_w \qquad (12)$$

The values of $\eta_{exp}(x_w, T)$ were fitted with a polynomial in the mole fraction of water, for each temperature, and the derivative $\left(\frac{\partial \eta}{\partial x_w}\right)_T$ could be calculated as a function $x_w$, eq. (S1) and Table S10. In previous cases the limiting value of this derivative was used, as the last water mole fraction measured was close to zero. However, in this case, the value of $x_w$ = 0.0643 was too high to use this methodology, especially at the lower temperatures, due to the fast change of slope. It was then decided to use the value of this derivative at this mole fraction and obtain $\eta_{wf}$ from eq. (12). The values obtained are also shown in Table 4 (line $x_{IL}$ = 1.0000) and are also plotted in Fig. 12, open dots. The coefficients of the viscosity fits as a function of the mole fraction of water are reported in Table S10 of SI, including their standard variation, and the standard deviation of the fit. Figure 12 also shows, for comparison, the previous published data for "pure" [$C_2$mim][$CH_3SO_3$], by Bioucas et al. (2018)[2], the lowest mole fraction in water. Deviations, at that mole fraction between the two sets of data are around 1.6% for temperatures up to 323.15 K, decreasing to less than 0.5% for higher temperatures, within the mutual uncertainty of the two sets of data.

All available data sets for the viscosity of [$C_2$mim][$CH_3SO_3$] and their mixtures with water, at $P$ = 0.1 MPa are presented in Table S11. In this table methods of measurement, temperature ranges, quoted uncertainty and sample purity are reported. There are 4 batches of data, data of Stark et al. (2011)[39], obtained with a cone and plate rheometer, for 298.15 < $T$/K < 353.15, have a claimed expanded relative uncertainty of 6 % ($k$=2), data of Krannich et al. (2016)[46], obtained with a capillary tube viscometer for 303.2 < $T$/K < 363.2 with a claimed expanded relative uncertainty of 3%, those of Marium et al. (2017)[18], obtained with a falling body viscometer for 293.15 < $T$/K < 313.15, with a claimed expanded relative uncertainty of 0.4%, and that of Chereches et al. (2020)[47], obtained with a parallel plate rheometer for 288.15 K < $T$/K < 333.15 , with a claimed expanded relative uncertainty of 6%. Results are presented in Fig. 13, which includes also the deviations of our data from the obtained fits. It is clear from this figure that these data are much lower than other data for all the isotherms, and with a different curvature. Data of Stark et al. (2017)[39] deviates very much from our data at the lower temperatures, especially for low water contents, and departures up to –30 % can be found for higher temperatures, as shown in Fig. 14. Data of and Krannich et al. (2016)[48] agree reasonably with our data, namely at the lower temperatures and low water contents, but seem to have a different slope with composition at the higher temperatures. This was already confirmed with the comparisons reported before[2]. The recent data of Chereches et al. (2020)[49], obtained only for a single composition of $x_w$ = 0.25, have a deviation of -56 % at 293.15 K, decreasing to -34 % at 333.15 K. Inset in this figure shows the values for the compositions $x_w$ > 0.88, to visualize better the deviations between the several sets of data in this region, namely for pure water.

The dependence on temperature of the viscosity for pure liquids can be approached in several ways, as currently discussed by us[10,26,45]. Two of these models, the Arrhenius model, eq. (13) and the Vogel-Fulcher-Tammann (VFT), eq. (14) have been applied here, for each composition studied.

$$\eta / \text{ mPa·s} = A(x_w)e^{\frac{E_a(x_w)}{R(T/\text{K})}} \quad (13)$$

where $A(x_w)$ is a collision frequency factor for each mole fraction of water and $E_a(x_w)$ is the activation energy for viscous flow,

$$\eta \text{ / mPa·s} = A_{VFT}e^{\frac{B_{VFT}}{(T/\text{K})-C_{VFT}}} \quad (14)$$

where $A_{\text{VFT}}$, $B_{\text{VFT}}$, $C_{\text{VFT}}$ are the coefficients of the VFT equation. The results obtained are displayed in Fig. 15, and it is clear that the experimental data obeys, for all the compositions an Arrhenius behavior, although there is a minor curvature for the data near the pure [$C_2$mim][$CH_3SO_3$], found previously for other ionic liquids.

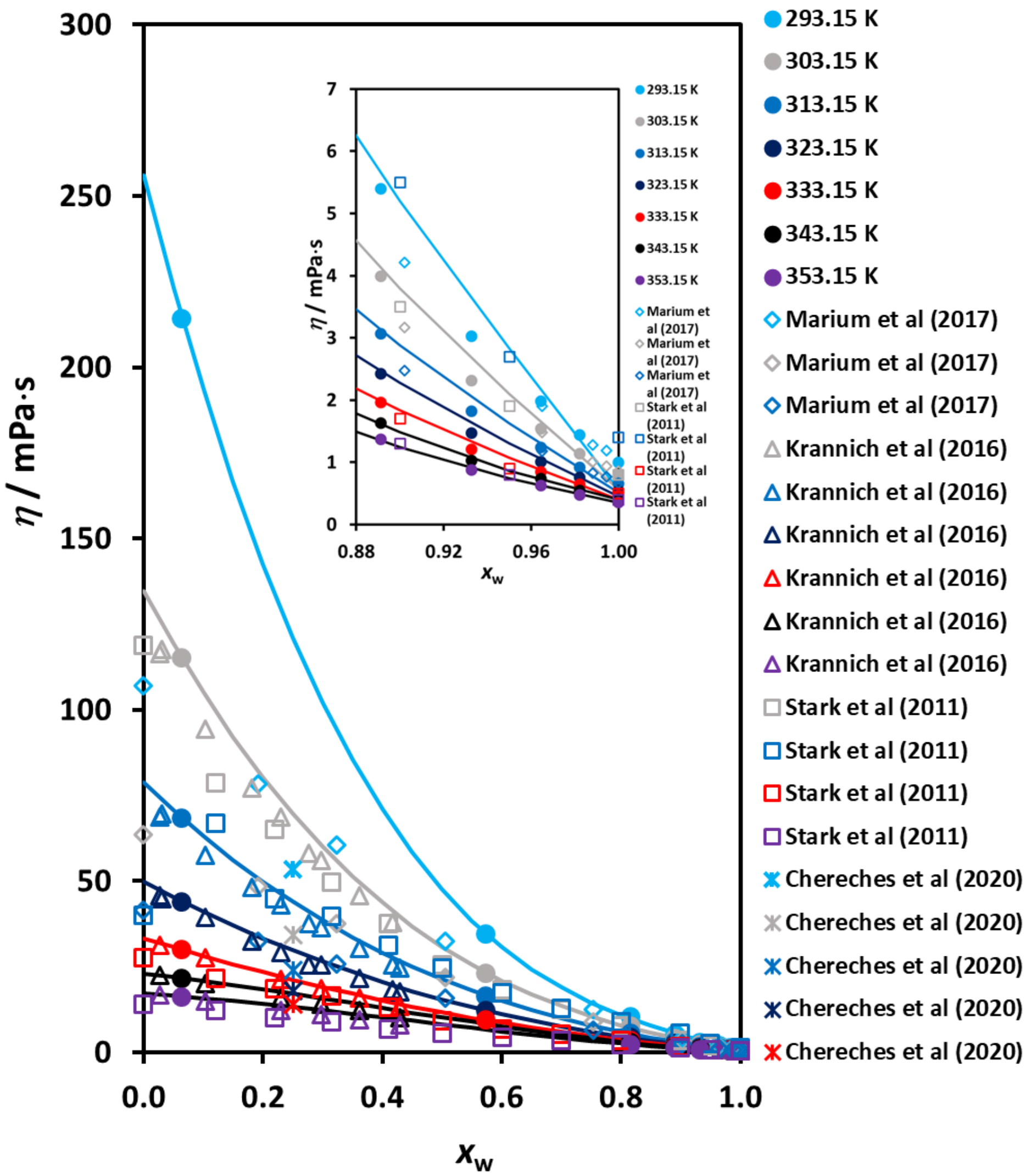

300
250
200
150
100
50
0
0.0
0.2
0.4
0.6
0.8
1.0
η / mPa·s
x_w
7
6
5
4
3
2
1
0
0.88
0.92
0.96
1.00
293.15 K
303.15 K
313.15 K
323.15 K
333.15 K
343.15 K
353.15 K
Marium et al (2017)
Marium et al (2017)
Marium et al (2017)
Krannich et al (2016)
Krannich et al (2016)
Krannich et al (2016)
Krannich et al (2016)
Krannich et al (2016)
Krannich et al (2016)
Stark et al (2011)
Stark et al (2011)
Stark et al (2011)
Stark et al (2011)
Chereches et al (2020)
Chereches et al (2020)
Chereches et al (2020)
Chereches et al (2020)
Chereches et al (2020)

Figure 13 – Comparison of our viscosity data obtained for the aqueous [$C_2$mim][$CH_3SO_3$] system, at the studied temperatures, as a function of the mole fraction of water, $x_w$, with values reported in the literature. Our data - full circles; Stark et al. (2011)[39], open squares; Krannich et al. (2016)[48], open triangles; Marium et al. (2017)[18], open diamonds; Chereches et al. (2020)[49], crosses. Each temperature has the same color. Full lines represent our polynomial fits for each temperature (Eq. (S1) and Table S10). Inset shows the values for the compositions $x_w$ > 0.88, to visualize better the deviations between the several sets of data in this region. **Table 3 - Kinematic viscosity, $\nu$, and dynamic viscosity, $\eta$, for the aqueous system [$C_2$mim][$CH_3SO_3$] from $T$ = (293.15 to 353.15) K, at $P$ = 0.1 MPa[a], as a function of the molar fraction of water, $x_w$. Water free values for the viscosity of pure [$C_2$mim][$CH_3SO_3$] are also presented. Water values[38].**

| $x_w$[c] | $T$ = 293.15 K | | $T$ = 303.15 K | | $T$ = 313.15 K | | $T$ = 323.15 K | | $T$ = 333.15 K | | $T$ = 343.15 K | | $T$ = 353.15 K | |
|---|---|---|---|---|---|---|---|---|---|---|---|---|---|---|
| | $\nu$[b] /mm$^2$.s$^{-1}$ | $\eta$[b] /mPa.s | $\nu$ /mm$^2$.s$^{-1}$ | $\eta$ /mPa.s | $\nu$ /mm$^2$.s$^{-1}$ | $\eta$ /mPa.s | $\nu$ /mm$^2$.s$^{-1}$ | $\eta$ /mPa.s | $\nu$ /mm$^2$.s$^{-1}$ | $\eta$ /mPa.s | $\nu$ /mm$^2$.s$^{-1}$ | $\eta$ /mPa.s | $\nu$ /mm$^2$.s$^{-1}$ | $\eta$ /mPa.s |
| 1.0000 | 1.0034 | 1.0016 | 0.8008 | 0.7974 | 0.6581 | 0.6530 | 0.5535 | 0.5469 | 0.4744 | 0.4664 | 0.4131 | 0.4039 | 0.3646 | 0.3544 |
| 0.9823 | 1.391 | 1.439 | 1.101 | 1.135 | 0.894 | 0.917 | 0.740 | 0.756 | 0.632 | 0.642 | 0.545 | 0.551 | 0.476 | 0.475 |
| 0.9643 | 1.862 | 1.980 | 1.455 | 1.541 | 1.174 | 1.237 | 0.965 | 1.012 | 0.813 | 0.848 | 0.709 | 0.735 | 0.610 | 0.625 |
| 0.9326 | 2.740 | 3.026 | 2.101 | 2.309 | 1.663 | 1.819 | 1.349 | 1.468 | 1.122 | 1.214 | 0.955 | 1.026 | 0.827 | 0.879 |
| 0.8911 | 4.723 | 5.395 | 3.515 | 3.994 | 2.715 | 3.068 | 2.161 | 2.428 | 1.759 | 1.965 | 1.468 | 1.629 | 1.246 | 1.370 |
| 0.8152 | 8.923 | 10.55 | 6.446 | 7.577 | 4.833 | 5.649 | 3.761 | 4.370 | 3.006 | 3.473 | 2.465 | 2.831 | 2.060 | 2.346 |
| 0.5737 | 28.19 | 34.52 | 19.09 | 23.250 | 13.63 | 16.504 | 10.14 | 12.22 | 7.830 | 9.379 | - | - | - | - |
| 0.0643 | 172.08 | 214.35 | 92.85 | 115.04 | 55.53 | 68.423 | 35.95 | 44.06 | 24.59 | 29.98 | 17.82 | 21.61 | 13.49 | 16.30 |
| 0.0000[d] | 203.75 | 253.54 | 107.99 | 133.66 | 63.51 | 78.18 | 40.51 | 49.60 | 27.31 | 33.25 | 18.98 | 22.98 | 14.25 | 17.17 |

[a] Standard uncertainties, $u$, are: $u(T)$ = 0.01 K, $u(P)$ = 1 kPa

[b] Expanded relative uncertainty $U(\nu)$ = 1.1 %; $U(\eta)$ = 1.2%, at a 95 % confidence level ($k$=2);

[c] Relative combined standard uncertainty, $u_{r,c}(x_w)$ = 0.002

[d] Extrapolated water-free values; kinematic viscosity calculated from density and viscosity values

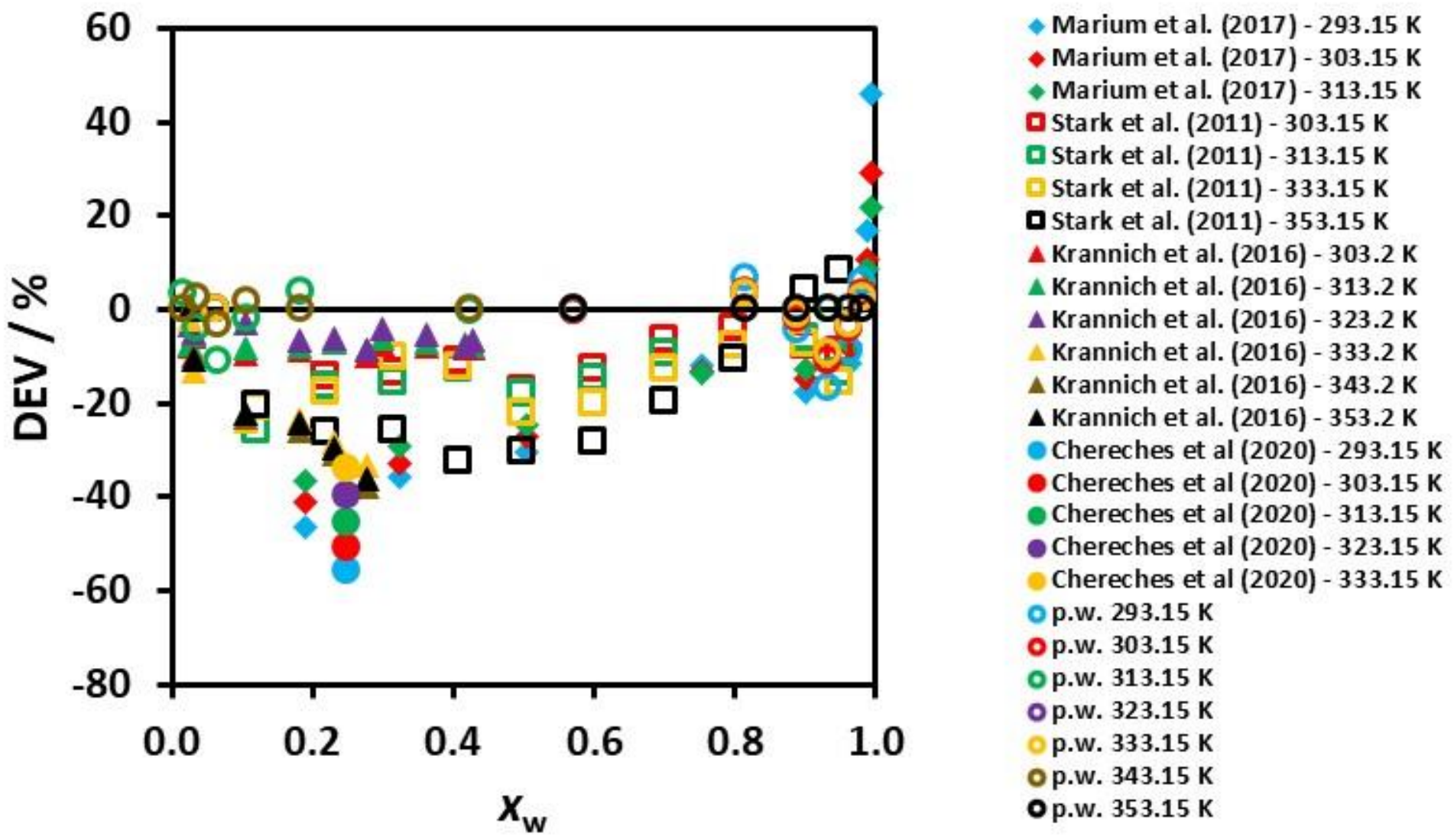

Figure 14 – Deviations of the available literature data for dynamic viscosity, $\left(\frac{\eta_{\text{exp}}-\eta_{\text{fit}}}{\eta_{\text{fit}}}\right) \times 100\%$, from our polynomial fits, Table S11 and eq. (S1) of SI. Present work is also included.

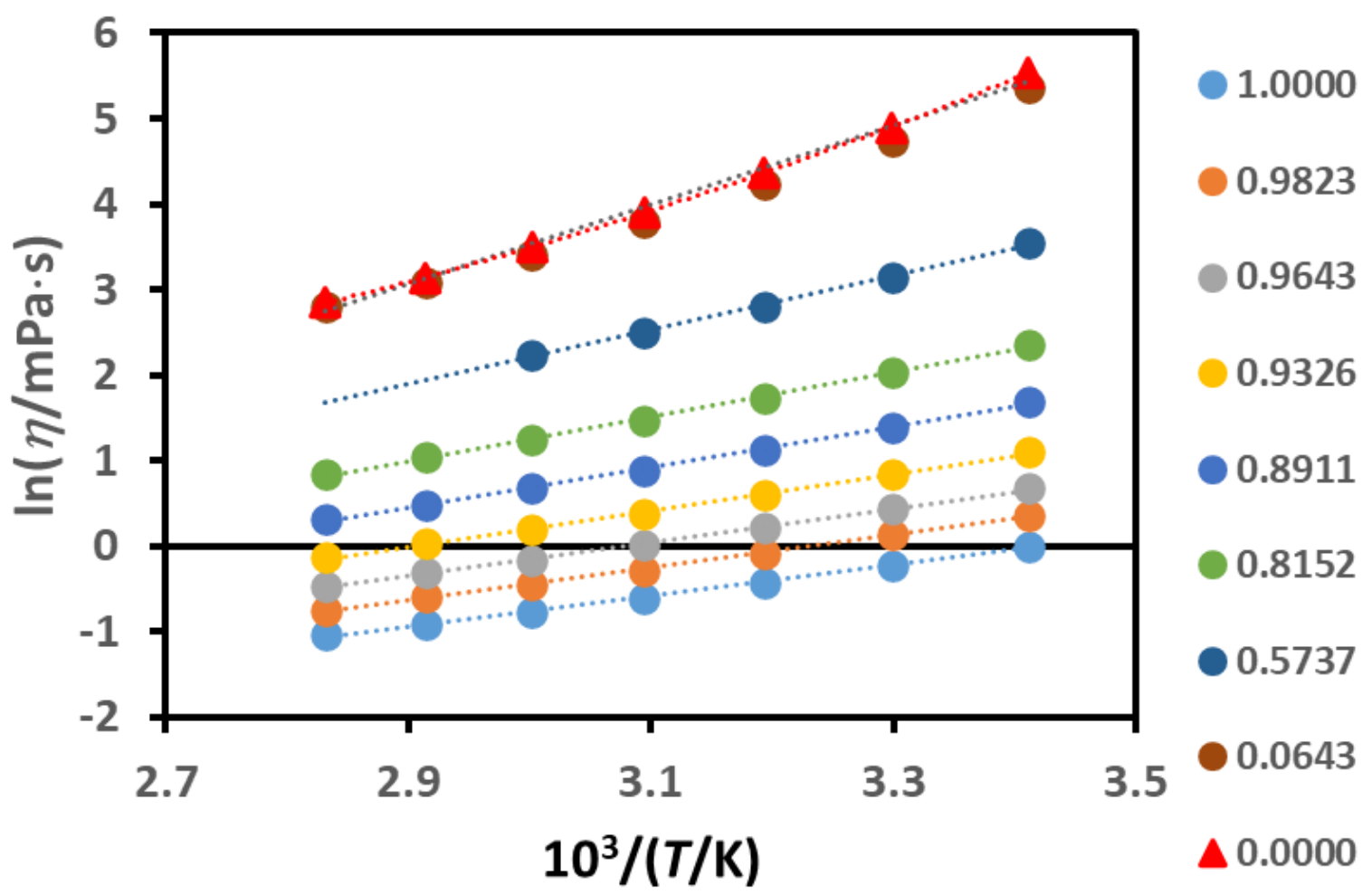

Figure 15 – Arrhenius plots for the dependence on temperature and mole fraction of water of the viscosity of the aqueous [$C_2$mim][$CH_3SO_3$] system. The dashed lines are the Arrhenius fits.

Values of the parameters $A(x_w)$ and $E_a(x_w)$, their standard variation and the standard variation of the fits are presented in Table S12. The activation energy decreases from the pure ionic liquid (38.50 J·mol$^{-1}$) to that of water (14.86 J·mol$^{-1}$), a logical consequence of the decrease in viscosity and density of the media, a fact already noted for other ionic liquid + water mixtures. Figure 16 shows the variation of the activation energy with the mole fraction of water, a quadratic shape, with a very small curvature. Values of the parameters $A(x_w)$, $B(x_w)$, and $C(x_w)$, for the VFT equation and their standard variation and the standard variation of the fits are also presented in Table S12.

Figure 17 shows the deviations of our data, for the different mole fractions of water, of our viscosity experimental data and the “water-free” values herein calculated. It can be seen that the fits are good, and only 1 experimental point has departures greater than the combined uncertainty of our data at a 95% confidence level ($k$ = 2), 0.7%, while the “water-free result at 343.15 K, seems to have a higher deviation, -1.7%, probably because we are lacking data for $x_w$ = 0.5737, for this temperature, which conditions the curvature of the VFT fit.

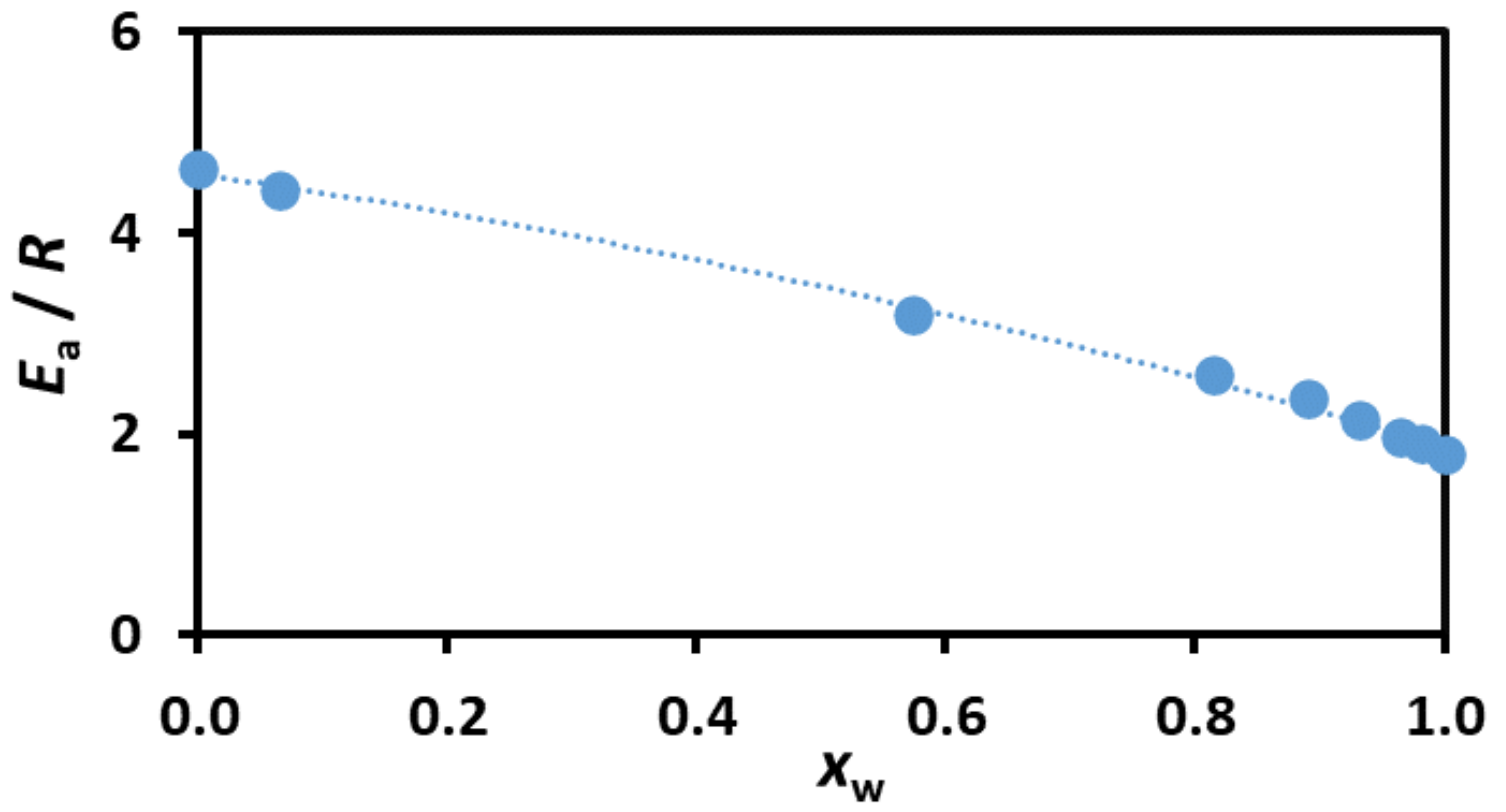

Figure 16 – The dependence of the activation energy for flow of the aqueous [$C_2$mim][$CH_3SO_3$] system on the mole fraction of water. Dotted line is just a quadratic fit trend line.

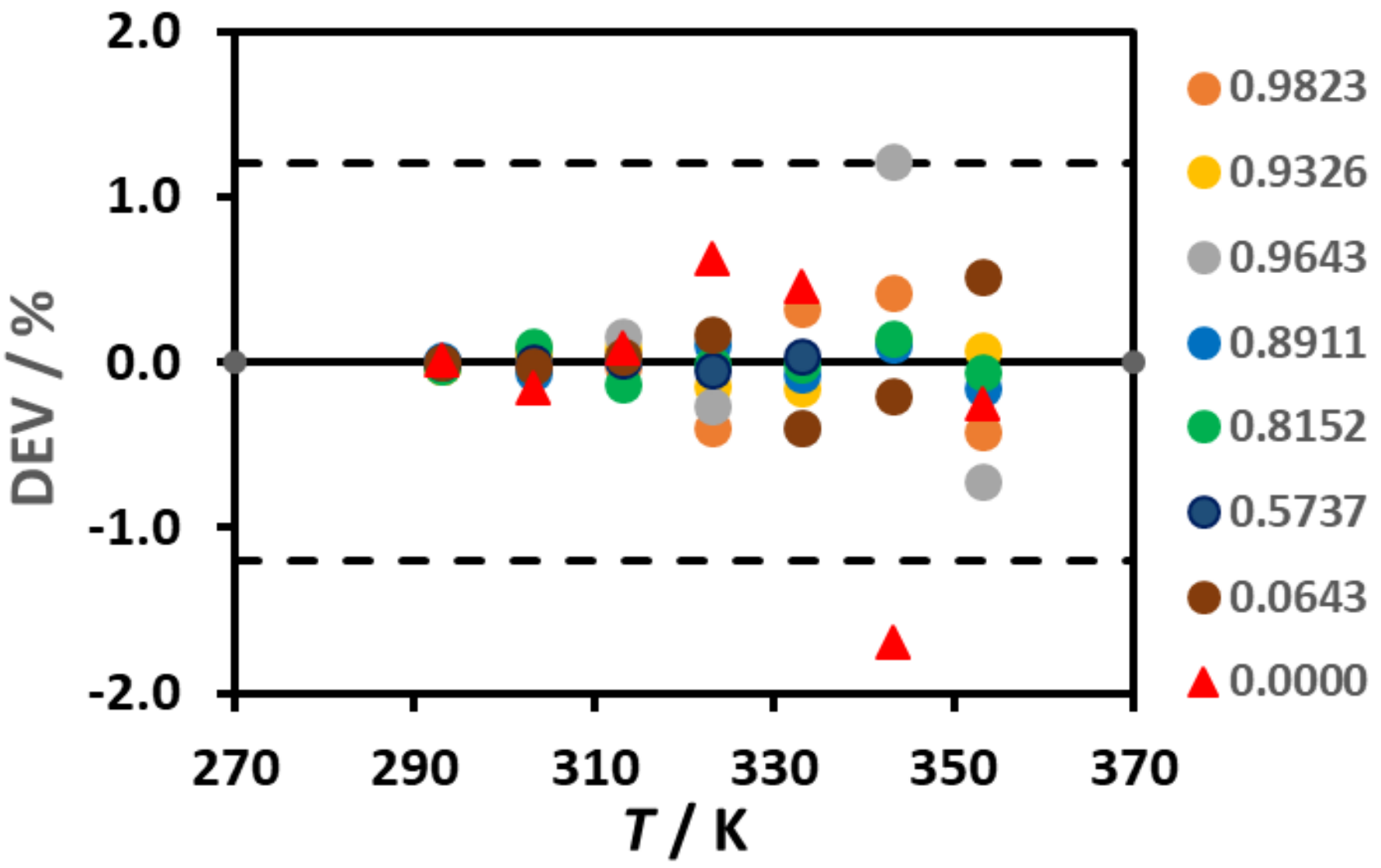

Figure 17 – Deviation plot of our data from the VFT fits, $\left(\frac{\eta_{exp} - \eta_{VFT}}{\eta_{VFT}}\right) \times 100\%$, for each molar fraction of water. The dashed lines represent the combined uncertainty of our data at a 95% confidence level ($k$ = 2).

The excess viscosity, $\eta^E$, was calculated from the viscosity of the mixture, $\eta_{mix}$, the viscosity of the pure ionic liquid, $\eta_{IL}$, and that of pure water, $\eta_w$,[8], at the same temperature according to:

$$\eta^{E} = \eta_{mix} - x_w\eta_w - x_{IL}\eta_{IL} \quad (15)$$

Our results are presented in Fig. 18. The curves are parabolic for all temperatures, with a minimum around $x_w$ = 0.43, not depending on temperature, and all negative, meaning that the ions of the ionic liquid and water attract themselves, in a privileged way, stronger than homomolecular or ion-ion interactions. This explains the complete miscibility of the system, agreeing with the results for the excess volumes, and will be analysed below.

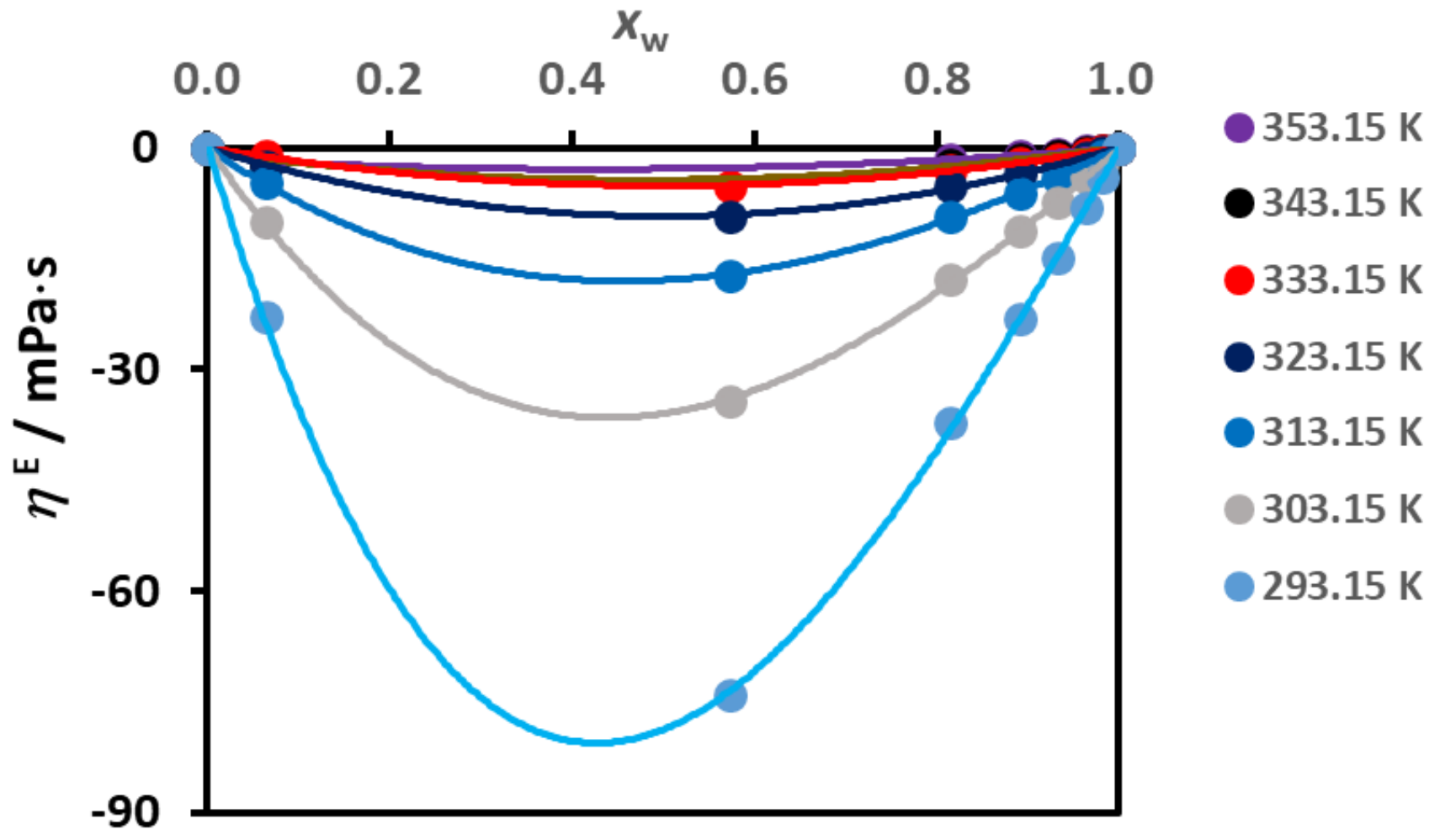

Figure 18 – Excess viscosity for the aqueous [$C_2$mim][$CH_3SO_3$] system, at the studied temperatures, as a function of the mole fraction of water. Lines are polynomial fits to the data.

**3.3. Electrical conductivity.** Table 4 and Fig. 19 shows the results obtained for the electrical conductivity, $\kappa$, as a function of composition, for each temperature. It also includes the data published before by the authors, for the "pure" [$C_2$mim][$CH_3SO_3$], containing the same amount of water ($x_w$ ~ 0.066)[2]. These results agree within 0.02%, or 0.005 S·m$^{-1}$, a negligible difference compared to the mutual uncertainty of the data. As in other ionic liquids mixtures with water, the electrical conductivity of the mixture will pass through a maximum for $x_w$ > 0.9 and then starts to decrease to the value of the pure water. For this system, the maximum of electrical conductivity varies from $x_w$ ~ 0.95 at 293.15 K to $x_w$ ~ 0.935 at 343.15 K, indicating a change of the structure of the mixture, as found for other properties described before. Data obtained was regressed with an analytical function, valid for all the temperatures, in the form of eq. (16):

$$\kappa(x_w, T) = \frac{a_0(T) + a_1(T)x_w^{0.5} + a_2(T)x_w + a_3(T)x_w^{1.5}}{1 + a_4(T)x_w^{0.5} + a_5(T)x_w + a_6(T)x_w^{1.5}} \quad (16)$$

where $\kappa(x_w, T)$ is the electrical conductivity at a given water mole fraction and temperature $T$, and $a_i$ are the different non-linear regression coefficients, obtained using the Levenberg-Marquardt algorithm. Their values are displayed in Table S13, and the standard deviations of the fits are less than 0.08 S·m$^{-1}$. The value of $a_0$ is the electrical conductivity of the pure ionic liquid ($x_w$ = 0), calculated with this function.

**Table 4 – Electrical conductivity, $\kappa$, for the aqueous system [$C_2$mim][$CH_3SO_3$] from $T$ = (293.15 to 353.15) K, at $P$ = 0.1 MPa[a,b], as a function of the molar fraction of water, $x_w$, at 3kHz. Water free values for the electrical conductivity of pure [$C_2$mim][$CH_3SO_3$] are also presented. Last line shows the calculated values for the limiting molar conductivity of the ionic liquid in water. Values for water measured in this work also shown.**

| $x_w$ [c] | $T$ = 293.15 K | $T$ = 303.15 K | $T$ = 313.15 K | $T$ = 323.15 K | $T$ = 333.15 K | $T$ = 343.15 K | $T$ = 353.15 K |
|---|---|---|---|---|---|---|---|
| | $\kappa$ / S·m$^{-1}$ | $\kappa$ / S·m$^{-1}$ | $\kappa$ / S·m$^{-1}$ | $\kappa$ / S·m$^{-1}$ | $\kappa$ / S·m$^{-1}$ | $\kappa$ / S·m$^{-1}$ | $\kappa$ / S·m$^{-1}$ |
| 1.0000 | 0.0000 | 0.0001 | 0.0002 | 0.0005 | 0.0015 | 0.0018 | 0.0023 |
| 0.9980 | 0.6929 | 0.8518 | 1.0208 | 1.1978 | 1.3838 | 1.5865 | 1.8096 |
| 0.9835 | 3.2506 | 4.0310 | 4.8664 | 5.7270 | 6.6591 | 7.5909 | 8.6060 |
| 0.9720 | 4.0728 | 5.0881 | 6.1821 | 7.3331 | 8.5274 | 9.7830 | 11.086 |
| 0.9511 | 4.5515 | 5.7472 | 7.0378 | 8.4243 | 9.8527 | 11.428 | 12.974 |
| 0.9300 | 4.3905 | 5.6082 | 6.9540 | 8.4250 | 9.9642 | 11.560 | 13.232 |
| 0.8878 | 3.6219 | 4.7410 | 5.9980 | 7.2773 | 8.8618 | 10.438 | 12.062 |
| 0.7966 | 2.2855 | 3.1286 | 4.0763 | 5.1493 | 6.3269 | 7.5811 | 8.8981 |
| 0.6248 | 1.1887 | 1.6716 | 2.2520 | 2.9173 | 3.6693 | 4.4966 | 5.3193 |
| 0.0661 | 0.2010 | 0.3509 | 0.5580 | 0.8268 | 1.1581 | 1.5525 | 2.0068 |
| 0.0661[d] | 0.2014 | 0.3512 | 0.5578 | 0.8257 | 1.1562 | 1.5490 | 2.0022 |
| 0.0000[e] | 0.201 | 0.351 | 0.558 | 0.826 | 1.156 | 1.549 | 2.002 |
| $\Lambda_m^\infty / \mathrm{S \cdot cm^2 \cdot mol^{-1}}$ [f] | 77.04 | 94.59 | 113.4 | 133.6 | 154.9 | 179.0 | 208.1 |

[a] Standard uncertainties: $u(T)$ = 0.01 K, $u$(P) = 1 kPa

[b] Expanded relative uncertainty $U(\kappa)$ = 2 %, at a 95 % confidence level ($k$=2)

[c] Relative combined standard uncertainty, $u_{r,c}(x_w)$ = 0.002

[d] Bioucas et al. (2018)[2]

[e] Extrapolated water-free values – pure ionic liquid. Expanded relative uncertainty $U(\kappa)$ = 5 %, at a 95 % confidence level ($k$=2)

[f] Expanded relative uncertainty $U(\Lambda_m^0)$ = 2 %, at a 95 % confidence level ($k$=2).

The regression curves are also shown, for the different temperatures, in Fig. 19. The uncertainty of the fitting, and the almost constant value of the electrical conductivity near infinite dilution of water, permit the extrapolation of the value to ionic liquid water-free results, as only 2 points in this range were measured. These results also presented in Table 4, do not deviate from the values at $x_w \sim 0.066$, by more than 0.005 $S \cdot m^{-1}$. The extrapolated water-free values – pure ionic liquid expanded relative uncertainty is slightly higher and assumed to be $U(\kappa) = 5$ %, at a 95 % confidence level ($k$=2). These data points do not deviate from the fitting curve by more than -0.02%. Further measurements are needed in the region $0.06 < x_w < 0.6$, to obtain a less uncertain value for the extrapolated water-free values. The variation of the electrical conductivity with temperature, at constant mole fraction of water is quadratic (eq. S2), Fig. S3, and coefficients, their standard deviations and the standard deviation of the fits are shown in Table S13.

Figure 20 shows the deviations of our data and the data of Stark et al. (2011)[39], from the fitted equations, the agreement between the two sets of data being between their mutual uncertainty, except for the values of the pure ionic liquid, probably due to the different purities of the starting material or deficient cell calibration, already found by Bioucas et al. (2018) and Harris and Kanakubo (2016)[69].

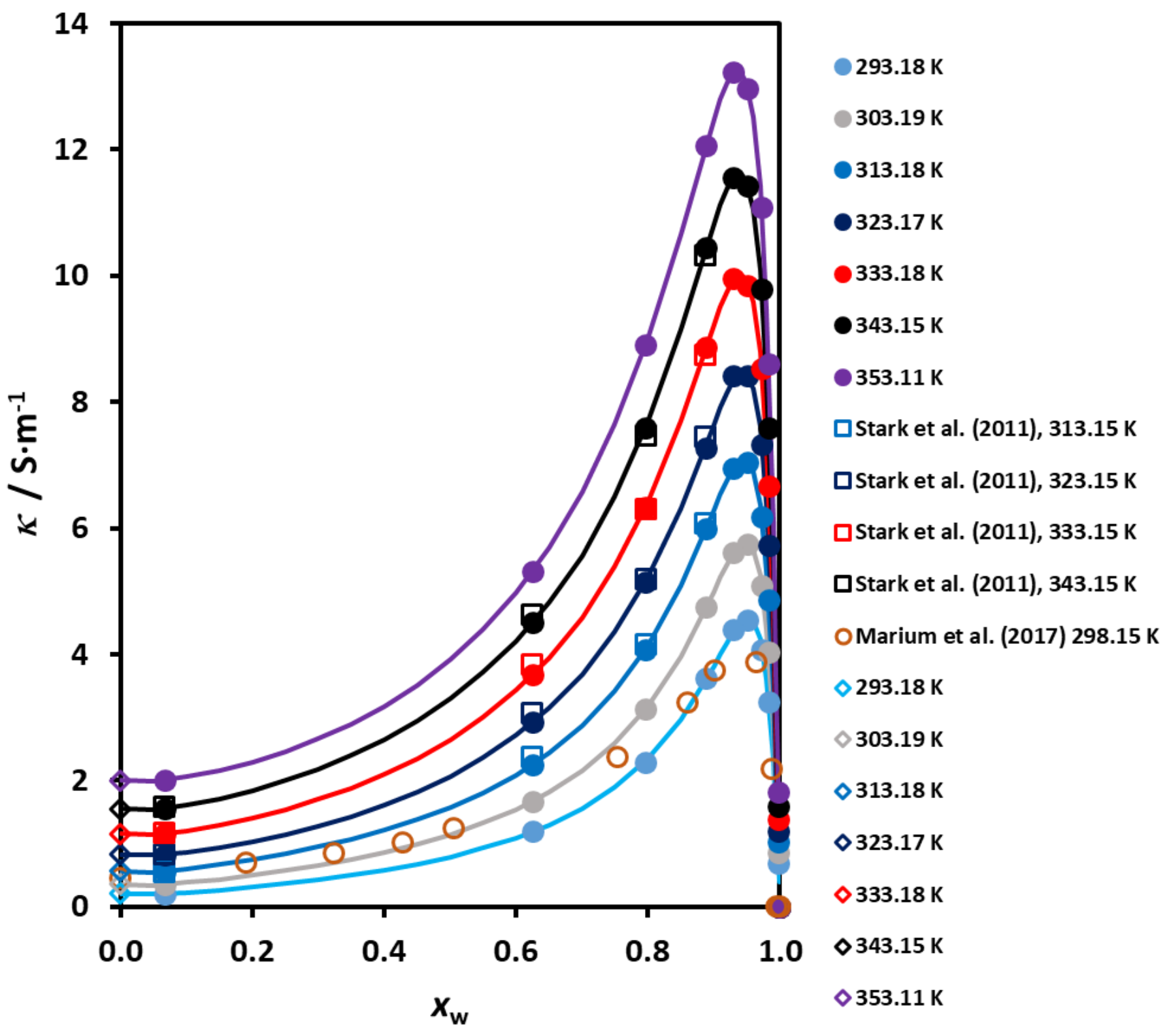

Figure 19 – The electrical conductivity of the aqueous $[C_2mim][CH_3SO_3]$ system, at the studied temperatures, as a function of the mole fraction of water, full dots. Lines are fits to the data, following the model described in the text, eq. (16), at the corresponding temperatures. Also shown data of other authors, open squares – Stark et al. (2011)[39], open circles, Marium et al. (2017)[18] and extrapolated water-free values for the ionic liquid, open diamonds.

As reported in our study of the electrical conductivity of [C2mim][CH3COO] + water mixtures[8], one of the best ways to understand the interionic/molecular forces in ionic liquid systems is the use of the Walden plot, proposed by Angell and co-workers[48,49,50], to classify the ionic power of the ionic liquid, followed by the concept of ionicity proposed by Watanabe and co-workers[51,52,53,54].

Figure 21 shows the molar electrical conductivity for the aqueous [$C_2$mim][$CH_3SO_3$] system, $\Lambda_m$, defined as $\Lambda_m = \kappa/c_{\mathrm{IL}}$, where $c_{\mathrm{IL}}$ is the molar concentration of the ionic liquid, given by $c_{\mathrm{IL}} = x_{\mathrm{IL}}\left(\frac{\rho}{M}\right)$ and, it can be seen that the molar conductivity of the ionic liquid increase smoothly up to the molar fraction $x_{\mathrm{w}} \approx 0.6$, showing that the mobility of the ions becomes greater around this composition. and then starts to increase fast for all the temperatures, the increase being greater for the highest temperature, probably due to an initial restriction of mobility of the ions when few molecules of water are present, which occupy holes in the IL structure.

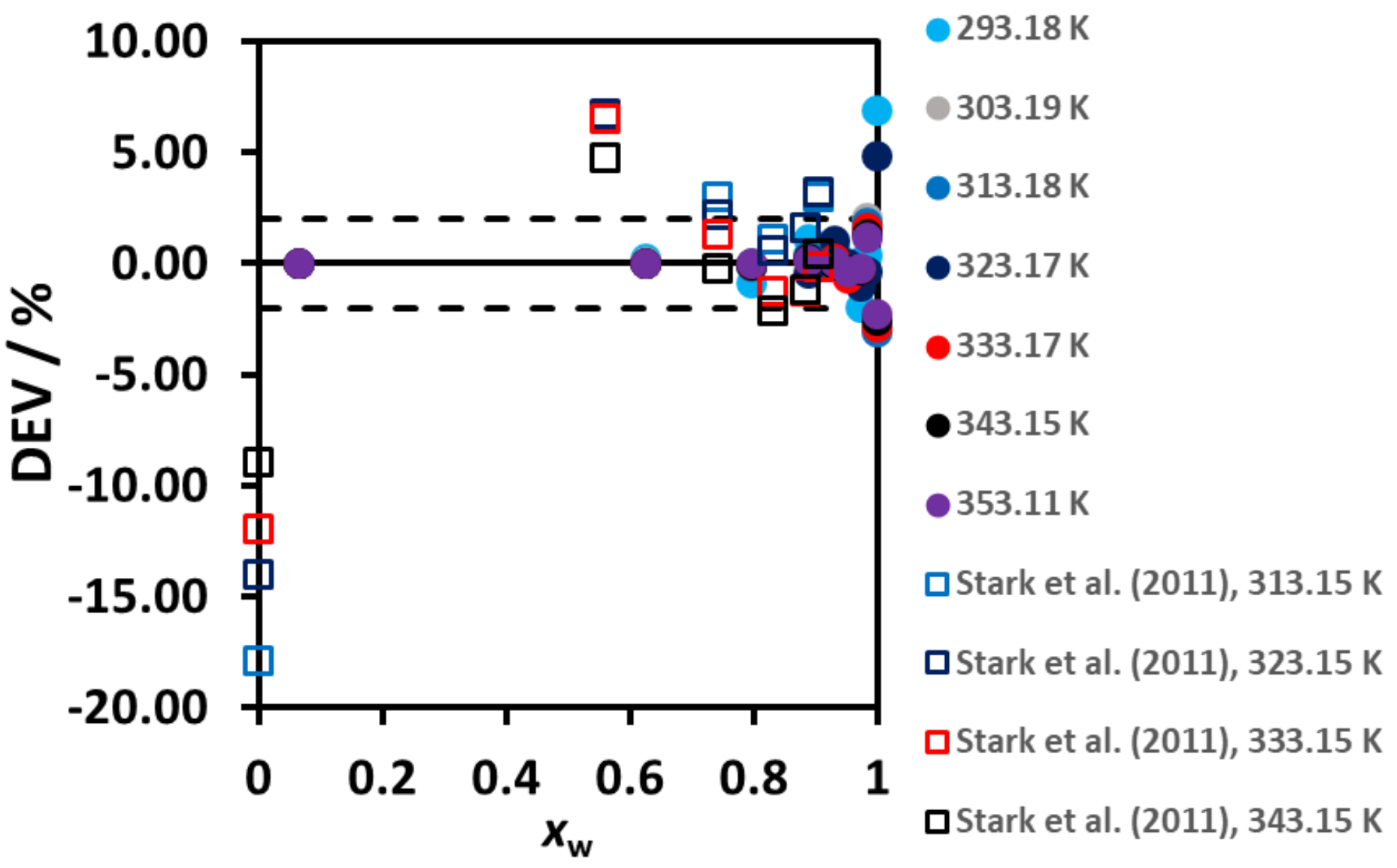

Figure 20 – Deviations of our data (filled dots) and data of Stark et al. (2011)[39], open squares from the fitted equations for the electrical conductivity, eq. (16), $\left(\frac{\kappa_{exp} - \kappa_{fit}}{\kappa_{fit}}\right) \times 100\%$. Dotted lines represent the expanded relative uncertainty of our data, $U_{\mathrm{r}}$ = 2%.

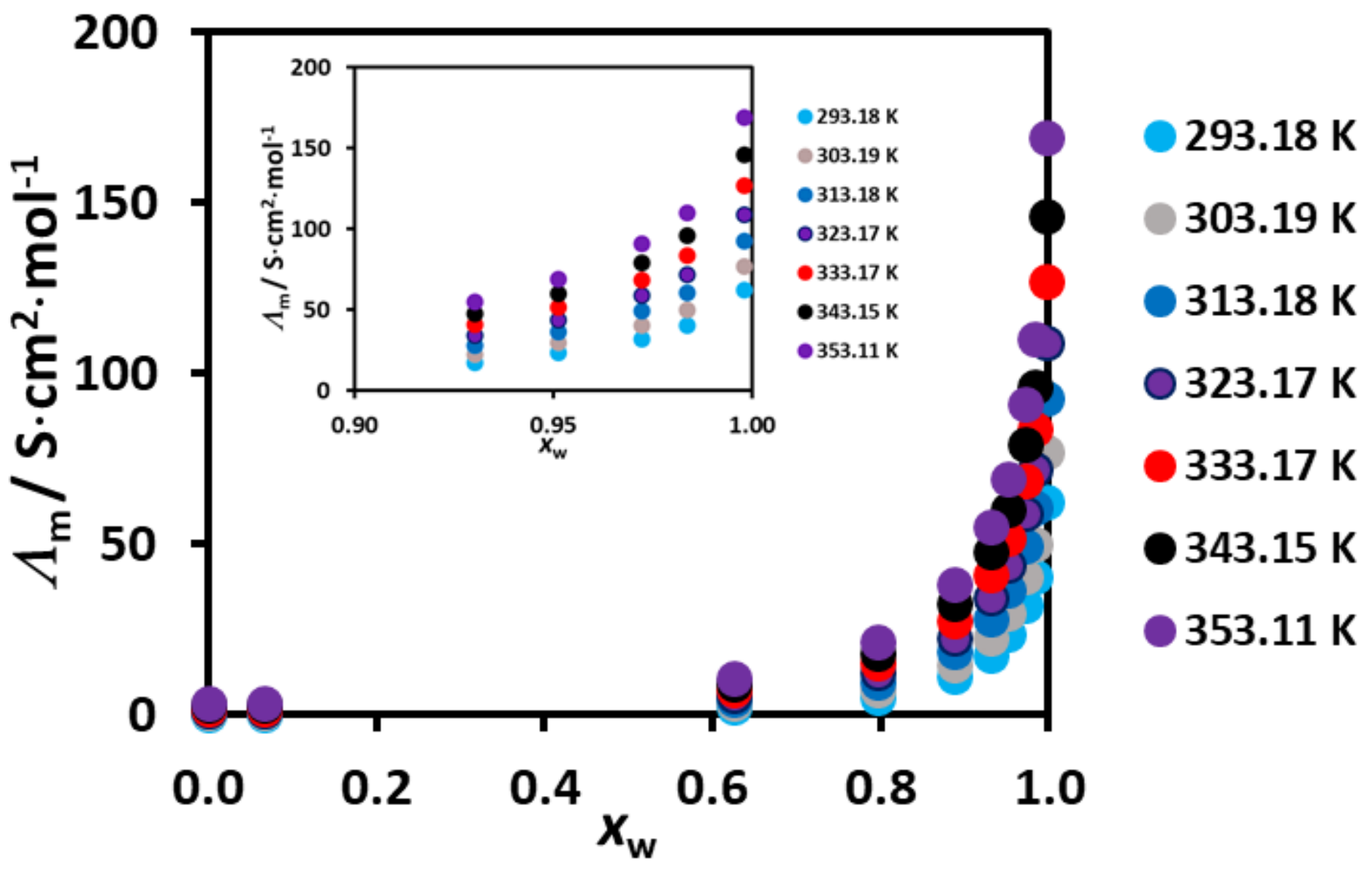

Figure 21 – The molar electrical conductivity for aqueous $[C_2mim][CH_3SO_3]$ system, as a function of the mole fraction of water, for the temperatures studied. The inset shows the variation for $0.90 < x_w < 1.00$.

The current data permits the calculation of the ionic liquid limiting molar conductivity, $\Lambda_m^\infty$, at all temperatures studied. This was done by extrapolation with a quadratic function the molar electrical conductivity as a function of $c_{IL}^{1/2}$ for zero concentration of the ionic liquid, as explained in our previous work, Bioucas et al. (2018)[2]. Table 4 shows the value obtained for each temperature. The value obtained at 293.15 K agree with the value previously obtained[2], within their mutual uncertainty. Results for the other temperatures are new, and no data exists for comparison.

The Walden rule relates limiting molar conductivity, $\Lambda_m^0$ , of the ionic liquid system, with the viscosity, $\eta$, of the pure ionic liquid, in the form:

$$\Lambda_m^0 \times \eta^\alpha = C' \tag{17}$$

where $\alpha$ was originally proposed as 1 ("ideal" Walden rule) but found to be of the order 0.8 ± 0.1, by several workers, well analyzed by Schreiner et al. (2010)[55], and $C'$ a constant. By definition, for a pure ionic liquid,

$$\Lambda_m^0 = \kappa_{wf}\left(\frac{\rho_{wf}}{M_{IL}}\right) \tag{18}$$

and

$$\log \Lambda_m^0 = \log C' + \alpha \log(\eta^{-1}) \tag{19}$$

Figure 22 displays the Walden plot for the pure ionic liquid and its mixtures, for the mole fractions in the ionic liquid rich side. We represent $\log \Lambda_m^0$ as a function of $\log(\eta^{-1})$, using our data for density, electrical conductivity and viscosity, as a function of temperature, between 293.15 and 353.15 K. For the pure ionic liquid, using water-free data, we obtain a straight line, with slope 0.858 ± 0.005, giving α = 0.86, agreeing for this liquid with the general trend observed for other ionic liquids, although [$C_6$mim][($CF_3SO_2)_2N$] obeys the "ideal" Walden rule[29]. However, there is a current discussion about the definition of Angell and co-workers of the ionicity concept, namely about the use of deviations to an arbitrarily chosen 0.01 M KCl aqueous solution as a reference, not thermodynamically ideal (activity coefficients greater than unity), and the consideration of the "ideal" Walden rule[52,56]. In fact, the aqueous solution of KCl was chosen because potassium and chloride ions have similar hydrodynamic radii in aqueous solutions.

The values of $\Delta W$, deviations to the KCl line, can be calculated, as the procedure described by MacFarlane et al (2009)[57], and are presented in Table S16. These values were calculated for a reference solvent with a viscosity of 10 mPa·s (vertical dashed line), for each composition Equation 19 is used for this value of viscosity and $\Delta W = 1 - (\alpha + \log C')$, the difference between the value of the log ($\Lambda_m$) of the 0.01 M KCl reference solution (=1) and that of the mixture, for $\eta$ = 10 mPa·s. Ionicity was calculated to obey the scale developed by these authors, a value of Ionicity of 10% corresponding to $\Delta W$ = 1, meaning that the ionic liquid is exhibiting only 10% of the 'ideal' molar conductivity that it should possess for a given viscosity. A value of $\Delta W$ = 0 correspond to an ionicity of 100%, equal to that of the reference line. Conversion between $\Delta W$ and % Ionicity can be obtained through the relation $\text{Ionicity (\%)} = 100 \times 10^{-\Delta W}$.

The pure ionic liquid has a value of $\Delta W$ = 0.26, a very good sign of its good electrical conductivity, 74 % of that of 0.01M KCl, and with an ionicity of 55%. The ionicity starts to decrease as water molecules are added, up to a value of 53 % for $x_w$ = 0.0661 ($x_{IL}$=0.9339), and then increases up to 69 % for $x_w$ = 0.8878 ($x_{IL}$=0.1122). This could signify that the cation and the anion of the ionic liquid can move better in the presence of water than in the pure ionic liquid, which does not support that ionicity is a sign of cation/anion interaction. However, we must remember that the number of ions decreases significantly in relation to the pure ionic liquid, and the viscosity of the mixture also decreases, increasing the values of the diffusion coefficients and transport numbers of the ions. The balance of all these factors explains an increase in the molar conductivity as a function of the molar fraction of water.

It is very interesting to note that D'Agostino et al (2018) [58] measured the self-diffusion of the [$C_2$mim]$^+$, [($CF_3SO_2)_2N$]$^-$ and [$CH_3COO$]$^-$ ions, using a 1H and 19F PFG NMR, at 353.15 K, and found that strong degree of ion pairing was observed for [$C_2$mim][$CH_3COO$], while the substitution of the [$CH_3COO$]$^-$ anion by the [($CF_3SO_2)_2N$]$^-$ anion reduced the pairing between the ions, attributed to a lower electric charge density on this anion, hence a weaker electric interaction with the [$C_2$mim]$^+$ cation. The self-diffusion of the [$C_2$mim]$^+$ in [$C_2$mim][$CH_3COO$], and in [$C_2$mim][($CF_3SO_2)_2N$] was found to be 12.8 and 28x$10^{-11}$ $m^2 \cdot s^{-1}$, respectively, while the value obtained for [($CF_3SO_2)_2N$]$^-$ was 16.5x$10^{-11}$ $m^2 \cdot s^{-1}$ and that of [$CH_3COO$]$^-$ ions was 12.5x$10^{-11}$ $m^2 \cdot s^{-1}$. Unfortunately, no data is available for our system, namely for the methane sulfonate anion, where only infinite dilution diffusion coefficient in water is determined below, and discussed in comparison with other anions data.

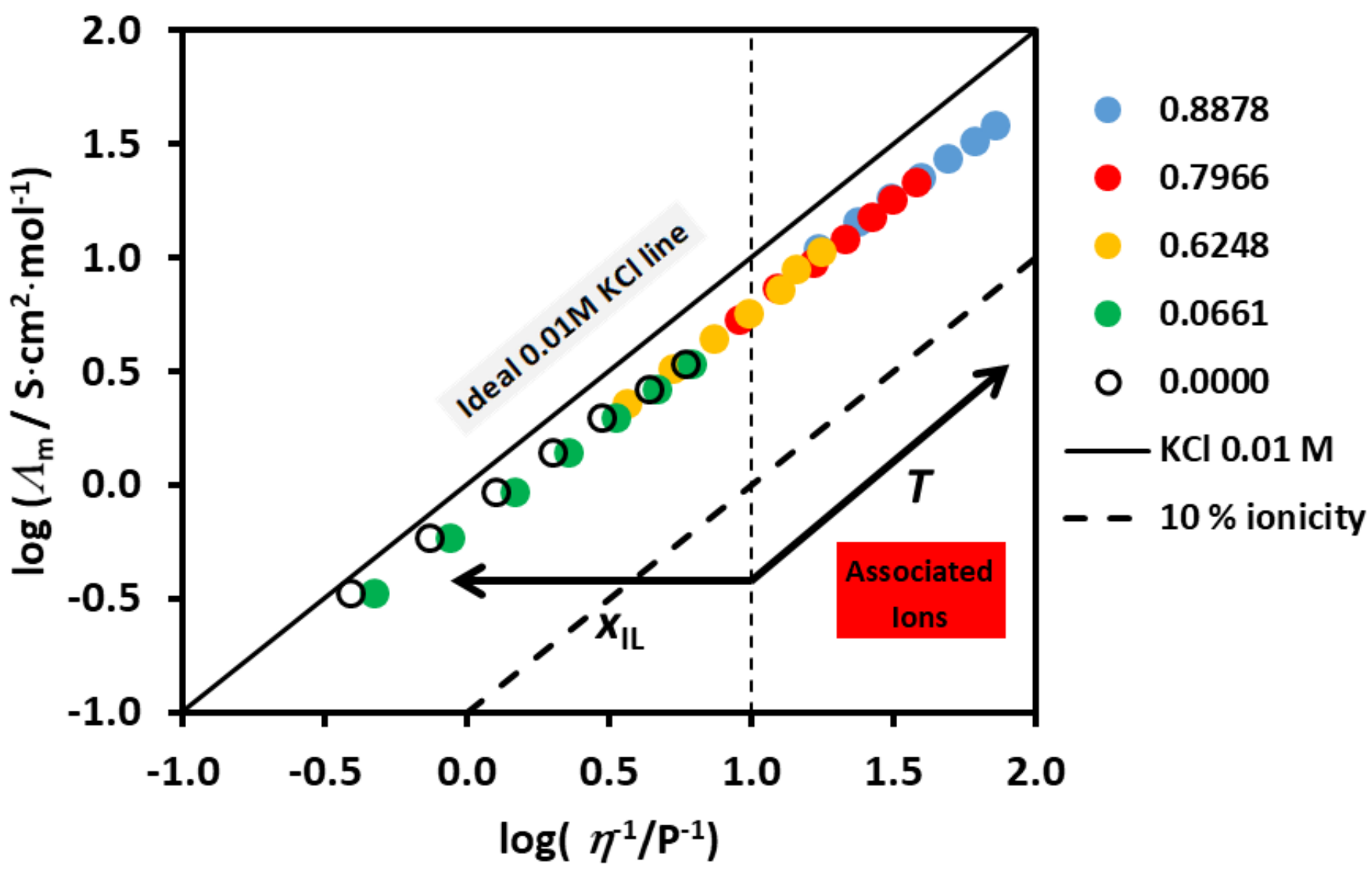

Figure 22 – Walden plot for the aqueous [$C_2$mim][$CH_3SO_3$] system. Points in the figure correspond to the different water mole fractions. $x_w$ = 0.0000 corresponds to the water-free ionic liquid. The dashed line corresponds to $\Delta W = 1$, the liquid exhibiting only 10% of the 'ideal molar conductivity' for the viscosity of 10 mPa.s (vertical dashed line)[47]. Zone of preference for ion clusters (ion pairs/aggregates, any type of ion association), below the dashed line, is also identified.

In our previous paper for [$C_2$mim][$CH_3$COO] + water mixtures[8], we also calculated the deviations $\Delta W$, to the 0.01M KCl line, used as a reference. However, these values have to be corrected to the same viscosity of the media, 10 mPa.s, and not function of temperature, which changes the viscosity of the media. The corrected values are also presented in Table S16. From the corrected data it can be seen that the pure ionic liquid as an ionicity of 47 % ($\Delta W$ = 0.33), and that the addition of water increased the ionicity of the system up to a value of 72 % ($\Delta W$ = 0.14), for $x_{IL}$ = 0.2003.

In conclusion, in comparison with [$C_2$mim][$CH_3$COO], [$C_2$mim][$CH_3SO_3$] has an electrical conductivity twice as big and a higher ionicity. This will favor its use in many applications, like new battery electrolytes.

**3.4. Thermal conductivity.** Table 5 and Fig. 23 show the thermal conductivity of the aqueous system [$C_2$mim][$CH_3SO_3$] for 293.15 < $T$/K < 353.15, at $P$ = 0.1 MPa, for the different mole fractions of water, including the extrapolated water-free values (pure ionic liquid), obtained recently by Lozano et al. (2020)[27] and data of Bioucas et al. (2018)[2] for the "pure" ionic liquid ($x_w$ = 0.0040). If these data are extrapolated for water-free values, they agree with the Lozano et al (2020)[27] to within -1.64 % at 293.15 K and -3.55 % at 353.15 K, within the mutual uncertainty of these two sets of measurements. Data for the molar fractions 0.9925 and 0.9333 were taken, for consistency tests described below, at different temperatures (see Fig. 24), and corrected for the nominal temperatures, according to procedure described before[27].

The thermal conductivity of the mixture increases very slowly with composition up to $x_w$ ~ 0.8, and then starts to increase very fast, for the water rich mixtures. In addition, the temperature dependence is very weak, being clearer also in the water rich region, as found for other properties. This is related with the structure of the mixture, starting to be dominated by the ionic liquid anion and cations interactions, changing to IL clusters interactions with water, approaching the structure of a dilute electrolyte solution, with ions completely solvated by water. This change of structure can also be identified in the inset of Fig.23, were values are shown for $0.985 < x_w < 1$, by a maximum for temperatures above 323.15 K. In order to confirm the existence of this maximum we selected one composition (identified by BT(K), crosses) $x_w = 0.9925$ and use the other 4 highest mole fractions of water to devise a trend line for the variation. The selected composition agrees completely with the expected trend, a sign of the real existence of this maximum. The points indicated by LT(K), triangles, in the main graphic correspond to a further consistency test, with a composition $x_w = 0.9333$, also measured at 353.15 K.

For each mole fraction, the experimental data was regressed linearly, at each temperature, in the form of eq. (20):

$$\lambda\,(x_w, T) = d_0(x_w) + d_1(x_w)T \qquad (20)$$

where $\lambda\,(x_w, T)$ is the thermal conductivity at a given water mole fraction and temperature $T$, and $d_i$ are the different non-linear regression coefficients, shown in Table S17, with their standard deviations, and with the standard deviation of the fits, varying between 0.0004 and 0.0052 $W \cdot m^{-1} \cdot K^{-1}$, a maximum of 1%. The variation of the thermal conductivity as a function of temperature can be seen in Fig. 24, having a linear dependence, but changing slope, especially for the higher mole fractions of water, near the maximum shown in Fig. 23. Figure 25 shows the deviation plot of the experimental data points from the regression lines. It can be seen that most of the points lie within the estimated expanded uncertainty of our measurements.

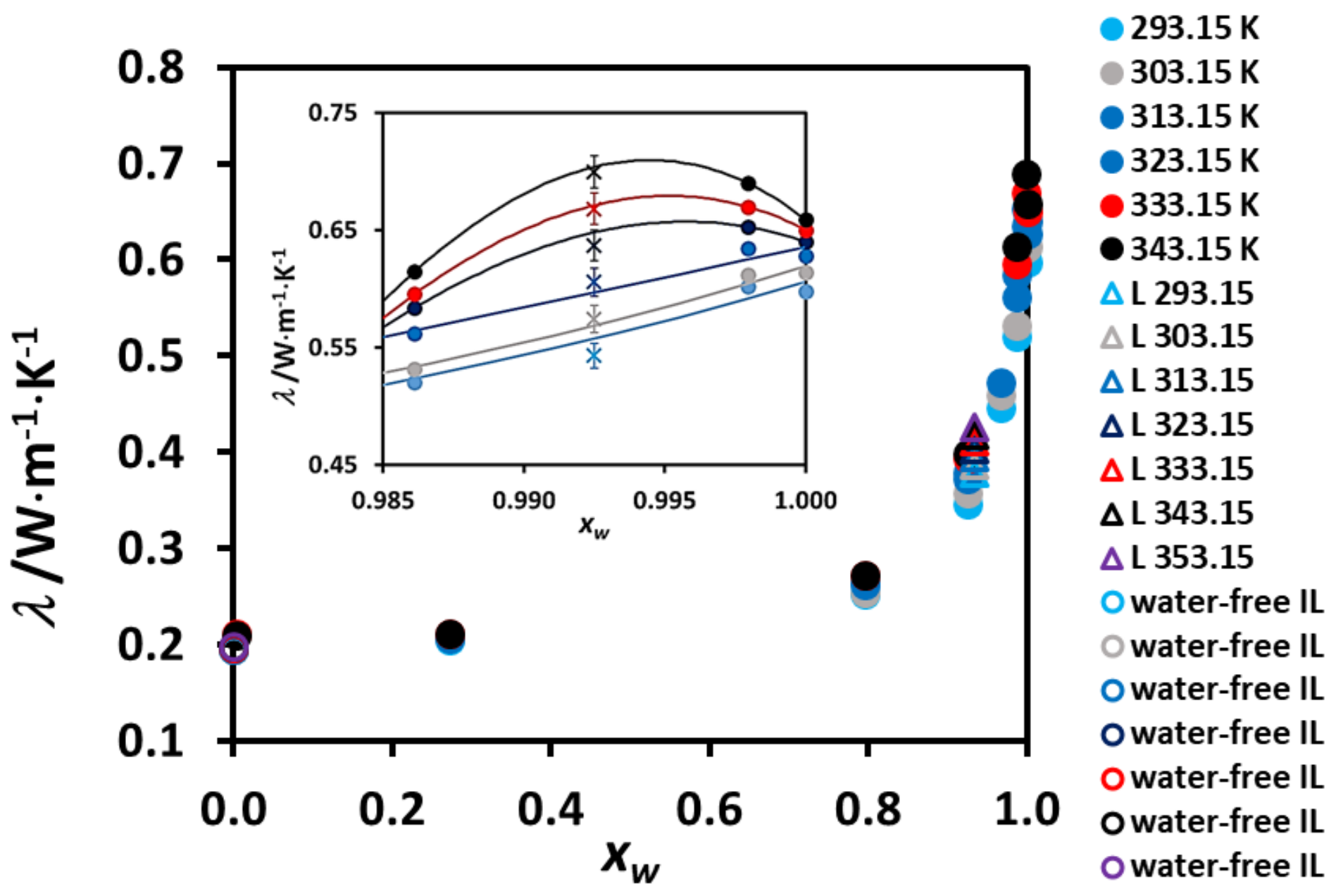

Figure 23 – The thermal conductivity for the aqueous [$C_2$mim][$CH_3SO_3$] system, as a function of the mole fraction of water, for the temperatures studied. The inset shows the maximum in the thermal conductivity found for temperatures above 323.15 K. Lines show, for each temperature, guides for the eye, for the four highest mole fractions of water, with the composition indicated by crosses, BT(K), $x_w$ = 0.9925, which fall perfectly in the trend, and within the uncertainty of the lines and the measurements (error bars). Points indicated by LT(K), triangles, in the main graphic correspond to a further consistency test, with a composition $x_w$ = 0.9333, also measured at the highest temperature, 353.15 K.

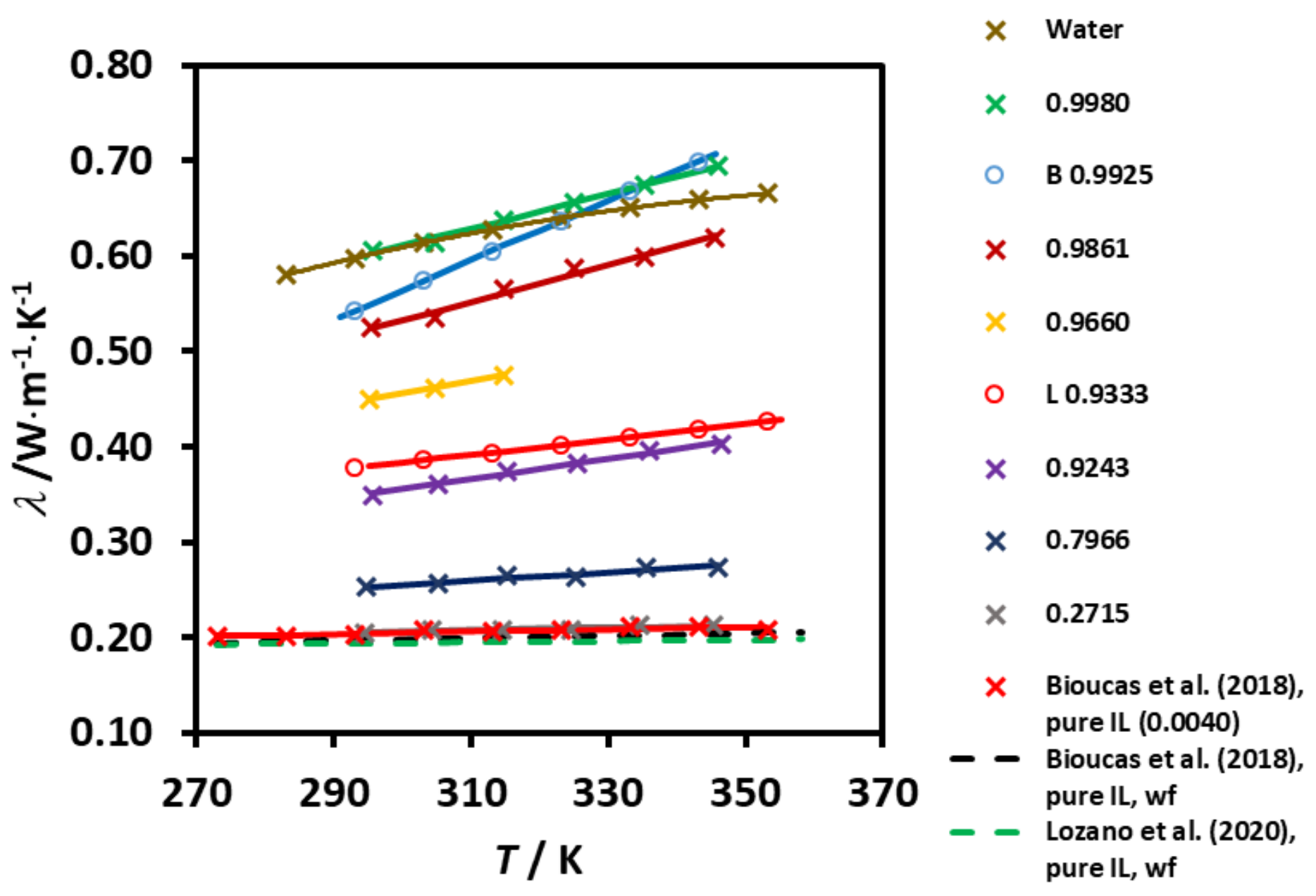

Figure 24 – The variation of the thermal conductivity for the aqueous [$C_2$mim][$CH_3SO_3$] system, as a function of temperature, for each mole fraction of water. The lines represent the linear regression fits. Points shown as B 0.9925 and L 0.9333 were additionally taken for the consistency tests explained in this section, obtained at different temperatures. Also shown the values for pure water, present work, and for the extrapolated water-free, pure ionic liquid, obtained by Bioucas et al. (2018)[2] and Lozano et al. (2020)[27], dashed lines.

Table 5 also shows that thermal conductivity measurements of water obtained with the current instrument by Lozano et al. (2020)[27], corrected to nominal temperatures, do not deviate from the IUPAC standard reference data[19] by more than 0.4%, and from NIST Database REFPROP[59] by more than 0.8%. Deviations are plotted in Fig. 26.

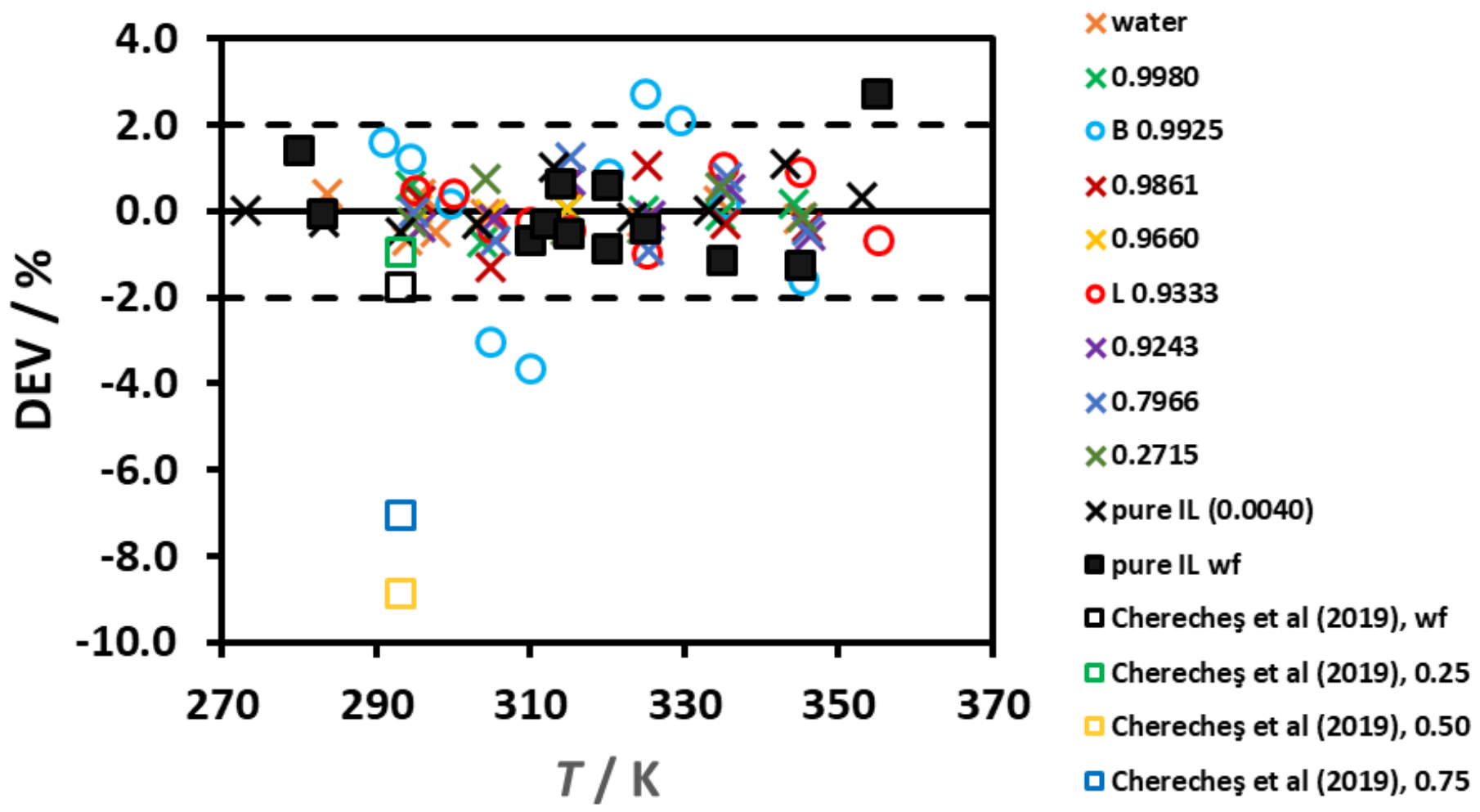

Figure 25 – Deviation of the thermal conductivity data from our correlations $\left(\frac{\lambda_{exp} - \lambda_{fit}}{\lambda_{fit}}\right) \times 100\%$, eq. (20), for the several mole fractions of water, including the data of Lozano et al. (2020)[27]. Dashed lines represent the uncertainty of the present data, $U_r(\lambda) = 2\%$. Also included the data of Chereches et al. (2019)[38], at 293.15 K, open squares.

**Table 5 – Thermal conductivity[a] $\lambda$ for the aqueous system [$C_2$mim][$CH_3SO_3$] from $T$ = (293.15 to 343.15) K, at $P$ = 0.1 MPa[a], as a function of the molar fraction of water, $x_w$ . Extrapolated water free values for the thermal conductivity of pure [$C_2$mim][$CH_3SO_3$] are also presented[27]. Values for water measured in this work also shown.**

| | $T$ = 283.15 K | $T$ = 293.15 K | $T$ = 303.15 K | $T$ = 313.15 K | $T$ = 323.15 K | $T$ = 333.15 K | $T$ = 343.15 K | $T$ = 353.15 K |
|---|---|---|---|---|---|---|---|---|
| $x_w$[b] | $\lambda$ / $W \cdot m^{-1} \cdot K^{-1}$ | $\lambda$ / $W \cdot m^{-1} \cdot K^{-1}$ | $\lambda$ / $W \cdot m^{-1} \cdot K^{-1}$ | $\lambda$ / $W \cdot m^{-1} \cdot K^{-1}$ | $\lambda$ / $W \cdot m^{-1} \cdot K^{-1}$ | $\lambda$ / $W \cdot m^{-1} \cdot K^{-1}$ | $\lambda$ / $W \cdot m^{-1} \cdot K^{-1}$ | $\lambda$ / $W \cdot m^{-1} \cdot K^{-1}$ |
| 1.0000[c] | 0.5802 | 0.5978 | 0.6137 | 0.6277 | 0.6398 | 0.6501 | 0.6586 | 0.6652 |
| 0.9980 | - | 0.6017 | 0.6118 | 0.6346 | 0.6525 | 0.6696 | 0.6896 | - |
| 0.9925 | - | 0.5428 | 0.5741 | 0.6054 | 0.6367 | 0.6681 | 0.6994 | - |
| 0.9861 | - | 0.5203 | 0.5315 | 0.5616 | 0.5836 | 0.5952 | 0.6142 | - |
| 0.9660 | - | 0.4468 | 0.4592 | 0.4726 | - | - | - | - |
| 0.9333 | - | 0.3778 | 0.3859 | 0.3939 | 0.4020 | 0.4101 | 0.4182 | 0.4263 |
| 0.9243 | - | 0.3468 | 0.3581 | 0.3721 | 0.3797 | 0.3929 | 0.3991 | - |
| 0.7966 | - | 0.2521 | 0.2547 | 0.2640 | 0.2626 | 0.2715 | 0.2724 | - |
| 0.2715 | - | 0.2041 | 0.2078 | 0.2067 | 0.2083 | 0.2118 | 0.2118 | - |
| 0.0040[d] | - | 0.2032 | 0.2070 | 0.2058 | 0.2074 | 0.2108 | 0.2104 | 0.2082 |
| 0.0000[e] | - | 0.1936 | 0.1942 | 0.1948 | 0.1955 | 0.1961 | 0.1967 | 0.1973 |

[a] Standard uncertainty $u(T)$ = 0.02 K, $u(P)$ = 1 kPa; expanded relative uncertainty $U(\lambda)$ = 2 %, at a 95 % confidence level ($k$=2);

[b] Relative combined standard uncertainty, $u_{r,c}(x_w)$ = 0.002

[c] Present work

[d] Values from Bioucas et al. (2018)[2]

[e] Extrapolated water-free values – pure ionic liquid, and calculated for nominal temperatures, using the data of Lozano et al. (2020)[27]

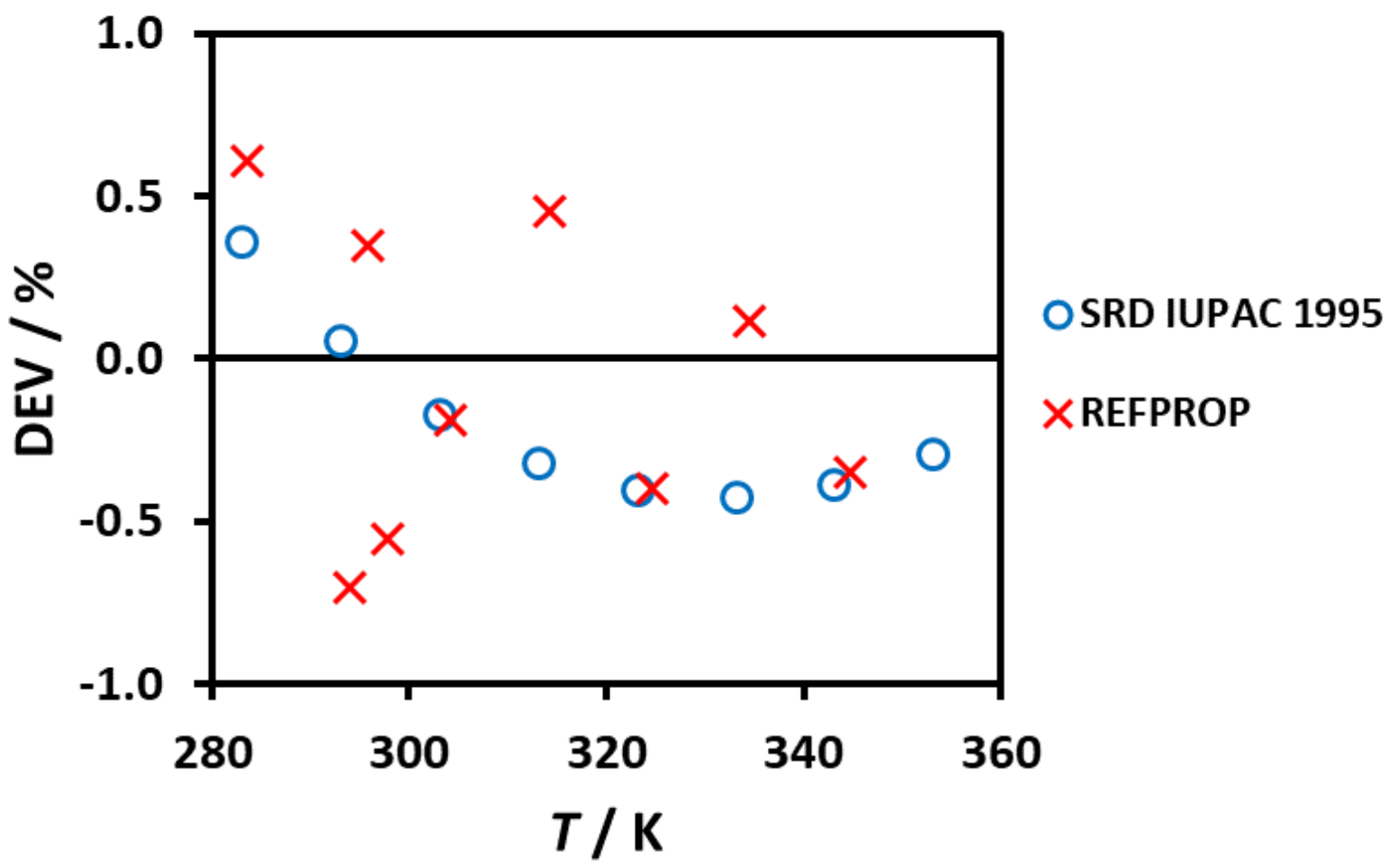

Figure 26 - Deviations between our thermal conductivity experimental data for water (nominal values) from Ramires et al. (1995)[28] and from REFPROP[59], raw data, $\left(\frac{\lambda_{exp}-\lambda_{corr}}{\lambda_{corr}}\right) \times 100\%$ .

Regarding the binary mixtures with water, there is only one set of data available, those of Chereches et al. (2019)[42], using a sample of $[C_2mim][CH_3SO_3]$ with a mass fraction of water of 0.029 %, and 95 % purity. These authors measured the "pure" ionic liquid and 3 mixtures with $x_w$ = 0.25, 0.50 and 0.75, for temperatures at 298.15, using a KD2 Pro Thermal Properties Analyzer, with an estimated uncertainty of 6 %, at 95% confidence level. Their data for the "pure" ionic liquid was corrected to extrapolated water-free value. The comparison for the different mole fractions was done, by extrapolating quadratically our experimental values to their mole fractions. Values for the pure ionic liquid and for the water mole fraction 0.25 deviates from our data -1.8 % and -0.96 %, in excellent agreement, but for the mole fractions of water 0.5 and 0.75, deviations are around – 8%, which is, however, within the mutual uncertainty of the data. These data are also presented in Figure 25, as deviations from our correlations.

The excess thermal conductivity of the mixture, $\lambda^E$, at a given temperature, can be defined as:

$$\lambda^E = \lambda_{mix} - x_w\lambda_w - x_{IL}\lambda_{IL} \tag{21}$$

where $\lambda_{mix}$, is the thermal conductivity of the mixture, $\lambda_{IL}$, the thermal conductivity of the pure ionic liquid, and, $\lambda_w$, that of pure water, at the same temperature. Figure 27 shows the results obtained, as a function of the mole fraction of water, for all the temperatures measured. It can be seen, an in accordance of the structure of the curves found for the thermal conductivity of the mixture, that the excess function has a weak temperature dependence up to $x_w$ ~ 0.8, this dependency increasing for water rich mixtures. The inset shows that the excess function, being negative for most of the composition range, becomes positive near $x_w$ ~ 0.99, for the three highest temperatures, a very interesting sign of the change in structural arrangement in the mixture.

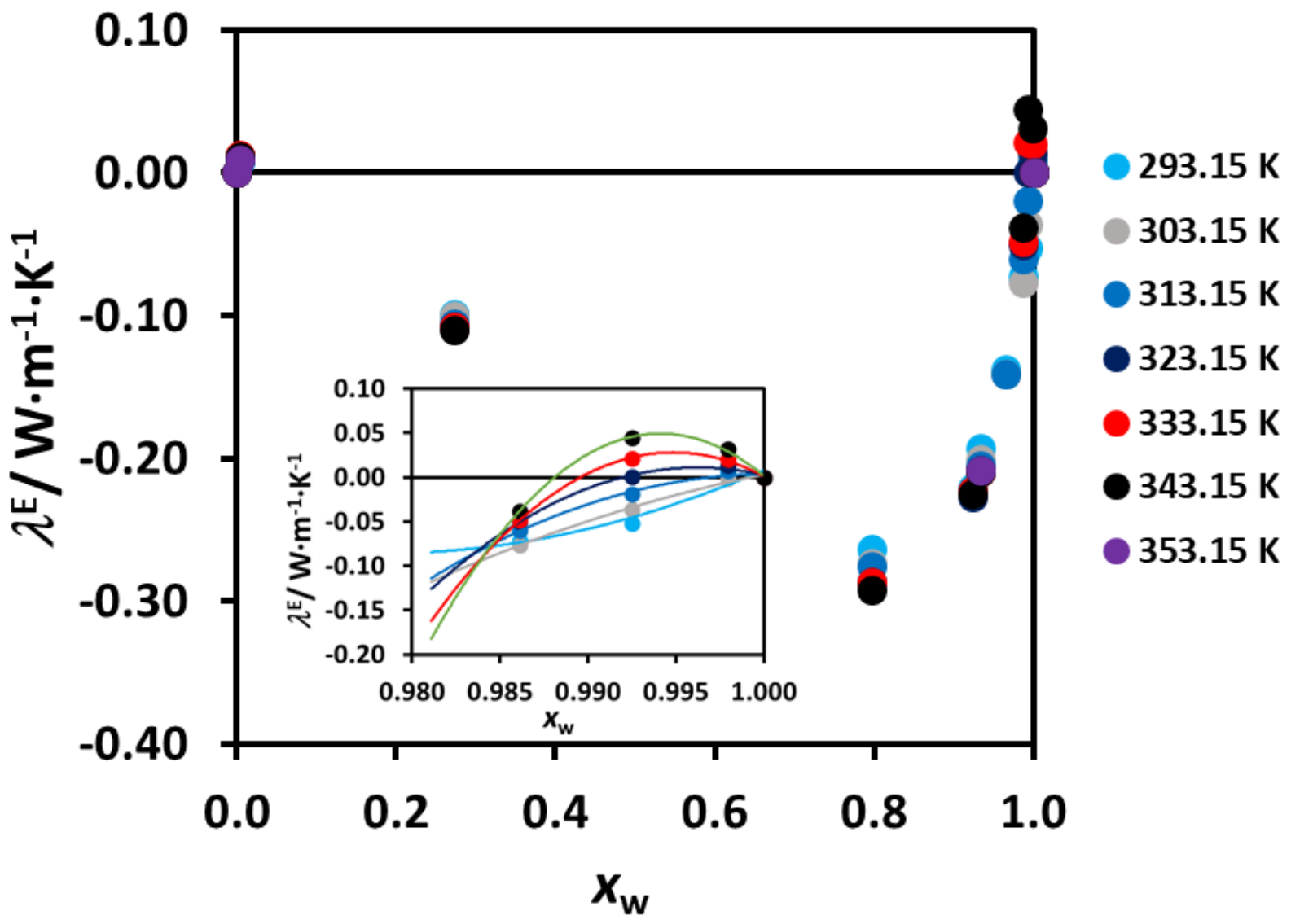

Figure 27 – Excess thermal conductivity for the aqueous [$C_2$mim][$CH_3SO_3$] system, as a function of the mole fraction of water, at the different temperatures. The inset shows the variation in the very narrow range $0.98 < x_w < 1.00$. Lines are just guiding for the eye.

**3.5 Refractive index.** The measured values for the refractive index of the [$C_2$mim][$CH_3SO_3$] + water mixtures are displayed in Table 6. The values for the "pure" ionic liquid were corrected to water free values, also presented in Table 6, by evaluating $\left(\frac{\partial n^{\mathrm{D}}}{\partial x_w}\right)_T$, a positive quantity, at each measuring temperature, and extrapolating for $x_{\mathrm{w}} = 0$. These values were fitted to a polynomial of the type:

$$n^{\mathrm{D}} = g_0 + g_1 T(\mathrm{K}) \tag{22}$$

with a standard deviation of 0.00021, at a 95% confidence level ($k = 2$). No data point deviates from the line by more than 0.0001. The coefficients and their standard deviations are $g_0$ = 1.580893± 0.0005914, $g_1$ = (-2.81342 ± 0.00185) x$10^{-3}$. The corrections for extrapolated water-free values were smaller than 0.0028, about 15 times the uncertainty of the measured values, justifying the corrections done. A value of $n^{\mathrm{D}}$ (298.15 K) = 1.49701 ± 0.00021 is recommended for the pure ionic liquid.

All the available data in the literature are shown in Table S18. Five data sets are available in the literature for the refractive index of [$C_2$mim][$CH_3SO_3$], Hasse et al. (2009)[60], Freire et al. (2011)[61], Seki et al. (2012)[62], and Shah et al. (2013)[63], measured with standard Abbé refractometers. All these authors used samples with small amounts of water, except Shah et al. (2013)[60], where samples with 0.38 % of water were used.

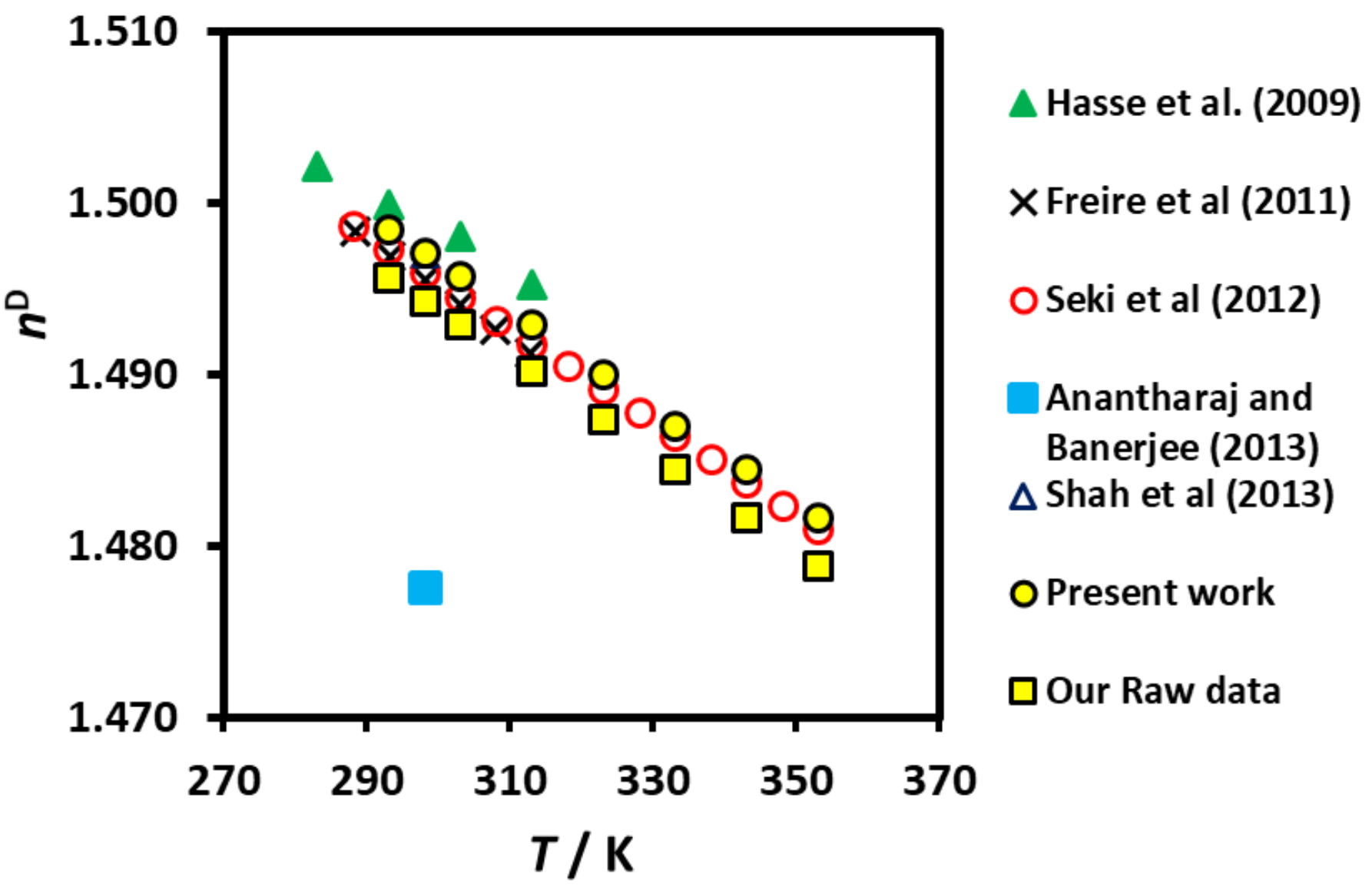

Figure 28 – Data for the refractive index of [$C_2$mim][$CH_3SO_3$] available in the literature, and present work. Our raw data is also presented, corresponding to water content of 0.997-1.945 wt %.

Figure 28 shows all the published data, corrected for the water content, our "water-free data", and our raw data, before the corrections for the water content. It is clear that the comparison between "water-free" data is very conclusive. Figure 29 shows deviations of these data, corrected for extrapolated water-free values, from our fit, eq. (22). All data deviates from our "water-free" data within their mutual uncertainty (max. 0.0008), the deviations never amounting to more than 0.0023. With the exception of Hasse et al. (2009)[57], all authors found approximately the same temperature coefficient (around $-1.8\times10^{-4}$), the difference being in the absolute values, possibly due to the calibration of the refractometers used[8].

For the mixtures of [$C_2$mim][$CH_3SO_3$] there is only one set of data available in the literature, at 298.15 K (see Table S17), that of Anantharaj and Banerjee (2013)[35]. Figure 30 shows the comparison of our data and it can be seen that the results agree within their mutual uncertainty, higher for these authors data ($U$ ($n$) = 0.026). Data of Queirós et al (2020)[8] for the refractive index of water agrees with present data also within their mutual uncertainty. The refractive index of the [$C_2$mim][$CH_3SO_3$] + water mixtures is displayed, as a function of the molar fraction in Fig. 31, for the different temperatures. The variation with temperature is greater on the IL side. The refractive index decreases smoothly up to $x_w \sim 0.6$, the derivative starting to be more negative afterwards, until it is very steep near pure water. The experimental data was fitted as quadratic function of the mass fraction for each temperature, for the ease of interpolation, eq. (S4). The coefficients, their standard deviations, and the deviations of the fit, varying between 0.00126 and 0.00203 RI units, are presented in Table S19, and Fig. S5 shows the deviations of the experimental points from the fit, at each temperature, not amounting by more than 0.02%. These fits were used to evaluate the derivative $\lim_{x_w\to 0}\left(\frac{\partial n^D}{\partial x_w}\right)_T = \lim_{w_w\to 0}\left(\frac{\partial n^D}{\partial w_w}\right)_T \times \frac{M_w}{M_{IL}}$, and the values obtained are also shown in Table S19 for further reference.

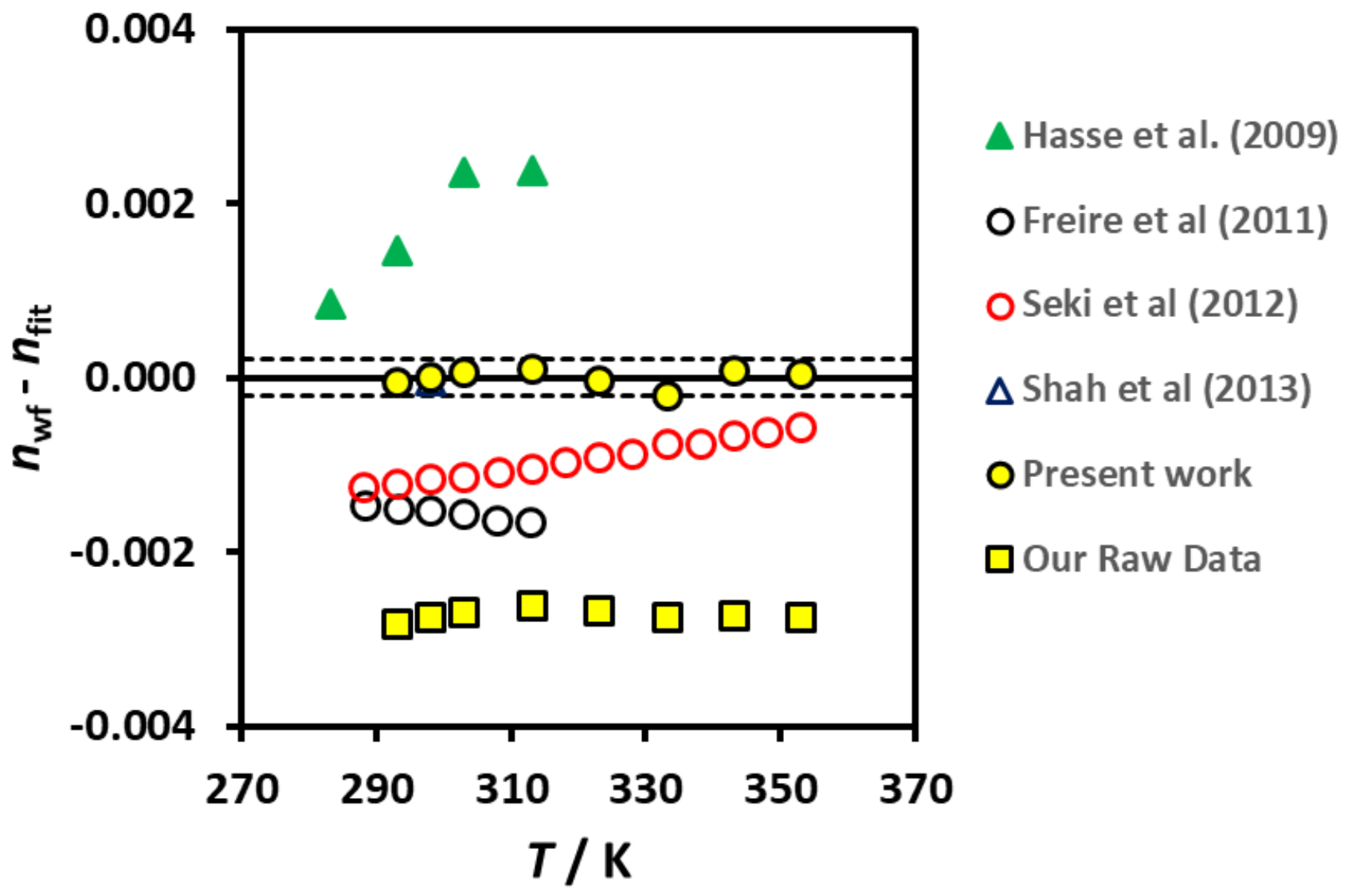

Figure 29 – Deviations from the refractive index experimental data, extrapolated to water-free values, of $[C_2mim][CH_3SO_3]$ from eq. (22), as a function of temperature. Also included are our raw data, not corrected for water content. Dashed line represents the standard deviation of eq. (22) at a 95% confidence level.

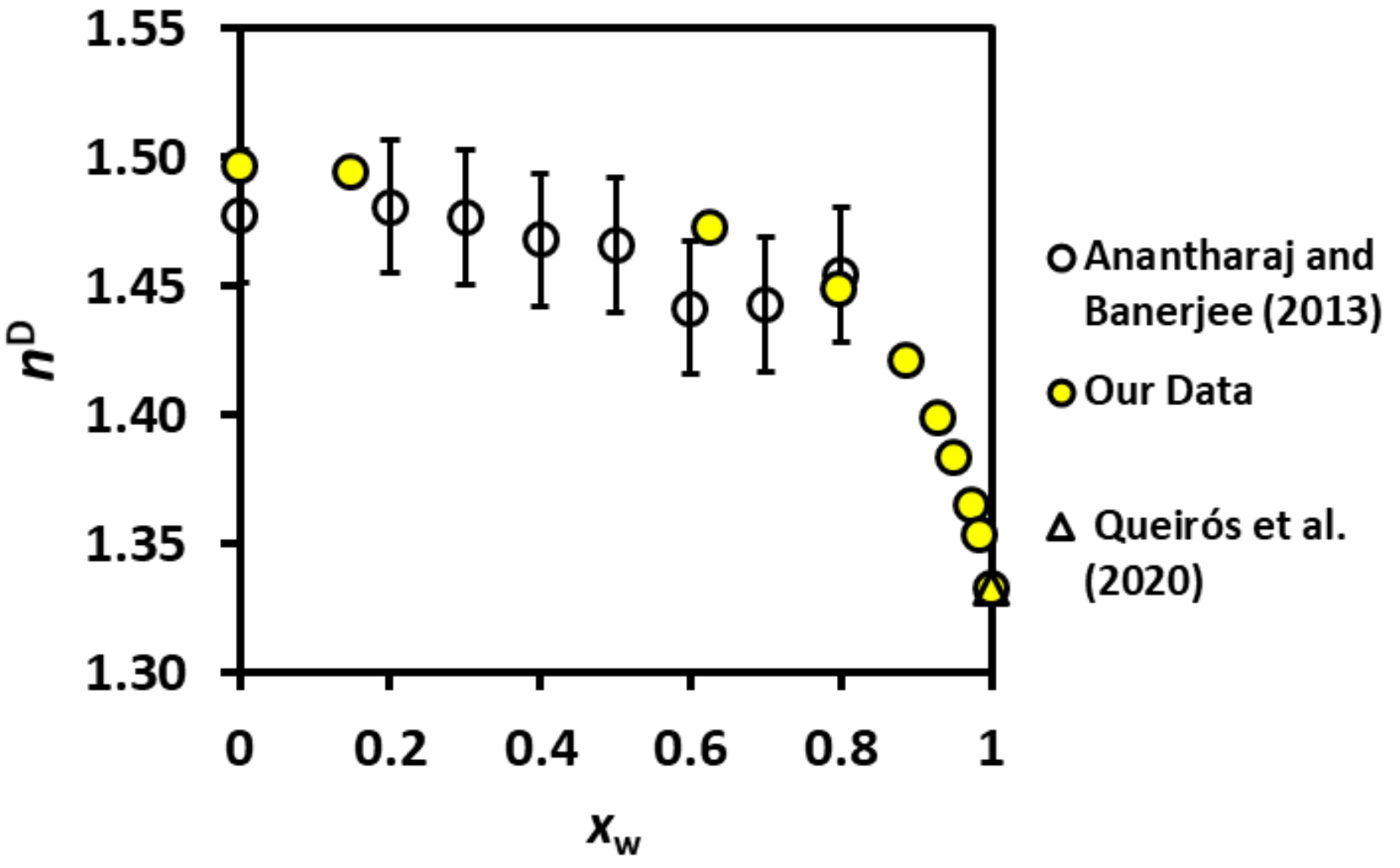

Figure 30 – The refractive index of the $[C_2mim][CH_3SO_3]$ + water mixture at 298.15 K. Data of Anantharaj and Banerjee (2013)[40] shows error bars, with the uncertainty at 95% confidence level. Also shown the data for pure water, Queirós et al. (2020)[8].

The refractive index of binary mixtures of ionic liquids can be interpreted by using the molecular polarization theory[8]. In the limit of zero frequency (which is not the case of our data), there is some possibility of considering it as a thermodynamic property and use solution thermodynamics theory as in density and thermal properties. Therefore, it can be assumed the existence of an excess refractive index, defined as:

**Table 6 – Refractive index (D-sodium line)[a,b], $n^{D}$, for $[C_2mim][CH_3SO_3]$ + water mixtures, in the temperature range of 293 K to 353 K, at $P$ = 0.1 MPa, as a function of the molar fraction of water, $x_w$ and calculated excess molar refractions, $R^{E}$**

| | $T$ = 293.15 K | | $T$ = 298.15 K | | $T$ = 303.15 K | | $T$ = 313.15 K | | $T$ = 323.15 K | | $T$ = 333.15 K | | $T$ = 343.15 K | | $T$ = 353.15 K | |
|---|---|---|---|---|---|---|---|---|---|---|---|---|---|---|---|---|
| $x_w$ | $n^{D}$ | $R^{E}$ / $cm^3$/mol | $n^{D}$ | $R^{E}$ / $cm^3$/mol | $n^{D}$ | $R^{E}$ / $cm^3$/mol | $n^{D}$ | $R^{E}$ / $cm^3$/mol | $n^{D}$ | $R^{E}$ / $cm^3$/mol | $n^{D}$ | $R^{E}$ / $cm^3$/mol | $n^{D}$ | $R^{E}$ / $cm^3$/mol | $n^{D}$ | $R^{E}$ / $cm^3$/mol |
| 1.0000[d] | 1.33299 | 0.00000 | 1.33249 | 0.00000 | 1.33192 | 0.00000 | 1.33052 | 0.00000 | 1.32887 | 0.00000 | 1.32703 | 0.00000 | 1.32500 | 0.00000 | 1.32268[e] | 0.00000 |
| 0.9834 | 1.35401 | -0.00472 | 1.35336 | -0.00449 | 1.35271 | -0.00373 | 1.35152 | 0.00153 | 1.35071 | 0.01379 | 1.35053 | 0.03554 | - | - | - | - |
| 0.9724 | 1.36584 | -0.00528 | 1.36510 | -0.00498 | 1.36436 | -0.00411 | 1.36321 | 0.00389 | 1.36256 | 0.02069 | 1.36208 | 0.04181 | - | - | - | - |
| 0.9511 | 1.38449 | -0.01922 | 1.38382 | -0.01585 | 1.38294 | -0.01483 | 1.38162 | -0.00475 | 1.38143 | 0.02382 | 1.38199[f] | 0.06515 | - | - | - | - |
| 0.9273 | 1.39977 | -0.05720 | 1.39870 | -0.05743 | 1.39751 | -0.05938 | 1.39563 | -0.05342 | 1.39459 | -0.03177 | 1.39419 | 0.00199 | 1.39429[f] | 0.04279 | - | - |
| 0.8872 | 1.42226 | -0.06360 | 1.42108 | -0.06312 | 1.41990 | -0.06252 | 1.41760 | -0.05860 | 1.41514 | -0.05475 | 1.41263 | -0.05022 | 1.41085 | -0.03513 | 1.40975 | 0.01438 |
| 0.7967 | 1.45009 | -0.10765 | 1.44877 | -0.10754 | 1.44748 | -0.10703 | 1.44489 | -0.10415 | 1.44223 | -0.09955 | 1.43964 | -0.09095 | 1.43710 | -0.08831 | 1.43461 | -0.06375 |
| 0.6248 | 1.47382 | -0.12322 | 1.47247 | -0.12464 | 1.47101 | -0.13013 | 1.46818 | -0.13548 | 1.46542 | -0.13188 | 1.46278 | -0.12097 | 1.46039 | -0.11454 | 1.45805 | -0.09720 |
| 0.1484 | 1.49559 | -0.07730 | 1.49425 | -0.07695 | 1.49291 | -0.07690 | 1.49016 | -0.07676 | 1.48729 | -0.07318 | 1.48440 | -0.06707 | 1.48162 | -0.08683 | 1.47879 | -0.15919 |
| 0.0000[c] | 1.49837 | 0.00000 | 1.49701 | 0.00000 | 1.49566 | 0.00000 | 1.49289 | 0.00000 | 1.48995 | 0.00000 | 1.48696 | 0.00000 | 1.48444 | 0.00000 | 1.48157 | 0.00000 |

[a] Standard uncertainty $u(T)$ = 0.01 K, $u(P)$ = 1 kPa, expanded uncertainty $U(n)$ = 0.0002, at a 95 % confidence level ($k$=2)

[b] Relative combined standard uncertainty, $u_{r,c}(x_w)$ = 0.002

[c] Extrapolated water-free values – pure ionic liquid

[d] Water values obtained in this work

[e] Extrapolated from our data

[f] Expanded uncertainty $U(n)$ = 0.0005, at a 95 % confidence level ($k$=2)

$$n^{\mathrm{E}} = n_{\mathrm{mix}} - n^{\mathrm{id}} = n_{\mathrm{mix}} - x_{\mathrm{w}}n_{\mathrm{w}} - x_{\mathrm{IL}}n_{\mathrm{IL}} \quad (23)$$

where $n^{\mathrm{id}}$ is the refractive index of an ideal mixture, assumed to be given by eq. (24):

$$n^{\mathrm{id}} = x_{\mathrm{w}}n_{\mathrm{w}} + x_{\mathrm{IL}}n_{\mathrm{IL}} \quad (24)$$

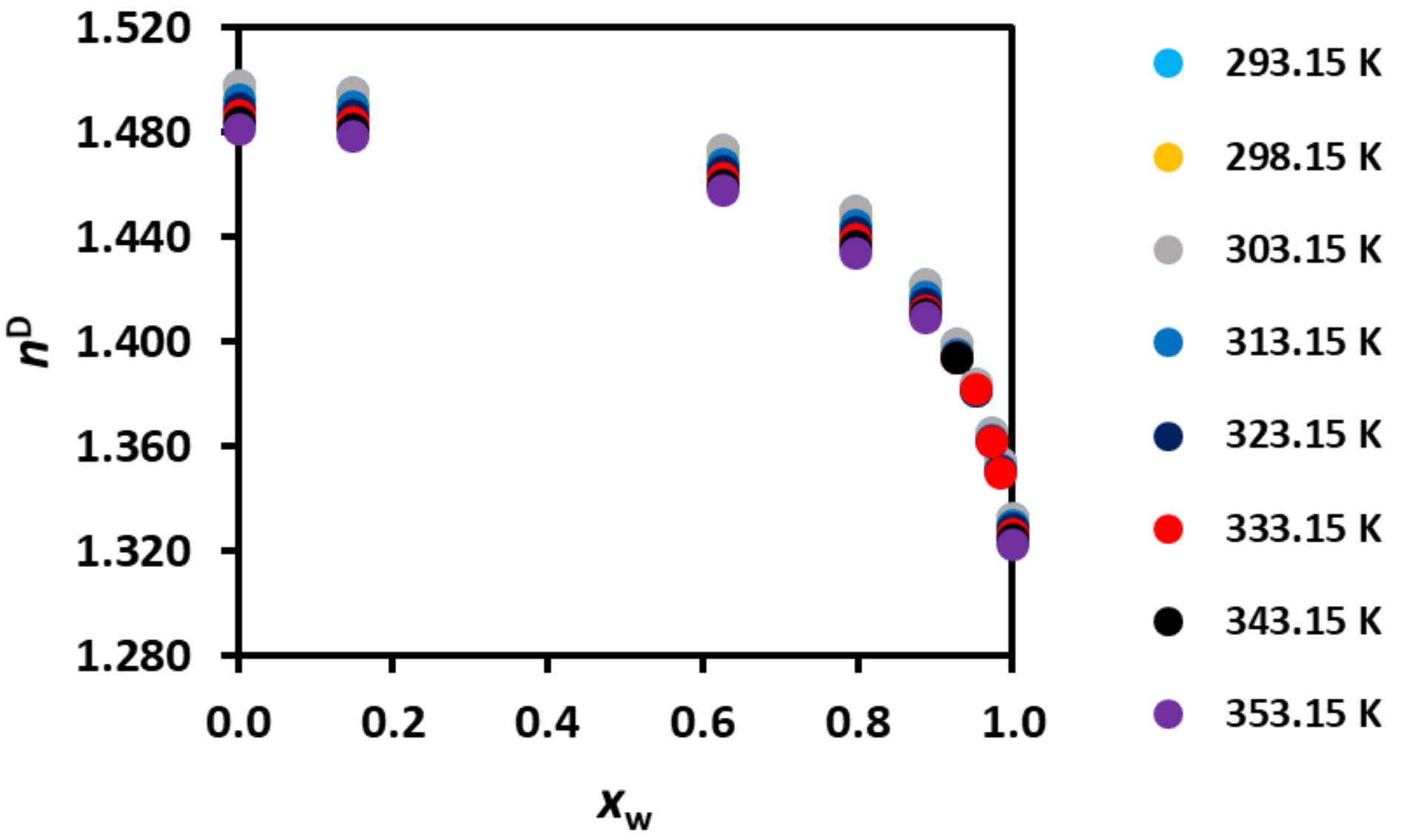

Figure 31 – The refractive index of the aqueous [$C_2$mim][$CH_3SO_3$] system as a function of the water mole fraction of the mixture, for the different temperatures studied.

Figure 32 shows the variation of the refractive index of each mixture, as a function of temperature. The variation is linear, as expected, but minor anomalies were found for the compositions ($x_{\mathrm{w}}$ = 0.9273 and higher). Trend linear lines were extended up to 353.15 K, and points at the temperature above 323.15 K, for these mixtures seem to have produced values slightly higher than the trend. For this reason, the expanded uncertainty of 2 data points, in Table 6 has been increased to $U(n)$ = 0.0005, at a 95 % confidence level ($k$=2), as explained in section 2.3.5. Further experimental points will be necessary to clarify this issue in a near future.

The excess refractive index, shown in Fig. 33, is very weakly dependent on temperature, as noted for other properties of this mixture, showing only a very small discrimination for $x_{\mathrm{w}} > 0.8$, at the higher temperatures.

As explained in our paper with results for the thermophysical properties of the aqueous [$C_2$mim][CH3COO][8], by using the concept of molar refraction, $R$, it is possible to define a refraction of the mixture, as $R$ is proportional to the volume of the mixture, as shown in eq. 25[64]:

$$R = \left(\frac{n^2 - 1}{n^2 + 2}\right)V_m \quad (25)$$

The molar refraction for the ideal mixture is additive, and can be calculated as:

$$R^{\mathrm{id}} = x_{\mathrm{w}}R_{\mathrm{w}} + x_{\mathrm{IL}}R_{\mathrm{IL}} \tag{26}$$

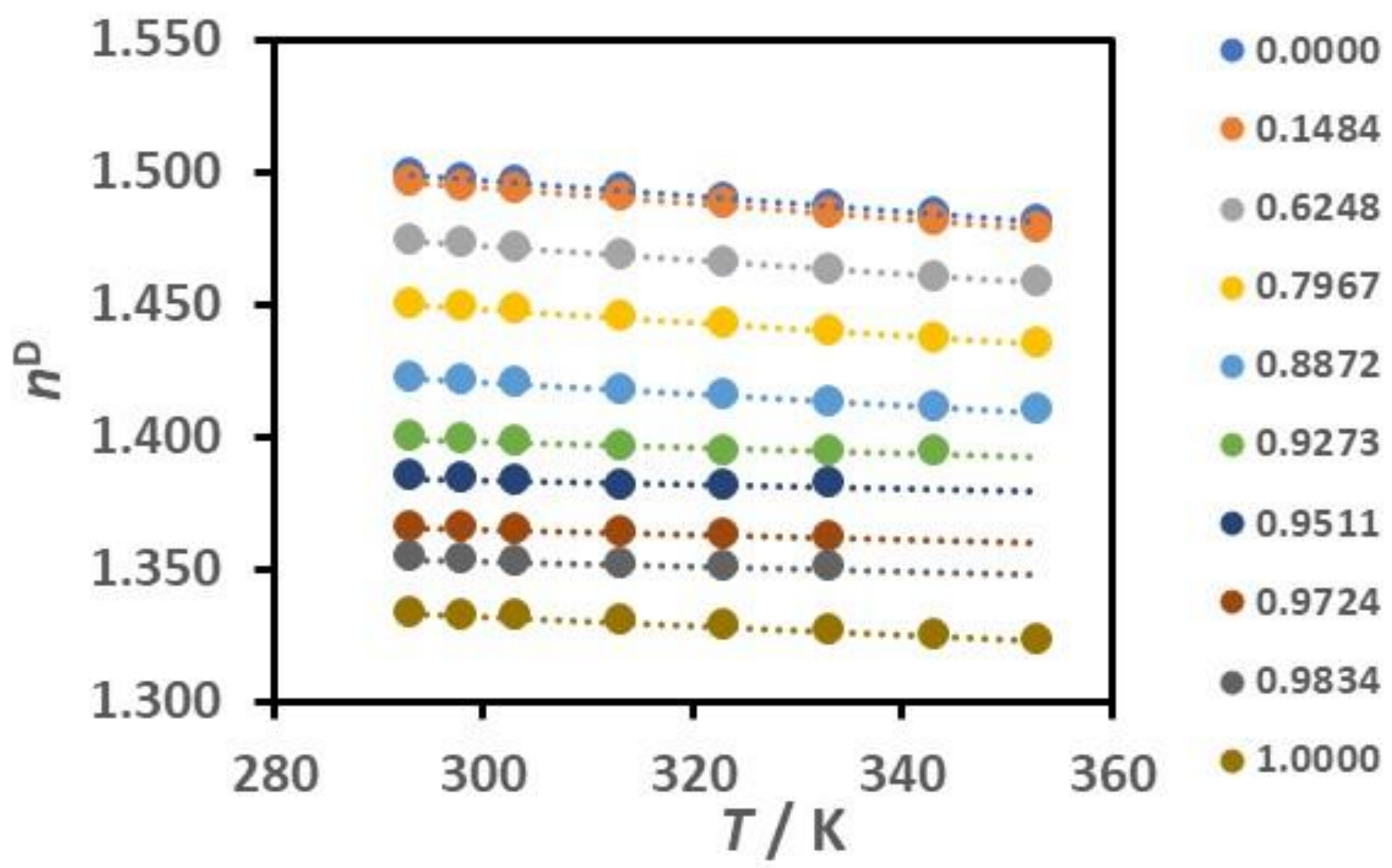

Figure 32 - The refractive index of the aqueous [$C_2$mim][$CH_3SO_3$] system as a function of temperature, for each water mole fraction studied.

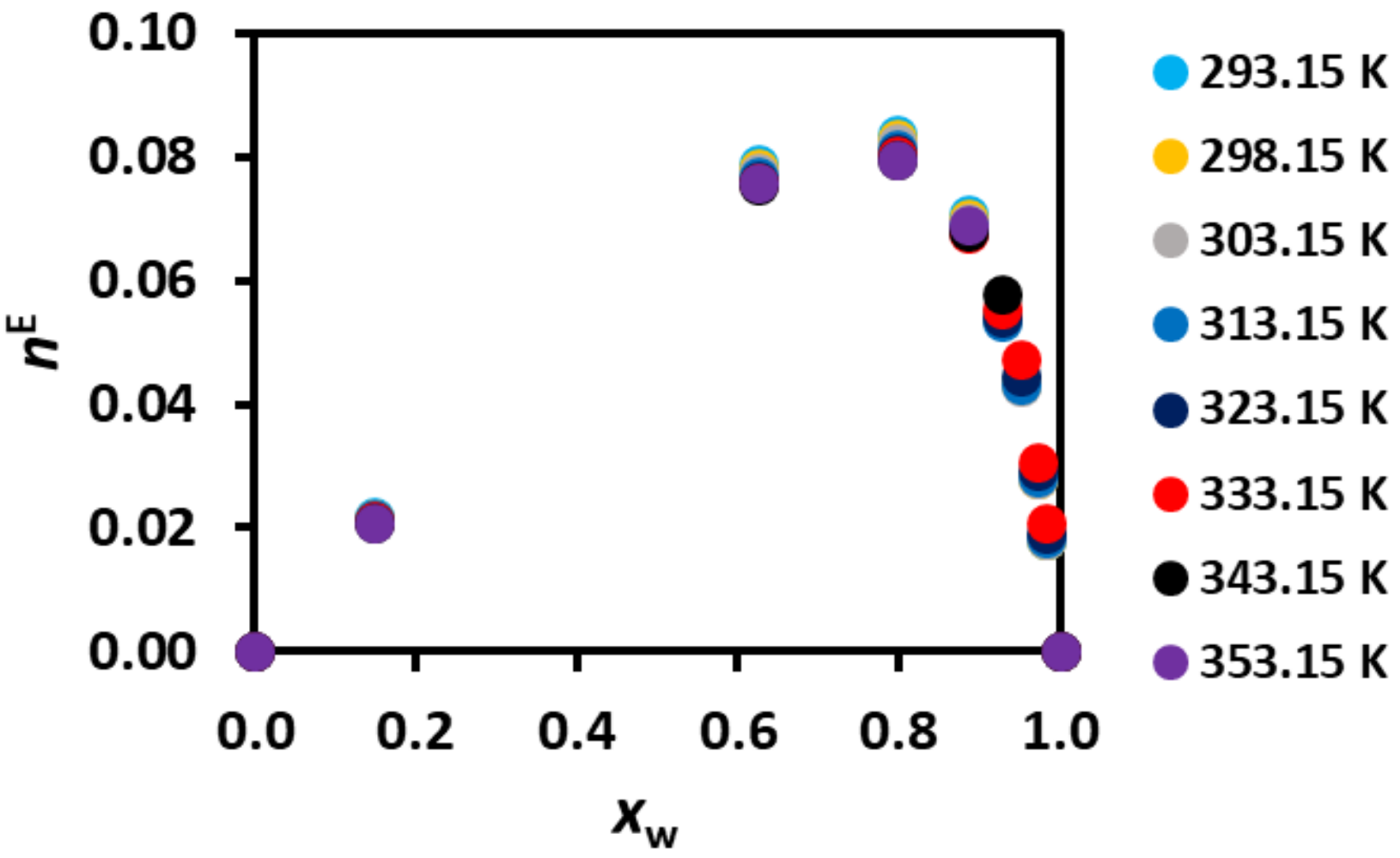

Figure 33 – The excess refractive index for the of the aqueous [$C_2$mim][$CH_3SO_3$] system as a function of the water mole fraction of the mixture, for the different temperatures studied.

The excess molar refraction of the mixture is then written as:

$$R^{\mathrm{E}} = R_{mix} - R^{\mathrm{id}} = R_{mix} - x_{\mathrm{w}}R_{\mathrm{w}} - x_{\mathrm{IL}}R_{\mathrm{IL}} \tag{27}$$

Figure 34 represents the excess molar refraction of the [$C_2$mim][$CH_3SO_3$] + water mixtures for the studied temperatures. Only 3 temperatures are displayed, including the extremes, since the different curves due to the collapse of on top of each other the different curves to due to the weak temperature dependence, especially for the IL liquid rich mixtures. Table 6 also displays the calculated values of the excess molar refraction in the studied temperature interval. Values for the refractive index of water at 353.15 had to be extrapolated from our data, by fitting the measured values up to 343.15 K to a quadratic dependence on the temperature, with a standard deviation of 0.00006 RI units, at a 95% confidence level ($k$ = 2). The coefficients of this equation, identical to eq. (22), with their standard deviations are $g_0$ = 1.263221 ± 0.011183, $g_1$ = (5.7888 ± 0.3293) x$10^{-4}$ and $g_2$ = (-1.1624 ± 0.0518) x$10^{-6}$. This table shows that the same behavior for rich water content is found, at the other temperatures. The excess molar refraction is negative in the whole concentration range, except for $T \geq 323.15$ K, where it becomes positive. The inset in the figure shows this behavior for $x_w$ > 0.90. This was found also for thermal conductivity, as explained above.

Although the definition of the ideal refractive index for a binary mixture, as a thermodynamic variable, has been the subject of several discussions, especially when the mixture components are molecularly very different[64], and a new definition for the refractive index of the ideal solution in terms of the volume fractions, based on its analogy with the permittivity of the ideal solution, was introduced by Reis et al. (2010)[65], the excess refractive index, calculated in terms of volume fractions, shows a maximum around $x_w$ = 0.3, for the temperatures up to 333.14, not being possible to confirm for the higher temperatures due to the lack of data. After $x_w \approx 0.5$ the excess refractive index tends to zero, except for the higher temperatures were the data seams to generate a maximum. However, the values found are always smaller than the combined uncertainty of the calculation of the excess refractive index, found to be 0.005, as shown in Fig. S6, except for the composition $\phi_w$ = 0.6778 ($x_w$ = 0.9511), probably due to unforeseen experimental errors As a whole, these facts also demonstrate a change of structure in the solution, complementing the results obtained for the other properties.

**3.6 Infinite dilution diffusion coefficient in water.** The infinite dilution diffusion coefficient of [$C_2$mim][$CH_3SO_3$] in water was determined at 298.15 K. As explained in previous papers, and according to the theory of the method[30], flow conditions were tested to assure that no convective effect arising from the diffusion tube curvature influenced the results for the molecular diffusion coefficient, as explained in previous works[8,31-35]. A selection of the measurements are shown in Table 7. In this table $\bar{t}$ and $\overline{\sigma}$ are, respectively, the corrected first and second moments of the distribution[30], and **Re** is the Reynolds number. Figure 35 shows the results obtained as a function of the flow rate, $\dot{V}$, for the mixture with an injection mole fraction, $x_{IL,inj}$ = 0.005, and we can see that almost all the points fall with the uncertainty bars, at 95% confidence level, and no significant dependence of the measured values with the volume flow rate. The value of the infinite dilution diffusion coefficient of [$C_2$mim][$CH_3SO_3$] in water, $D_{21}^{\infty}$ , was found to be $D_{21}^{\infty} = (1.022 \pm 0.049) \times 10^{-9}\ \mathrm{m^2 \cdot s^{-1}}$**.** The uncertainty range corresponds to the expanded uncertainty at 0.95 level of confidence, for 19 points ($k$ = 2.093), $U_r$ ($D$) = 4.8 %. These values are of the same order of magnitude than those obtained by us with other ionic liquids[8,35].

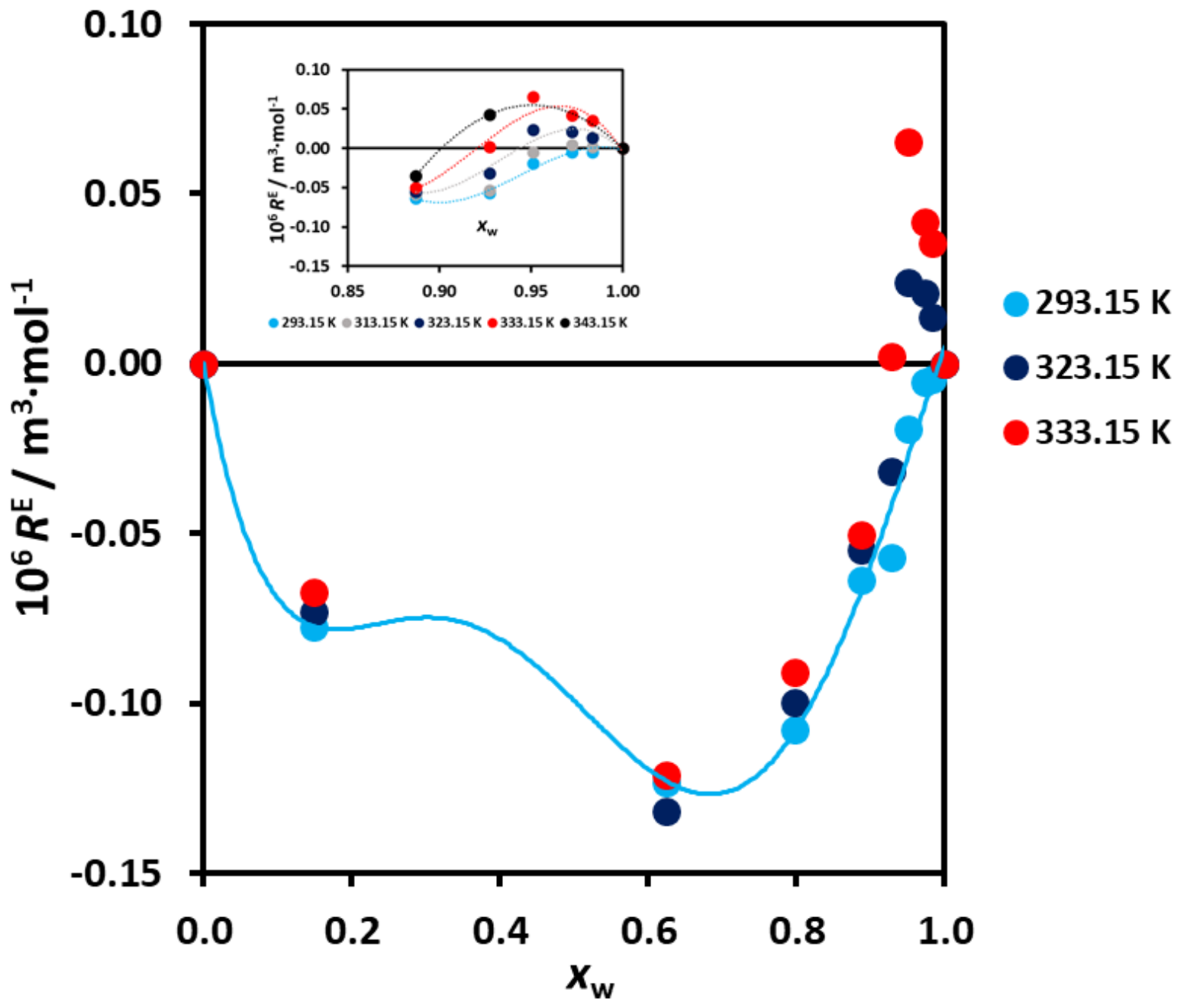

Figure 34 – The excess molar refraction for the of the aqueous [$C_2$mim][$CH_3SO_3$] system as a function of the water mole fraction of the mixture. Due to the collapse of the different curves to weak temperature dependence, only 3 temperatures are displayed. The inset shows more clearly the behavior of the system for $0.85 < x_w < 1.00$. Line is a tendency curve, to emphasize the composition dependence at 293.15 K.

**Table 7 - Infinite dilution diffusion coefficient[a], $D_{21}^{\infty}$, of [$C_2$mim][$CH_3SO_3$] in water at 298.15 K and 0.1 MPa[b], for $x_{IL,inj}$ = 0.005[c]. Values of the flow rate, first and second moments of the refractive index distribution and Reynolds number, are also presented.**

| Test | $\dot{V}$ / ml·h⁻¹ [d] | $\bar{t}$ /s[e] | $\bar{\sigma}$ /s[e] | $10^9 D_{21}^{\infty}$/ m²· s⁻¹ | Re |
|---|---|---|---|---|---|
| 1.2 | 10.0 | 2203.78 | 113.87 | 1.043 | 4 |
| 2.2 | 10.0 | 2204.36 | 116.52 | 1.047 | 4 |
| 4.2 | 7.0 | 2633.46 | 125.61 | 1.047 | 3 |
| 6.2 | 7.0 | 2706.71 | 125.82 | 1.032 | 3 |
| 9.2 | 7.0 | 2252.71 | 116.94 | 1.042 | 4 |
| 10.2 | 8.5 | 2208.81 | 115.54 | 1.045 | 4 |
| 11.2 | 8.5 | 2589.75 | 125.49 | 1.041 | 3 |
| 12.2 | 3.5 | 3127.40 | 137.60 | 0.990 | 3 |
| 13.2 | 5.0 | 3134.02 | 138.74 | 1.008 | 3 |
| 14.2 | 3.5 | 3145.23 | 138.28 | 1.025 | 3 |
| 16.2 | 5.0 | 4155.46 | 197.90 | 0.975 | 2 |
| 17.2 | 5.0 | 4399.75 | 164.36 | 0.995 | 2 |
| 18.2 | 3.5 | 6307.14 | 167.44 | 1.012 | 1 |
| 19.2 | 5.0 | 4417.50 | 198.91 | 1.011 | 2 |
| 20.2 | 3.5 | 6321.63 | 166.48 | 0.999 | 1 |
| 23.2 | 6.0 | 3712.47 | 152.80 | 1.008 | 2 |
| 24.2 | 6.0 | 3690.44 | 151.80 | 1.015 | 2 |
| 25.2 | 6.0 | 3692.31 | 148.22 | 1.065 | 2 |
| 26.2 | 6.0 | 3647.11 | 150.77 | 0.995 | 2 |

[a] Expanded relative uncertainty at 0.95 level of confidence ($k$=2.093) $U_r$ ($D$) = 4.9 %

[b] Standard uncertainty $u(T)$ = 0.02 K, $u(P)$ = 1 kPa

[c] Relative combined standard uncertainty, $u_{r,c}(x_{IL})$ = 0.0005

[d] Relative standard uncertainty of flow rate $u_r$ ($\dot{V}$) = 0.35 %

[c] Relative standard uncertainty first and second moments of the refractive index distribution, $u_r(\bar{t}) < 0.01$ % and $u_r(\bar{\sigma}) < 0.1$ %, respectively

The diffusion in ionic liquids and their water solutions is a very interesting subject in physical chemistry, as we have at least 3 molecular entities, the cation, the anion and the water molecule. With respect to a pure ionic liquid, a complete discussion has been performed in a recent paper[25] and in the papers by Harris (2019,2016)[56,66] and Harris and Kanakubo (2015)[67]. Ionic liquids cations can associate in ion-pairs and ion clusters/aggregates, which makes the ion movement very complicated to study, affecting all the transport properties, namely electrical conductivity and diffusion, the ionic liquid cannot be considered as a completely dissociated electrolyte (without solvent). The application of the Nernst-Einstein equation to relate the molar conductivity and the diffusion coefficient of the anions and the cations, permitting to quantify the decrease of conductivity by self-aggregation/clustering of the ions, cannot be applied, as no data for the binary diffusivities or self-diffusion coefficients of the ions are available, namely to methanesulfonate.

We are then restricted to analyze a water predominant fluid, like the nearly infinite dilution of the ionic liquid. We can consider the mixture an electrolyte solution, with completely dissociated ions, and relate the infinite dilution diffusion with the electrical conductivity, via the Nernst-Haskell equation[68]:

$$D_{12}^{\infty} = \frac{RT}{F^2}\left(\frac{|z_+| + |z_-|}{|z_+ z_-|}\right)\frac{\Lambda_+^{\infty} \times \Lambda_-^{\infty}}{\Lambda_+^{\infty} + \Lambda_-^{\infty}} \tag{28}$$

In this equation $R$ is the universal gas constant, $T$ the absolute temperature in K, $F$ the Faraday constant, $z_+$ and $z_-$ the ions electrical charge, and $\Lambda_+^{\infty}$ and $\Lambda_-^{\infty}$ are, respectively the infinite dilution electrical conductivities of the cation and the anion.

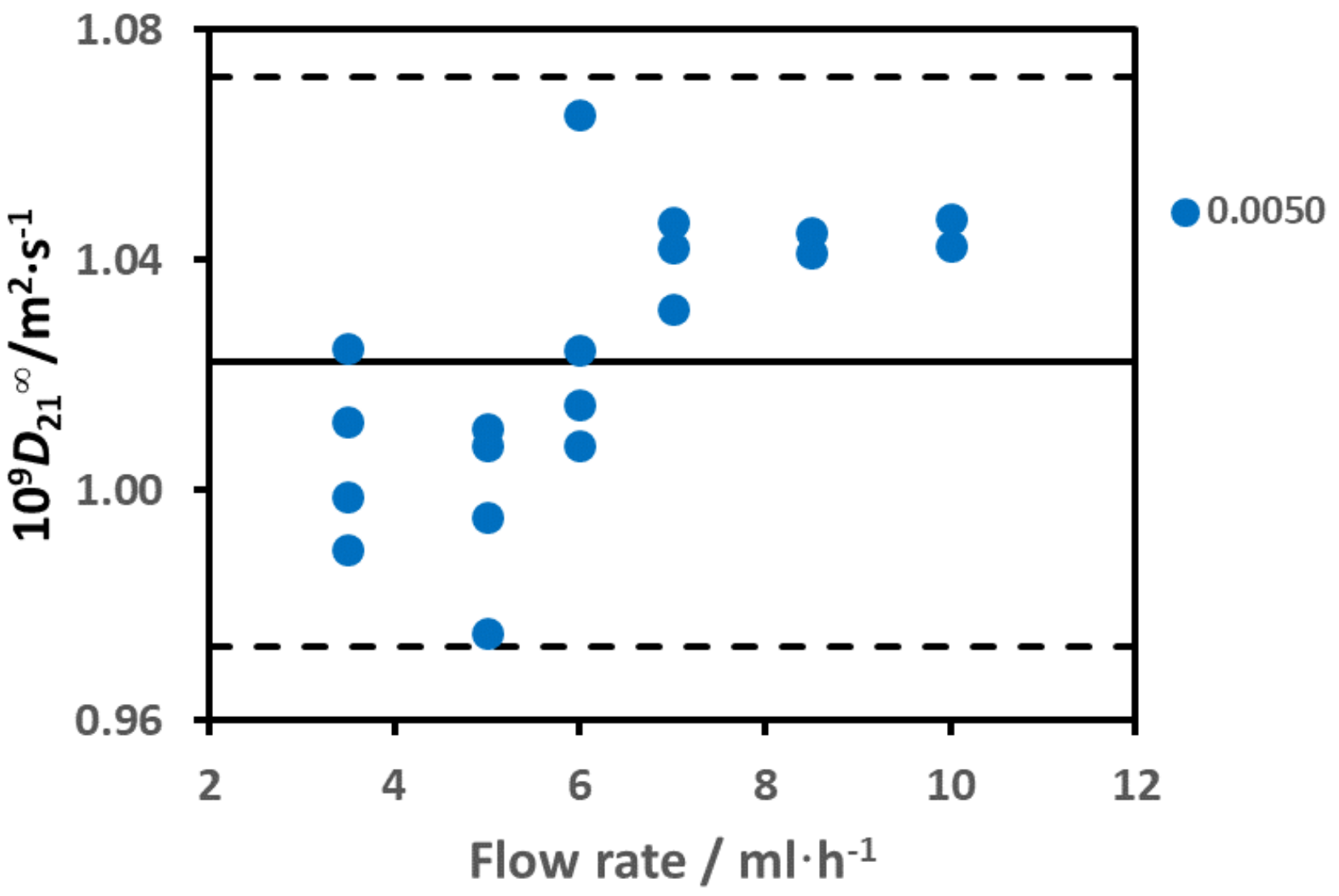

Figure 35 - Experimental measurements of the infinite dilution coefficient of $[C_2mim][CH_3SO_3]$ in water, as a function of the volume flow rate, at 298.15 K. The average value is denoted by the full line, while the dashed lines represent the limits of the standard deviation, at the 95% confidence level ($k$ = 2.093).

The value of $[C_2mim]^+$ was obtained for the temperature of 293.15 K by Bioucas et al. (2018)[2], as $\Lambda_+^\infty = 38.9 \pm 2.4$ S·cm$^2$·mol$^{-1}$, and corrected to 298.15 K, using Wong et al. (2008) data[69] between 303.2 and 323.2 K and ours to interpolate. The value obtained was $\Lambda_+^\infty = 43.9$ S·cm$^2$·mol$^{-1}$, while that of $[CH_3SO_3]^-$ was found from Dawson et al. (1950)[70] as $\Lambda_-^\infty = 42.5$ S·cm$^2$·mol$^{-1}$. Using eq. (28), the theoretical calculation gives for the infinite dilution diffusion coefficient $D_{21}^\infty = 1.15 \times 10^{-9}\ \mathrm{m^2 \cdot s^{-1}}$, a value that is 10.9 % greater than the experimental value. Due to possible uncertainties in the value of the infinite dilution electrical conductivity of the anion[70], it is reasonable to conclude that the directly measured experimental diffusion coefficient agrees with the Nernst-Haskell equation within their mutual uncertainty. This allows us to conclude, that the cation and the anion of the ionic liquid are free to move, and that the mixture, at this IL concentration, can be considered a dilute electrolyte solution, where the ions are completely dissociated, but solvated by water molecules.

Assuming this last conclusion is true, we can obtain an estimate the cation, $D_+$, and the anion, $D_-$, self-diffusion coefficients, by using the Nernst-Hartley limiting expression, eq. (29)[64]:

$$D = \frac{D_+ D_- (q_+^2 + q_-^2)}{q_+^2 D_+ + q_-^2 D_-} = \frac{2 D_+ D_-}{D_+ + D_-} = D_{NH}^\infty \tag{29}$$

where $q_+$ and $q_-$ are, respectively, the charges of the cation and the anion, ±1 in the case of $[C_2mim][CH_3SO_3]$. Assuming $D_{NH}^\infty$ as the experimental determined coefficient, it is possible, if we know the value for the cation, to estimate that for the anion and vice-versa. For the cation $[C_2mim]^+$, Sarraute et. al. (2009)[71] calculated the cation self-diffusion, with $[(CF_3SO_2)_2N]^-$ in several solvents (water, acrylonitrile and methanol), having found a value of $D_-$= 0.89 x10$^{-9}$ m$^2$·s$^{-1}$ at 298 K, for the water mixture. Heintz et. al. (2011)[72] used also $[(CF_3SO_2)_2N]^-$ in water and methanol, obtaining $D_-$ = 0.92 x10$^{-9}$ m$^2$·s$^{-1}$ for the water mixtures, a very similar value. We can assume an average value of $D_-$ ($[(CF_3SO_2)_2N]^-$) = (0.91 ± 0.02) x10$^{-9}$ m$^2$·s$^{-1}$ at 298 K. Using our experimental result of $D_{21}^\infty = (1.022 \pm 0.049) \times 10^{-9}\ \mathrm{m^2 \cdot s^{-1}}$, we can obtain from eq. (29) a value of $D_-$ ($[CH_3SO_3]^-$) = (1.17 ± 0.04) x10$^{-9}$ m$^2$·s$^{-1}$ at 298 K. Using our previous data for $[C_2mim][C_2H_5SO_4]$[30], $D_{21}^\infty = (0.96 \pm 0.03) \times 10^{-9}\ \mathrm{m^2 \cdot s^{-1}}$, and for $[C_2mim][CH_3COO]$[8], $D_{21}^\infty = (1.00 \pm 0.10) \times 10^{-9}\ \mathrm{m^2 s^{-1}}$, we can obtain also the values for $D_-$ ($[C_2H_5SO_4]^-$) = (1.02 ± 0.04) x10$^{-9}$ m$^2$·s$^{-1}$ and $D_-$ ($[CH_3COO]^-$) = (1.12 ± 0.10) x 10$^{-9}$ m$^2$·s$^{-1}$. These results show an order for the infinite dilution diffusion coefficients of the anions, for $T$ = 298 K, $D_-$ ($[(CF_3SO_2)_2N]^-$) < $D_-$ ($[C_2H_5SO_4]^-$) < $D_-$ ($[CH_3COO]^-$) < $D_-$ ($[CH_3SO_3]^-$), a result that shows the higher mobility of the methane-sulphonate anion, compared to the bis(trifluoromethylsulfonyl)imide, ethylsulphate and acetate anions.

From data from other authors, we can obtain the anion infinite dilution diffusion coefficients for $[N(CN)_2]^-$, $[CF_3SO_3]^-$ and $[MDEGSO_4]^-$ from Wong et al. (2008)[73], $[PF_6]^-$ from Su et al. (2007)[74], for an approximate temperature, $T = 303.2$ K, and $[BF_4]^-$ and $[Cl]^-$ from Sarraute et. al. (2009)[71] at 298 K. Results are displayed in Table 8, and an extended order for them, around 300 K, is $D_-$ ($[(CF_3SO_2)_2N]^-$) < $D_-$ ($[C_2H_5SO_4]^-$) < $D_-$ ($[MDEGSO_4]^-$) ≈ $D_-$ ($[CH_3COO]^-$) < $D_-$ ($[CH_3SO_3]^-$) < $D_-$ ($[CF_3SO_3]^-$) < $D_-$ ($[BF_4]$) < $D_-$ ($[PF_6]^-$) ≈ $D_-$ ($[Cl^-]$) ≈ $D_-$ ($[N(CN)_2]^-$). A better view of the relative values of the diffusion coefficients of the different anions at infinite dilution in water can be seen in Fig. 36. Although there is some insight that heavier anions have smaller values of the limiting diffusion coefficient, there are exception. Attempts to correlate them with the ions molar volumes COSMO or van der Waals calculations[75] of the anions gave no evidence of such a correlation.

**Table 8 – Anion infinite dilution diffusion coefficients, calculated using the Nerst-Hartley eq. (29), in $10^{-9}$m$^2$·s$^{-1}$ [a]**

| $T$/K | $[CH_3SO_3]^-$ | $[CH_3COO]^-$ | $[C_2H_5SO_4]^-$ | $[BF_4]$ | $[Cl]^-$ | $[(CF_3SO_2)_2N]^-$ |
|---|---|---|---|---|---|---|
| 298 | 1.17 | 1.12[8] | 1.02[26] | 1.81[58] | 2.03[58] | 0.88[58]<br>0.94[59] |
| | $[N(CN)_2]^-$ | $[CF_3SO_3]^-$ | $[MDEGSO_4]^-$ | $[PF_6]^-$ | | |
| 303.2 | 2.04[60] | 1.60[60] | 1.10[60] | 2.02[61] | | |

[a] MDEGSO4 is diethylene glycol monoethylether, or 2-(2-Ethoxyethoxy)ethanol, CAS N# 111-90-0

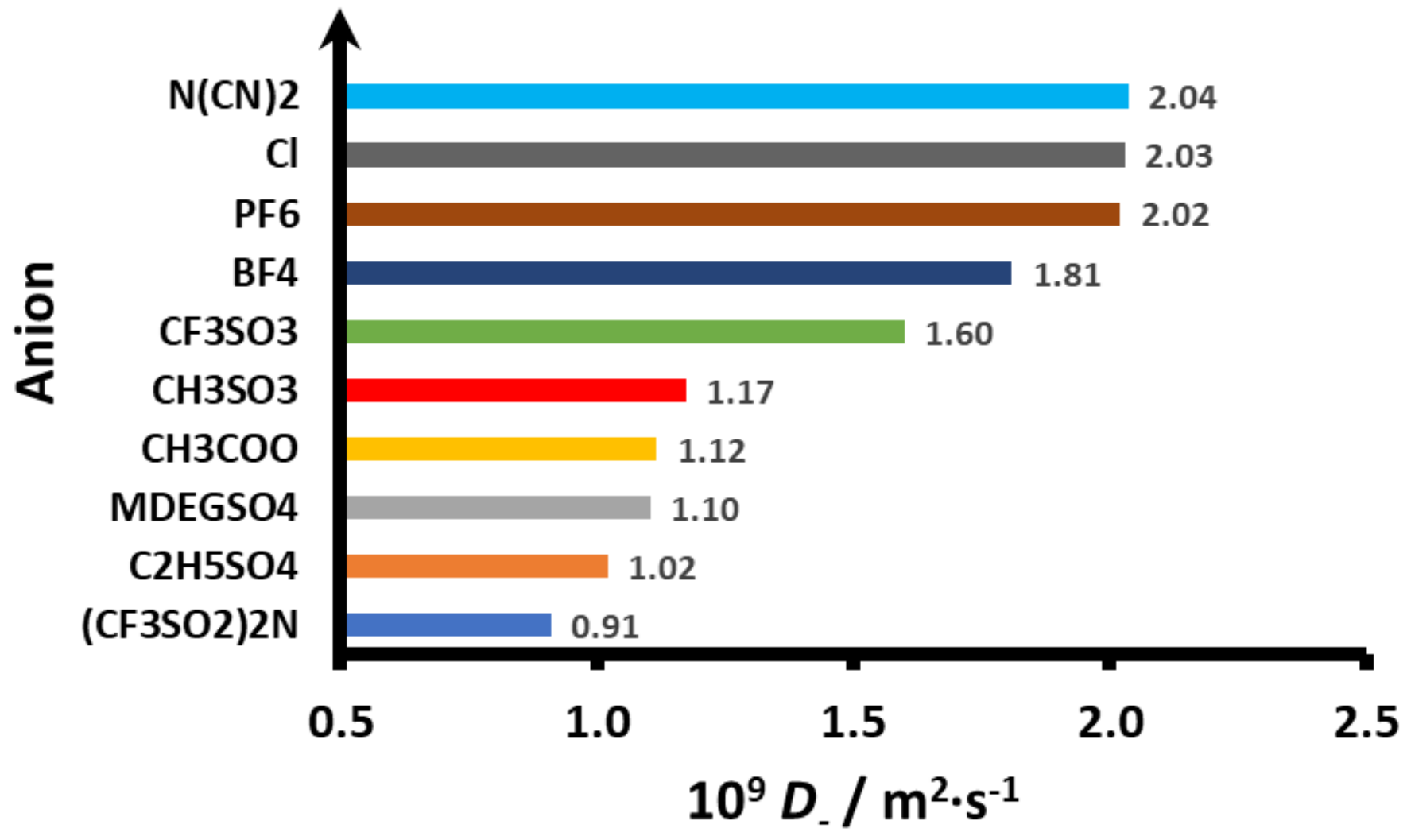

Figure 36 - Comparison of the diffusion coefficients of the different anions at infinite dilution in water, calculated by Nernst-Haskell equation, eq. (29).

The data obtained for the ionic liquid limiting molar conductivity also permits the use of the Nernst-Einstein equation[64] to calculate the electrical conductivity of the mixture. This equation reads, for a 1:1 electrolyte:

$$\Lambda_{\mathrm{NE}} = \frac{F^2}{RT}\left(q_+^2 D_+ + q_-^2 D_-\right) = \frac{F^2}{RT}\left(D_+ + D_-\right) \tag{30}$$

$$Y = \frac{\Lambda}{\Lambda_{\mathrm{NE}}} = (1 - \Delta)$$

In eq. (30) $F$ is the Faraday and $R$ is the gas constant. $Y$ is the ration between the experimental molar conductivity and the value given by the Nernst-Einstein equation, strictly valid only at the infinite dilution of the ionic liquid ions in water. The function $\Delta$ is a measure of the ion aggregation in the mixture, tending to be zero at infinite dilution. Figure 37 shows the variation of $Y$ and $\Delta$ with the molar fraction of water, the ion aggregation starting to be significant for $x_w < 0.8$, as expected (see following section and Fig. 37).

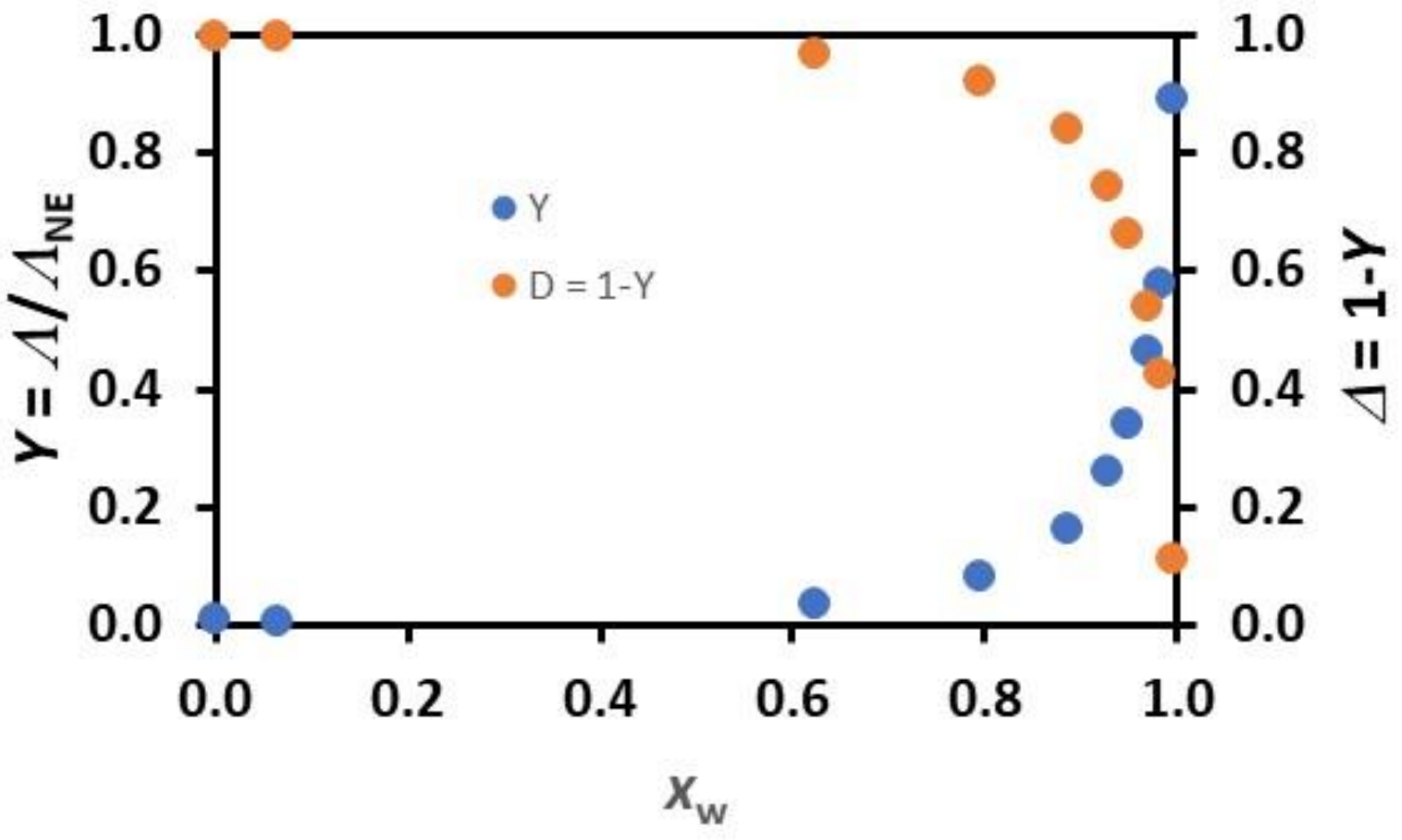

Figure 37– Ratio between the molar electrical conductivity and the Nernst-Einstein prediction as a function of the molar fraction of water. Secondary axis shows the function $\Delta$.

**3.7 New Insights in the Structure of IL + Water Mixtures.** It is clear from the previous results that the structure of the [$C_2$mim][$CH_3SO_3$] + water mixtures, whatever it may be, influences differently the various thermophysical properties of the mixture, either thermodynamic (equilibrium), transport or optical properties. The amount of data herein provided, in addition to the results obtained for the [$C_2$mim][$CH_3$COO] + water[8] , [$C_2$mpyr][N(CN)$_2$] + water[10], [$C_2$mim][N(CN)$_2$] and [$C_2$mim][$C_2H_5SO_4$], to be reported soon, in conjunction with data available based on correlation spectroscopy and computer simulations for this and similar mixtures of hydrophilic ionic liquids with water, like, [$C_2$mim][$C_2H_5SO_4$][4], and [$C_4$mim][$BF_4$][5,6] and [$C_2$mim][$CH_3$COO][7,8] and this ionic liquid[32], using surface tension and acoustic impedance and acoustic relaxation times.

The initial structure of the ionic liquid starts to break with the addition of water molecules, much smaller than the cation and anion, passes through a competition between ion clusters of different size, including cations, anions and water molecules, to finish in water strings embracing the cation/anion clusters, up to individual solvation of the ions by water molecules, as in a dilute electrolyte solution.

Zhang et al. (2008)[5], using 2D vibrational spectroscopy and the ionic liquid [$C_4$mim] [$BF_4$], arrived at the conclusion that, when water is added to the pure ionic liquid, having a structure dominated by an extended H-bond anion-cation 3D network, 3 domains appear: *First*, the few isolated water molecules start to interfere and destroy this network, originating a *second* zone of increasing size ionic clusters. This agrees with the observed fact that the density decreases slightly to accommodate the water molecules and cluster growing, the speed of sound and thermal conductivity increase slightly, isentropic compressibility also decreases, up to $x_w \approx 0.80$, and that the molar electrical conductivity starts to decrease. Further addition of water molecules will destroy these ionic clusters, giving rise to ion-pairs solvated by molecules of water, the *third* zone, until complete dissolution of the ionic liquid in water. This is well seen in the behaviour of the fall in density, speed of sound, increase in electrical and thermal conductivity, refractive index and excess molar isobaric expansion and molar refraction.

Bernardes et al. (2011)[4] studied the structure of aqueous solutions of 1-ethyl-3-methylimidazolium ethylsulfate, [$C_2$mim][$C_2H_5SO_4$], in the full concentration range, using molecular dynamics. They concluded that in the *first zone* the water molecules are isolated and dispersed in the IL structure ($x_w < 0.5$), in the *second zone*, small chain-like linear water aggregates form ($0.5 < x_w < 0.8$). These water molecules chains continue to grow with the addition of water molecules and start to engulf the ionic liquid network, their ion clusters, *third zone*, until the IL network collapses completely, $x_w > 0.95$, creating the known solvated cations and anions, in a *forth zone*. These conclusions agree qualitatively with Zhang et al. (2008), clarifying, in more detail, the difference in behavior for $x_w > 0.95$.

Zhong et al. (2012)[6] studied the self-diffusion coefficients of water, in the same system than Zhang et al. (2008)[5], the [$C_4$mim][$BF_4$] + water mixtures using molecular dynamics simulations. These authors found again three distinct regions, reflected in totally different slopes in the variation of ln $D_{H_2O}$ with the mole fraction of water, $x_w$, increasing with it. For $x_w < 0.2$, most water molecules are isolated from each other and experience a local environment in the polar network nearly the same as that in pure ionic liquid. In the concentration range of $0.2 < x_w < 0.8$, water molecules tend to form clusters by self-aggregation. The size of the clusters increases with the mole fraction of water, slightly disturbing the polar network of IL. When there are enough water molecules, small clusters are all connected with each other to form a single large cluster, and the IL is percolated by water molecules at $x_w = 0.8$ (complete solvation). In fact, it seems that percolation occurs later, in the *fourth zone* defined by Bernardes et al. (2011)[4]. However, these results support the initial conclusions of Zhang et al. (2008)[5].

Niazi et al. (2013)[76] studied the system [$C_2$mim] [$CH_3COO$] + water also by simulation, agreeing in general terms with the behaviour explained above, also supported by our investigation on the thermophysical properties of this system[8].

Marium et al. (2017)[18] where the only authors that studied the experimental properties of our system in an extensive way, and comparison with their data has been discussed above. They also report measurements of the surface tension, showing that this property decreased sharply as the ionic liquid was added to water (Fig. 11)[58], up to $x_{IL} \approx 0.014$ ($x_w \approx 0.986$), from the water value around 72 $mN \cdot m^{-1}$ to 42.5 $mN \cdot m^{-1}$, becoming almost constant, slightly decreasing, up to the value for the pure ionic liquid, found to be around 41 $mN \cdot m^{-1}$. Sung et al. (2005)[77] , investigated the surface molecular structure of [$C_4$mim][$BF_4$] + water mixture using surface tension measurement and surface SFG vibrational spectroscopy, and their study indicates that the liquid surface is mostly covered by the cations at very low bulk concentration of [$C_4$mim][$BF_4$], and anions start to appear at the surface from $x_{IL} \approx 0.016$ until they are populated about equally at $x_{IL} \approx 0.05$, suggesting that these anomalies could be related to some change in the bulk of [$C_4$mim][$BF_4$] and water. Their analysis of the IL + water mixture structure[32,68] agrees qualitatively with the preceding studies.

A new work on the possibility of CH···O hydrogen bonding in imidazolium-based ionic liquids using far-infrared spectroscopy measurements and DFT calculations shows that there is a possibility of formation of moderate H-bonds between the anion, the methanesulfonate, [$CH_3SO_3$], with the cation ring, namely at low temperatures[78]. These authors also state that H-bonded ion pairs are favored in energy over the dispersion-interaction-dominated pairs by ~8 kJ/mol for the sample with the methanesulfonate anion, in [$C_2$mim][$CH_3SO_3$]. This supports previous findings in Zone 1, and possibly helps to explain the metastability found for liquid phase of this compound below the melting point[18].

In conclusion, and complementing our previous study of the [$C_2$mim][$CH_3COO$] + water system[8], the properties of IL + water mixtures are very interesting, as the different cation-anion-water interactions play a different role that depends on the relative amount of IL and water present. This structural behaviour can be illustrated in the scheme of Fig. 37, where the different zones of IL + water interaction for our system discussed above are illustrated.

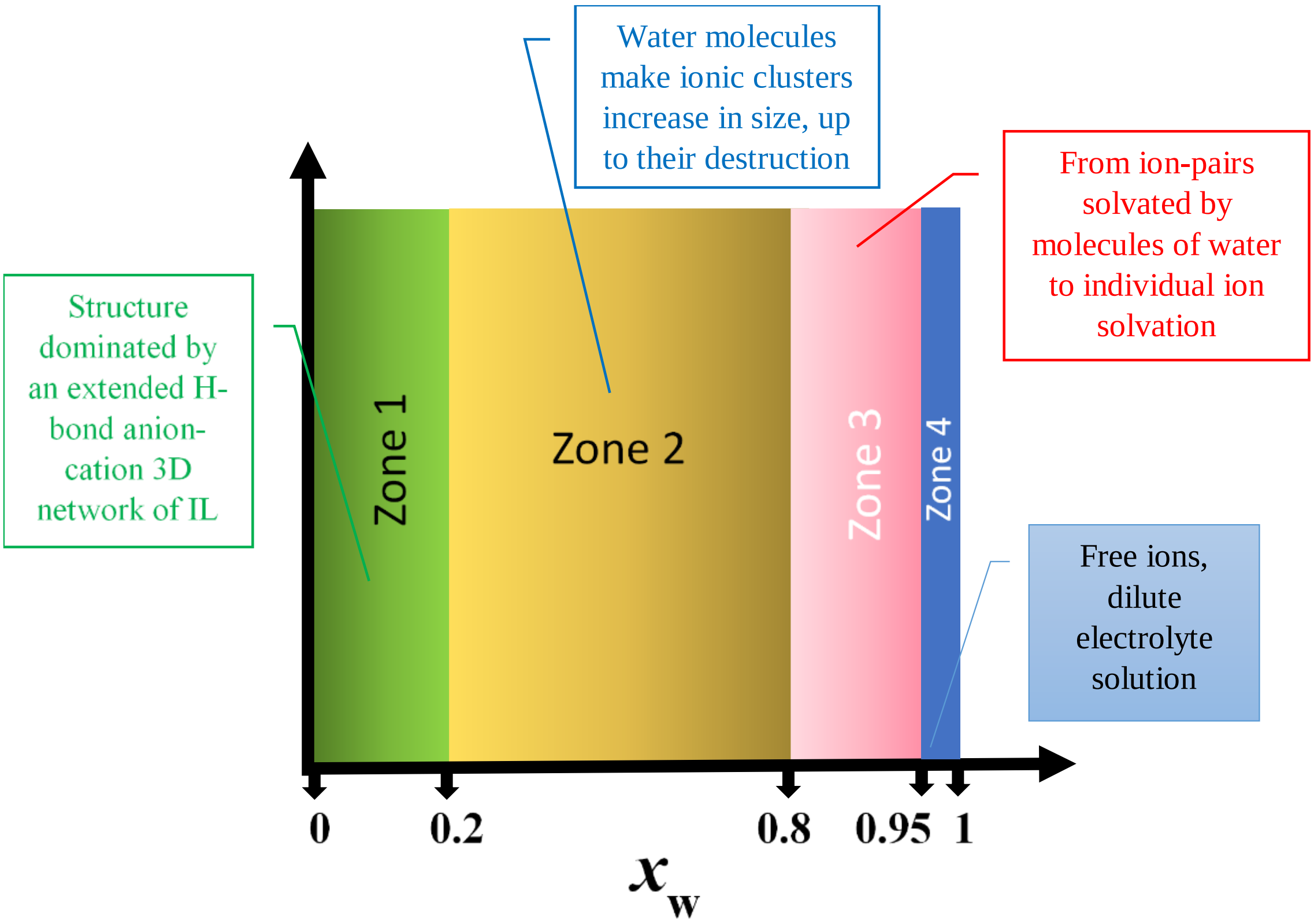

Figure 37 – Schematic representation for the different structural zones of the [$C_2$mim][$CH_3SO_3$] + water mixture. Scale in molar fraction is not proportional.

## 4. CONCLUSIONS

Extensive measurements of the thermophysical properties of [$C_2$mim][$CH_3SO_3$] + water mixtures have been performed, for 293.15 < T/K < 343.15 for density and speed of sound, for 293.15 < T/K < 353.15 for viscosity, electrical conductivity, thermal conductivity and refractive index), at $P$ = 0.1 MPa, in whole concentration range, complementing the previous data published for the pure ionic liquid[2]. Data for the infinite dilution coefficient of [$C_2$mim][$CH_3SO_3$] in water was obtained at 298.15 K. Extrapolated water-free values for the properties of the ionic liquid were obtained using methodologies previously applied. This work permitted to analyse, when appropriate, the quality of all published data for this system. From the principal measurements done on density, speed of sound, viscosity, electrical and thermal conductivity, refractive index and infinite dilution diffusion coefficients, it was possible to calculate several thermodynamic properties and cation and anion diffusion coefficients, while analysing the ionicity of the ionic liquid + water mixtures.

The results obtained demonstrate, as previously observed for [$C_2$mim][$CH_3COO$] + water system[8], uncommon behaviour of the properties, as the water content increases and the molar ratio varies, particularly near the diluted solution side of the ionic liquid. In the case of density, it decreases with the incorporation of water, as mostly ionic liquids, while the speed of sound increases up to a maximum value, decreasing then sharply to pure water values, and having a temperature coefficient changing from negative to positive for $x_w$ ~ 0.975. Viscosity and refractive index decrease with the mole fraction of water, while electrical conductivity increases enormously, by a factor of six, up to $x_w$ ~ 0.975, depending slightly on temperature, and then sudden decrease up to water levels. Thermal conductivity starts do have a small increase with water mole fraction, not varying very much with temperature, but having a sharp increase for $x_w$ ~ 0.9, and a maximum, around $x_w$ ~ 0.992, for temperatures 323.15 K and higher. The infinite dilution coefficient obtained at 298.15 K permitted the calculation of the cation value, and the values for 10 anions, including $[CH_3SO_3]^-$, the methane sulfonate anion, showing that $D_-$ ($[(CF_3SO_2)_2N]^-$) < $D_-$ ($[C_2H_5SO_4]^-$) < $D_-$ ($[MDEGSO_4]^-$) ≈ $D_-$ ($[CH_3COO]^-$) < $D_-$ ($[CH_3SO_3]^-$) < $D_-$ ($[CF_3SO_3]^-$) < $D_-$ ($[BF_4]$) < $D_-$ ($[PF_6]^-$) ≈ $D_-$ ($[Cl^-]$) ≈ $D_-$ ($[N(CN)_2]^-$), a very interesting result, that might explain the mobility of our anion, and the results obtained for the high ionicity of the ionic liquid studied (55 %), compared to other ionic liquids, as its ionic conductivity is only 26 % lower than that of 0.01M KCl, an excellent sign of its good electrical conductivity.

The excess thermodynamic and transport properties calculated, along with existing literature data, spectroscopic, properties and molecular simulation, permit to understand better the structure of the studied mixtures, and the influence of composition and temperature. The structural behaviour as a function of the mole fraction of water was discussed for the system studied, and an interpretation of the zones where there is a predominance of the IL structure, the existence of ionic clusters, less or more enveloped by water molecule chains, the destruction of the IL structure and the solvation by water, up to dilute electrolyte solution behaviour showed to be compatible with the experimental data obtained.

Up to the authors knowledge, this is the most comprehensive study of the thermophysical properties of $[C_2mim][CH_3SO_3]$ + water mixtures, as a function of composition and temperature, including high quality experimental measurements. It provides complete information for future industrial plant applications, namely in the area of new engineering fluids and new battery electrolytes, as $[C_2mim][CH_3SO_3]$ has an electrical conductivity twice as big and a much higher ionicity, in comparison with [C2mim][CH3COO]. This will favor its use in many applications, like heat transfer fluids and battery electrolytes.

This study will be complemented with heat transfer studies in pilot heat exchangers, to decide about its suitability as a new and more efficient heat transfer fluid, the ionic liquid possibly also promoting better lubrication and protection/pickling effect on the surface of the metallic plates (usually stainless steel) of the heat exchanger. Preliminary results show that selected $[C_2mim][CH_3SO_3]$ + water mixtures have thermophysical properties that contributes to energy and capital savings in cylindrical tube heat exchangers, results that will be presented in Paper 3 of this series.

## 5. ACKNOWLEDGMENTS

The authors would like to thank BASF for supplying Basionics® ST35. Daniel Lozano-Martín thanks the University of Valladolid for its mobility grant. This work was partially supported by Centro de Química Estrutural - UID/QUI/00100/2013, UID/QUI/00100/2019, and UIDB/00100/2020, funded by FCT – Fundação para a Ciência e Tecnologia, Portugal.

**6. CONFLICTS OF INTEREST -** There are no conflicts of interest. Information disclosed permitted by BASF.

**7. SUPPLEMENTARY INFORMATION** This file (*Supplementary Information for Publication (ST35+water).pdf*) the data found in the literature for the experimental properties of the mixtures studied in this article, namely density, viscosity, electrical conductivity, and refractive index. It also contains calculated values for the isobaric thermal expansion coefficient, isentropic compressibility, excess molar volumes, the apparent molar volumes, the apparent molar isentropic compressions, the thermodynamic properties of pure water and pure [$C_2$mim][$CH_3SO_3$] used to calculate the excess properties of the mixtures, namely excess molar isobaric expansions, excess molar isentropic compressions, the Walden rule deviations, ionicity and excess refractive index. Tables with available experimental data for the mixtures are also presented. The coefficients of the correlations presented for the different properties are also displayed therein. This information is available free of charge via the Internet at http://pubs.acs.org/.

**8. AUTHOR CONTRIBUTIONS -** The experimental measurements were performed by FB, CSGPQ, DL-M, MF, XP and MJVL. MJVL, AFS, MLML, FJVS and IMSL supervised the experimental work and participated with CANC in the final interpretations and data discussion. FB, CSGPQ and XP participated in data availability, property correlation and comparison of available literature data. CANC coordinated with MJVL and KM the project and links with BASF. Carlos Nieto de Castro validated the results, promoting the necessary discussions with all the authors, and wrote the manuscript.

[6] Zhong, X.; Fan, Z.; Liu, Z.; Cao, D. Local structure evolution and its connection to thermodynamic and transport properties of 1-butyl-3-methylimidazolium tetrafluoroborate and water mixtures by molecular dynamics simulations. *J. Phys. Chem. B* **2012**, 116, 3249-3263. DOI: https://doi.org/10.1021/jp3001543

[7] Niazi, A.A.; Rabideau, B.D.; Ismail, A.E. Effects of water concentration on the structural and diffusion properties of imidazolium-based ionic liquid–water mixtures. *J. Phys. Chem. B* **2013**, 117, 1378-1388. DOI: https://doi.org/10.1021/jp3080496

[8] Queirós, C.S.G.P.; Paredes, X.; Avelino, T.F.S.; Bastos, D.E.N.; Ferreira, M.; Santos, F.J.V.; Santos, A.F.; Lopes, M.L.M.; Lourenço, M.J.V.; Pereira, H.; Nieto de Castro, C.A. The Influence of Water on the Thermophysical Properties of 1-Ethyl-3-methylimidazolium Acetate. *J. Mol. Liq.* **2020**, 297, 111925. DOI: https://doi.org/10.1016/j.molliq.2019.111925

[9] *BASF Safety data sheet* according to Regulation (EC) No. 1907/2006. Revised: 05.09.2016 Version: 10.0

[10] Cumicheo, M.C.; Nobre, L.C.S.; Santos, A.F.; Lampreia, I.M.S.; Santos, M.S.C.S.; Santos, F.J.V.; Segura, H.; Nieto de Castro, C.A. Thermophysical properties of 1-butyl-1-methyl-pyrrolidinium dicyanamide +H2O mixtures. *J. Chem. Eng. Data* **2015**, 60, 3766–3775. DOI: https://doi.org/10.1021/acs.jced.5b00944

[11] Lampreia, I.M.S.; Nieto de Castro, C.A. A new and reliable calibration method for vibrating tube densimeters over wide ranges of temperature and pressure, *J. Chem. Thermodyn.* **2011**, 43, 537–545. DOI: https://doi.org/10.1016/j.jct.2010.11.002

[12] Nobre, L.C.S; Cristino, A.F.; Santos, A.F.S.; Nieto de Castro, C.A.; Lampreia, I.M.S. Ultrasound speed study of the ternary liquid mixture (water + ethanol + 1-propanol) at T = 293.15 K and P = 0.1 MPa", *J. Chem. Thermodyn.*, **2020**, 150, 1062262020. DOI: https://doi.org/10.1016/j.jct.2020.106226

[13] Nobre, L.C.S; Santos, A.F.S.; Moita, M-L.C.J.; Lampreia, I.M.S. Ultrasound speeds and molar isentropic compressions of (3-butoxypropan-1-amine + water) mixtures from T = (283.15 to 303.15) K, *J. Chem. Thermodyn*., **2017**, 132, 220-226. DOI: https://doi.org/10.1016/j.jct.2017.05.010

[14] Cristino, A.F.; Nobre, L.C.S; Bioucas, F. E.B.; Santos, A.F.S.; Nieto de Castro, C.A.; Lampreia, I.M.S. Volumetric and sound speed study of aqueous butanol mixtures at different temperatures. *J. Chem. Thermodyn*., **2019**, 134, 127–135. DOI: https://doi.org/10.1016/j.jct.2019.03.006

[15] Dzida, M; Zorębski, E; Zorębski, M;Zarska, M; Geppert-Rybczyńska, M; Chorążewski, M; Jacquemin, J; Cibulka, I. Speed of sound and ultrasound absorption in ionic liquids. *Chem. Rev.* **2017**, 117, 3883–3929. DOI: https://doi.org/10.1021/acs.chemrev.5b00733

[16] Nieto de Castro, C.A.; Ribeiro, A.P.C.; Figueiras, A.O.; Langa, E.; Vieira, S.I.C.; Lourenço, M.J.V.; Santos, A.F.S.; Santos, F.J.V.; Lampreia, I.M.S.; Goodrich, P.; Hardacre, C. Thermophysical Properties of 1-butyl-3-methylimidazolium tris(pentafluoroethyl)trifluorophosphate, [C4mim][(C2F5)3PF3] and of its IoNanofluid with multi-walled carbon nanotubes. *J. Chem. Eng. Data* **2021**, 66, 1717−1729. DOI: https://doi.org/10.1021/acs.jced.0c01017

[17] Musiał, M; Zorębski, M; Dzida, M; Safarov, J; Zorębski, E; Hassel, E. High pressure speed of sound and related properties of 1-ethyl-3-methylimidazolium methanesulfonate. *J. Mol. Liq.,* **2019**, 276, 885–896. DOI: https://doi.org/10.1016/j.molliq.2018.12.005

[18] Marium, M.; Auni, A.; Rahman, M.M.: Mollah, M.Y.A.; Susan, M.A.B.H. Molecular level interactions between 1-ethyl-3-methylimidazolium methanesulphonate and water: Study of physicochemical properties with variation of temperature. *J. Mol. Liq.* **2017**, 225, 621–630. DOI: https://doi.org/10.1016/j.molliq.2016.11.074

[19] Singh, M.P; Mandal, S.K; Verma, I.L; Gupta, A.K; Singh, R.K; Chandra, S. Viscoelastic, surface, and volumetric properties of ionic liquids [BMIM][OcSO4], [BMIM][PF6], and [EMIM][MeSO3]. *J. Chem. Eng. Data* **2014**, 59, 2349−2359. DOI: https://doi.org/10.1021/je5000617

[20] Rowlinson, J.S.; Swinton, F.L. *Liquids and Liquid Mixtures*, 3rd ed., Butterworth Scientific, London, **1982,** Chapter 2. ISBN 978-0-408-24193-9. DOI: https://doi.org/10.1016/C2013-0-04175-8

[21] Spieweck, F.; Bettin, H. Review: Solid and liquid density determination / Übersicht: Bestimmung der Dichte von Festkörpern und Flüssigkeiten, *Tm - Tech. Mess.* **1992**, 59285–292. DOI: https://doi.org/10.1524/teme.1992.59.78.285

[22] Landolt-Börnstein: Neue Serie, Band 5, *Molekul Akustik Temperatura Skala von 1990*; PTB-Mitt, **1990**; Vol. 100, (n.d.). New edition 2008. ISBN: 978-3-540-85250-6. https://link.springer.com/book/10.1007%2F978-3-540-85251-3

[23] ISO 3696:1987.Water for analytical laboratory use - Specification and test methods. https://www.iso.org/standard/9169.html. Confirmed in 2018.

[24] ISO/TR 3666:1998(E). Viscosity of water. https://www.iso.org/standard/28607.html. Confirmed in 2019.

[25] Nunes, V.M.B.; Lourenço, M.J.V.; Santos, F.J.V.; Lopes, M.L.S.M.; Nieto de Castro, C.A. Accurate measurement of physicochemical properties on ionic liquids and molten salts. In: M. Gaune-Escard, K.R. Seddon (Eds.), *Molten Salts and Ionic Liquids. Never the Twain?*, John Wiley & Sons, Inc.: Hoboken, NJ, USA, **2012**: pp. 229–263. DOI: https://doi.org/10.1002/9780470947777.ch17